\documentclass[a4paper,11pt]{article}
\usepackage{jheppub}
\usepackage[utf8]{inputenc} 

\usepackage{subcaption}
\usepackage{amsmath}
\usepackage{todonotes}
\usepackage{paralist}
\usepackage{afterpage}

\newcommand{\llbracket}{\mathopen{[\![}}
\newcommand{\rrbracket}{\mathclose{]\!]}}

\usepackage{tikz}
\usetikzlibrary{math}
\usetikzlibrary{cd}
\usetikzlibrary{arrows.meta} 
\usetikzlibrary{decorations.pathreplacing}
\usetikzlibrary{calc}
\usetikzlibrary{patterns,shapes}
\usetikzlibrary{decorations.markings}
\usetikzlibrary{positioning}
\tikzset{
  mid arrow/.style={
    postaction={decorate},
    decoration={markings, mark=at position 0.5 with {\arrow{Latex}}}
  }
}
\tikzset{bd/.style={circle, draw=black, inner sep=0pt, fill=black, minimum size=2mm}}
\tikzset{wd/.style={circle, draw=black, inner sep=0pt, fill=white, minimum size=2mm}}
\tikzstyle{ligne}=[draw, very thick] 
\tikzstyle{brane}=[draw] 
\tikzset{sevenb/.style={circle, draw,fill=white,inner sep=2.5pt}}
\tikzstyle{gridline}=[draw, black!20!] 
\tikzset{hasse/.style={circle, fill,inner sep=2pt}}   
\tikzset{
  fivebrane/.style={
    draw=black,
    line width=0.75pt,
    line cap=round
  }
}

\tikzset{focalPoint/.style={circle, fill, inner sep=0pt,minimum size=4pt}}
\tikzset{kink/.style={thick}}
\tikzset{broken/.style={dashed}}
\tikzset{wall/.style={thick,decoration={markings,mark=at position 0.5 with{\arrow{latex}}},postaction=decorate}}
\tikzset{every node/.style={font=\footnotesize}}
\colorlet{goodgreen}{green!55!black}
\colorlet{goodred}{red}
\colorlet{goodblue}{blue!65!black}
\colorlet{goodorange}{orange!85!black}

\usepackage[section]{placeins}
\usepackage{needspace}

\newlength{\exampleleftmargin}
\newenvironment{example}[1][Example]
  {\par\FloatBarrier\addvspace{\baselineskip}\Needspace{6\baselineskip}%
   \list{}{\leftmargin=\exampleleftmargin \rightmargin=0pt
           \topsep=0pt \partopsep=0pt \itemsep=0pt
           \parsep=\parskip \listparindent=\parindent}%
   \item\relax
   \noindent{\rule{\linewidth}{0.4pt}}\par\nobreak
   \smallskip\noindent{\sffamily\bfseries#1}\par\nobreak
   \smallskip\sffamily\noindent\ignorespaces}
  {\par\FloatBarrier\nobreak\smallskip
   \noindent{\rule{\linewidth}{0.4pt}}\par
   \endlist
   \addvspace{\baselineskip}\ignorespacesafterend}

\newenvironment{examplefigure}
  {\par\medskip\noindent\begin{minipage}{\linewidth}\centering
   \captionsetup{type=figure}}
  {\end{minipage}\par\medskip}

\preprint{ }

\title{Generalized Toric Threefolds from Scattering Diagrams and Broken Lines}

\author[a]{Antoine Bourget,}
\author[b,c]{Ignacio Carre\~no Bolla,}
\author[a]{Quentin Lamouret}
\affiliation[a]{Institut de Physique Théorique, CEA, CNRS, Université Paris-Saclay, France}

\affiliation[b]{
Departamento de F\'isica,
Universidad de Oviedo,\\
c/ Leopoldo Calvo Sotelo 18, ES-33007, Oviedo, Spain}

\affiliation[c]{
Instituto de Ciencias y Tecnolog\'ias Espaciales de Asturias (ICTEA),\\
c/ Independencia 13, ES-33004, Oviedo, Spain}
\emailAdd{antoine.bourget@ipht.fr}
\emailAdd{ignaciocarbolla@gmail.com}
\emailAdd{quentin.lamouret@ipht.fr}

\abstract{We present an explicit and practical description of Generalized Toric Calabi--Yau threefolds, which are associated with five-brane webs ending on seven-branes, using scattering diagrams and broken lines, based on ideas from \cite{Alexeev:2024bko}. Local theta-function algebras and their transition maps give explicit affine charts for opened (resolved) webs, extending familiar toric constructions to non-toric situations. We give a physical interpretation of the deformation parameters, in the case of one or several 7-branes, and deduce behavior under Hanany-Witten transitions and changes of tesselation.   }

\begin{document} 

\maketitle

\section{Introduction}

A remarkable feature of string theory is its ability to relate seemingly different physical and geometric constructions through dualities. A standard example is the correspondence between webs of 5-branes in type IIB string theory and toric Calabi--Yau threefolds: the five-dimensional supersymmetric theory engineered by a web $W$ can also be obtained from M-theory on a toric Calabi--Yau threefold $\mathcal X$ whose toric diagram is dual to $W$ \cite{Aharony:1997bh,Leung:1997tw}.

Allowing 5-branes to end on 7-branes greatly enlarges the class of theories accessible through this construction \cite{DeWolfe:1999hj,Benini:2009gi}. The corresponding threefold is generally non-toric, and its geometric description has long been less explicit than in the toric case. Recent progress includes the study of generalized toric polygons (GTPs) \cite{vanBeest:2020kou,VanBeest:2020kxw,Bourget:2023wlb}, polytope mutations, and T-cones \cite{Arias-Tamargo:2024fjt,CarrenoBolla:2024fxy,Franco:2026bie,Franco:2026hiz}. Locally, near a stack of 5-branes, the non-toric deformation induced by 7-branes can be derived directly through a chain of well-understood dualities \cite{Collinucci:2026kom}. Extending this local description across brane junctions and gluing the resulting patches over an entire web is more subtle. Recent work has related the obstruction to extending such local deformations to the $s$-rule \cite{Franco:2026gnm}. There are also many connections with the abundant literature connecting 5d SCFTs to (not necessarily generalized toric) geometry \cite{Collinucci:2020jqd,DeMarco:2022dgh,Collinucci:2021ofd,Collinucci_2022}.

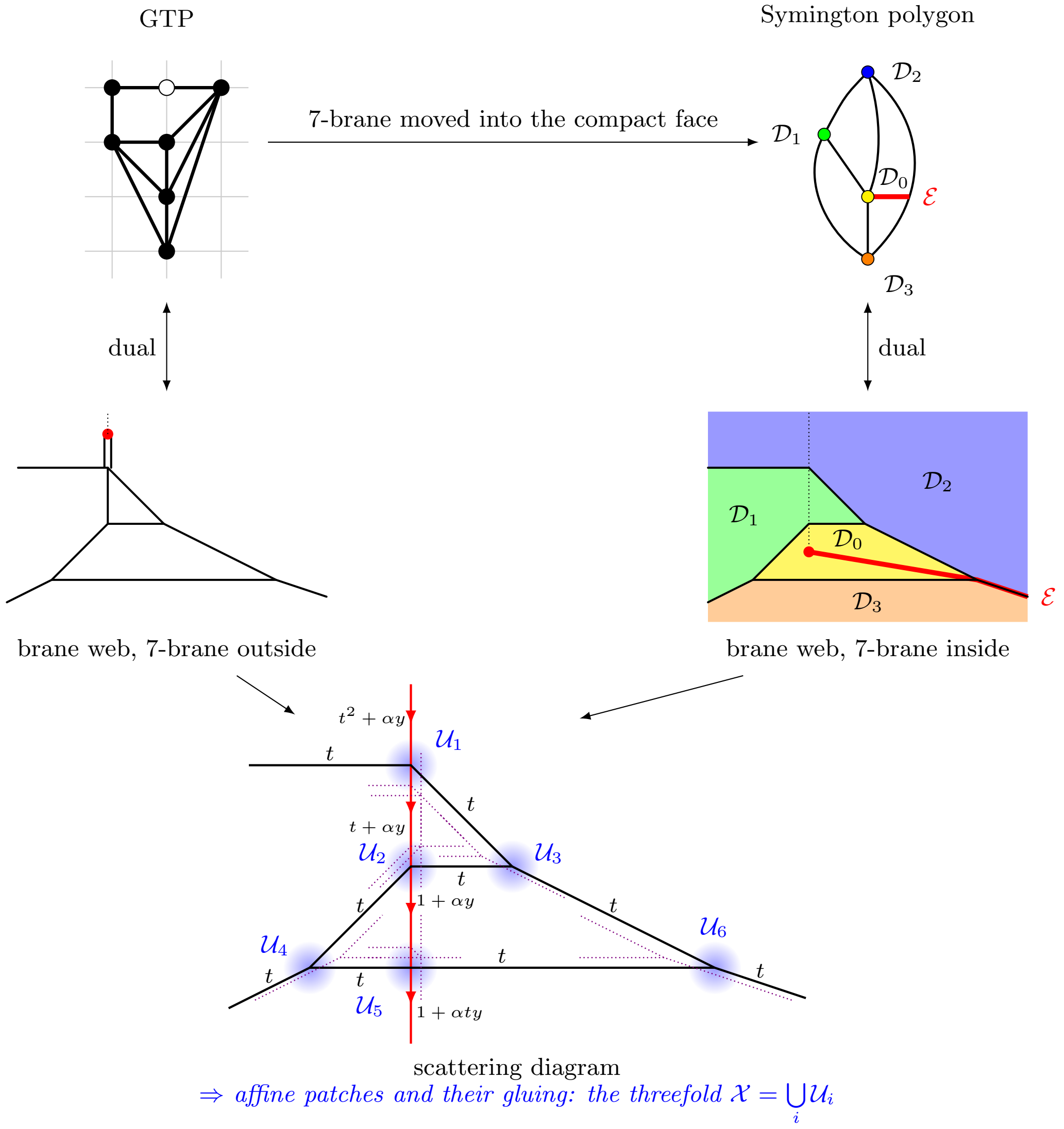
\begin{figure}[p]
    \centering
\begin{tikzpicture}[
    symdivisor/.style={circle,draw=black,line width=.35pt,inner sep=0pt,minimum size=4.5pt},
    exuberant/.style={red,line width=1.8pt},
    brokenline/.style={violet,densely dotted,line width=.45pt}
]
    \begin{scope}[shift={(-4.5,8.6)},x=.7cm,y=.7cm]
        \foreach \i in {-1,0,1} {\draw[gridline] (\i,-1.5)--(\i,2.5);}
        \foreach \j in {-1,0,1,2} {\draw[gridline] (-1.5,\j)--(1.5,\j);}
        \draw[ligne] (-1,2)--(1,2)--(0,-1)--(-1,1)--cycle;
        \draw[ligne] (-1,1)--(0,1)--(1,2);
        \draw[ligne] (0,1)--(0,0)--(0,-1);
        \draw[ligne] (-1,1)--(0,0)--(1,2);
        \foreach \p in {(-1,2),(1,2),(-1,1),(0,-1),(0,1),(0,0)} {\node[bd] at \p {};}
        \node[wd] at (0,2) {};
        \node at (0,3.27) {GTP};
    \end{scope}

    \begin{scope}[shift={(4.5,8.6)},x=.8cm,y=.8cm]
        \coordinate (SP2) at (0,2);
        \coordinate (SP1) at (-.7,1);
        \coordinate (SP0) at (0,0);
        \coordinate (SP3) at (0,-1);
        \draw[thick] (SP2) .. controls (-.37,1.62) .. (SP1)
            .. controls (-1.12,.25) and (-.62,-.65) .. (SP3);
        \draw[thick] (SP2) .. controls (1.06,1.1) and (.95,-.12) .. (SP3);
        \draw[thick] (SP2) .. controls (.24,1.28) and (.22,.48) .. (SP0);
        \draw[thick] (SP1) -- (SP0) -- (SP3);
        \draw[exuberant] (SP0) -- (.668,0);
        \node[red,right=1pt] at (.668,0) {$\mathcal E$};
        \node[symdivisor,fill=yellow] at (SP0) {};
        \node[symdivisor,fill=green] at (SP1) {};
        \node[symdivisor,fill=blue] at (SP2) {};
        \node[symdivisor,fill=orange] at (SP3) {};
        \node at (.42,.3) {$\mathcal D_0$};
        \node[left=4pt] at (SP1) {$\mathcal D_1$};
        \node[right=5pt] at (SP2) {$\mathcal D_2$};
        \node[below right=3pt] at (SP3) {$\mathcal D_3$};
        \node at (0,2.9) {Symington polygon};
    \end{scope}

    \begin{scope}[shift={(-5.256,4.4)},x=.72cm,y=.72cm]
        \draw[fivebrane] (-1.8,-1.4) -- (-1,-1) -- (0,0) -- (1,0) -- (3,-1) -- (3.9,-1.3);
        \draw[fivebrane] (-1.6,1) -- (0,1) -- (1,0);
        \draw[fivebrane] (-1,-1) -- (3,-1);
        \draw[fivebrane] (0,1) -- (0,0);
        \draw[fivebrane] (-.06,1) -- (-.06,1.6);
        \draw[fivebrane] (.06,1) -- (.06,1.6);
        \node[focalPoint,fill=red] at (0,1.6) {};
        \draw[densely dotted] (0,1.6) -- (0,2);
        \node at (1.05,-2.25) {brane web, 7-brane outside};
    \end{scope}

    \begin{scope}[shift={(3.744,4.4)},x=.72cm,y=.72cm]
        \fill[color=blue,fill opacity=.4] (-1.8,1) -- (0,1) -- (1,0) -- (3,-1) -- (3.9,-1.3) -- (3.9,2) -- (-1.8,2) -- cycle;
        \fill[color=green,fill opacity=.4] (-1.8,1) -- (0,1) -- (1,0) -- (0,0) -- (-1,-1) -- (-1.8,-1.4) -- cycle;
        \fill[color=yellow,fill opacity=.6] (0,0) -- (1,0) -- (3,-1) -- (-1,-1) -- cycle;
        \fill[color=orange,fill opacity=.4] (-1.8,-1.4) -- (-1,-1) -- (3,-1) -- (3.9,-1.3) -- (3.9,-1.75) -- (-1.8,-1.75) -- cycle;
        \draw[exuberant,line join=round] (0,-.5) -- (3,-1) -- (3.9,-1.3);
        \draw[fivebrane] (-1.8,-1.4) -- (-1,-1) -- (0,0) -- (1,0) -- (3,-1) -- (3.9,-1.3);
        \draw[fivebrane] (-1.8,1) -- (0,1) -- (1,0);
        \draw[fivebrane] (-1,-1) -- (3,-1);
        \draw[densely dotted] (0,-.5) -- (0,2);
        \node[focalPoint,fill=red] at (0,-.5) {};
        \node[red,right=1pt] at (3.9,-1.3) {$\mathcal E$};
        \node at (.7,-.28) {$\mathcal D_0$};
        \node at (-1.15,.15) {$\mathcal D_1$};
        \node at (2.3,.75) {$\mathcal D_2$};
        \node at (1.05,-1.4) {$\mathcal D_3$};
        \node at (1.05,-2.25) {brane web, 7-brane inside};
    \end{scope}

    \begin{scope}[shift={(-1.365,0)},x=1.3cm,y=1.3cm,
                  wallfn/.style={font=\tiny,inner sep=1.5pt,text=black},
                  kinklab/.style={font=\scriptsize,inner sep=1.5pt}]
        \foreach \p in {(0,1),(1,0),(0,0),(-1,-1),(0,-1),(3,-1)} {\shade[inner color=blue!80, outer color=white, opacity=.5] \p circle (.26);}
        \draw[wall,red] (0,1.8) -- node[pos=.4,left,wallfn] {$t^2+\alpha y$} (0,1);
        \draw[wall,red] (0,1) -- node[pos=.62,left,wallfn] {$t+\alpha y$} (0,0);
        \draw[wall,red] (0,0) -- node[pos=.35,right,wallfn] {$1+\alpha y$} (0,-1);
        \draw[wall,red] (0,-1) -- node[pos=.6,right,wallfn] {$1+\alpha t y$} (0,-1.75);
        \draw[kink] (-1.8,-1.4) -- node[above,kinklab] {$t$} (-1,-1)
            -- node[above,kinklab] {$t$} (0,0)
            -- node[below,kinklab] {$t$} (1,0)
            -- node[above,kinklab] {$t$} (3,-1)
            -- node[above,kinklab] {$t$} (3.9,-1.3);
        \draw[kink] (-1.6,1) -- node[above,kinklab] {$t$} (0,1)
            -- node[above right,kinklab] {$t$} (1,0);
        \draw[kink] (-1,-1) -- node[below,kinklab] {$t$} (0,-1)
            -- node[pos=.3,above,kinklab] {$t$} (3,-1);
        \tikzmath{\b = .1; \c = .42;}
        \draw[brokenline] (\b,1-3*\b+\c) -- (\b,1-3*\b) -- (-\c,1-3*\b);
        \draw[brokenline] (\b+\c,1-3*\b-\c) -- (\b,1-3*\b) -- (\b,1-3*\b-\c);
        \draw[brokenline] (\b,1-3*\b) -- (0,1-2*\b) -- (-\c,1-2*\b);
        \draw[brokenline] (\b,2*\b+\c) -- (\b,2*\b) -- (0,2*\b) -- (-\c,2*\b-\c);
        \draw[brokenline] (\c+\b,2*\b) -- (\b,2*\b) -- (\b-\c,2*\b-\c);
        \draw[brokenline] (\b,2*\b) -- (\b,2*\b-\c);
        \draw[brokenline] (1-3*\b-\c,\b) -- (1-3*\b,\b) -- (1-3*\b-\c,\b+\c);
        \draw[brokenline] (1-3*\b,\b) -- (1-3*\b+2*\c,\b-\c);
        \draw[brokenline] (-1+3*\b-2*\c,-1+\b-\c) -- (-1+3*\b,-1+\b) -- (-1+3*\b+\c,-1+\b+\c);
        \draw[brokenline] (-1+3*\b,-1+\b) -- (-1+3*\b+\c,-1+\b);
        \draw[brokenline] (\b,-1+\b+\c) -- (\b,-1+\b) -- (\b,-1+\b-\c);
        \draw[brokenline] (-\c,-1+\b) -- (\b,-1+\b) -- (\b+\c,-1+\b);
        \draw[brokenline] (\b,-1+\b) -- (0,-1+2*\b) -- (-\c,-1+2*\b);
        \draw[brokenline] (3-5*\b+3*\c,-1+\b-\c) -- (3-5*\b,-1+\b) -- (3-5*\b-2*\c,-1+\b+\c);
        \draw[brokenline] (3-5*\b,-1+\b) -- (3-5*\b-2*\c,-1+\b);
        \node[blue,right] at (.15,1.25) {$\mathcal U_1$};
        \node[blue,left] at (-.13,.13) {$\mathcal U_2$};
        \node[blue,right] at (1.13,.13) {$\mathcal U_3$};
        \node[blue,left] at (-1.1,-.78) {$\mathcal U_4$};
        \node[blue,left] at (-.16,-1.38) {$\mathcal U_5$};
        \node[blue,above] at (3,-.8) {$\mathcal U_6$};
        \node at (1.05,-2.0) {scattering diagram};
        \node[blue,font=\footnotesize\itshape] at (1.05,-2.35) {$\Rightarrow$ affine patches and their gluing: the threefold $\mathcal X = \bigcup\limits_i \mathcal{U}_i$};
    \end{scope}

    \draw[latex-latex] (-4.5,7.25) -- node[left] {dual} (-4.5,6.1);
    \draw[latex-latex] (4.5,7.25) -- node[right] {dual} (4.5,6.1);
    \draw[-latex] (-3.2,9.3) -- node[above] {7-brane moved into the compact face} (3.1,9.3);
    \draw[-latex] (-3.6,2.45) -- (-2.85,1.95);
    \draw[-latex] (2.9,2.45) -- (.8,1.9);
\end{tikzpicture}
    \caption{Schematic guide through the objects considered in the paper. All are illustrated on the example of local $\mathbb{F}_3$, dealt with in detail in Section \ref{sec:divisors}. On the top left, the GTP with a maximal tessellation, dual to a brane web with asymptotic 7-brane (bottom left). The 7-brane can be pulled in the (unique) compact face of the web (bottom right), a configuration which defines by duality the Symington polygon (top right). In these last two panels, one can read certain geometric features of the threefold, like divisors and their intersections, including exuberant divisors introduced in \cite{ABBGS}. The central object of our work is the central bottom \emph{scattering diagram}, and its \emph{broken lines} (dotted). The black lines are \emph{kinks}, the red arrows are \emph{walls} (some superimposed with kinks). The 7-branes can be thought of as \emph{delocalized on the red wall}. The deformation parameter $\alpha$ corresponds to the position of the 7-brane along this wall, see Figure \ref{fig:alphainterpretation}. Each vertex in the scattering diagram (blue blobs) gives an affine patch $\cal U$, defined by a theta function algebra specified by the broken lines. Note that there are more vertices than in either brane webs. The way the patches are glued to form the threefold is specified by the theta functions as explained in detail in Section \ref{sec:general_prescription}.  }
    \label{fig:summary}
\end{figure}
\afterpage{\clearpage}
 
Two years ago, Alexeev, Arg\"uz, and Bousseau gave a systematic construction of the M-theory geometries associated with consistent brane webs \cite{Alexeev:2024bko}, using mirror symmetry.  A particularly useful aspect of this construction is its \textit{practical, combinatorial description} in terms of scattering diagrams and broken lines \cite{kontsevich2004:asn,gross2009:msp,Gross_2010,Gross:2011af,Gross:2012du,gross2012:rgd,bridgeland:2020sdh,kontsevich:2013wdt,Bousseau:2019ift,Gross_2022:cws,arguz:2022fdt,arguz:2025qfl,Bousseau_2024,Le_Floch_2026}.\footnote{More precisely, the planar diagrams used here are the ``hearts'' of canonical scattering diagrams, with the 7-branes sent to infinity; see Remark~6.10 of \cite{Alexeev:2024bko}.} These tools are closely related to the wall-crossing formalism for BPS states \cite{Kontsevich:2008fj,Gaiotto:2010okc} and to the study of framed BPS states and line operators \cite{Gaiotto:2010be}; see also Section~1.3 of \cite{Bousseau:2020qgo} for a discussion of these connections. The core construction determines the algebra of global regular functions on $\mathcal X$. For fixed 7-brane data, this algebra depends only on the asymptotic brane web and is independent of the tessellation of the GTP. It therefore determines the affinization of $\mathcal X$, but does not by itself recover the compact cycles of a resolution. At the SCFT point, where all 5-branes meet at a single junction, $\mathcal X$ is affine and the algebra determines the full threefold. 

The present note has three main aims: 
\begin{enumerate}
    \item Make this description accessible to the physics community and give a physical interpretation of the deformation parameters entering the wall functions.  Section~\ref{sec:scattering-basics} therefore reviews the construction of Section~6.4 of \cite{Alexeev:2024bko} in an explicit, computational form.
    \item Extend this description to arbitrary points on the extended Coulomb branch of the 5d theory. This constructs the non-affine $\mathcal X$ by gluing affine patches.
    \item Use this construction to study geometrical and topological properties of these \emph{generalized toric} threefolds. 
\end{enumerate} 
 
When the faces of the brane web are opened, one can construct an initial scattering diagram $\mathfrak{D}^{\text{in}}$ with the same finite faces. Consistency is restored by adding outgoing walls locally and propagating them throughout the diagram. In this process, new crossings of walls may be introduced, and the consistency has to be checked at these new scattering points. We show on examples that this process nevertheless terminates. The theta functions of the full diagram reproduce the same global algebra as in the limit in which the faces are collapsed. We argue that the additional geometric information is captured by the theta functions of the local diagrams $\mathfrak D_p$, obtained by restricting $\mathfrak D$ to formal neighborhoods of its points $p$: these define affine charts in $\mathcal X$. Their embeddings in the full diagram determine compatible gluing maps between the charts, giving a practical description of the full geometry of $\mathcal X$.

The reader can't fail to notice that the above description closely mimics that of normal toric varieties, with fans replaced by scattering diagrams. In this sense, the naming of \emph{generalized toric} varieties is justified; and this is indeed the logic we are following -- rephrase toric geometry in terms of scattering diagrams involving only \emph{kinks}, and add walls. Although much of the combinatorics is close to its toric counterpart, we show that one should be careful about transposing certain statements from the toric to the generalized toric setting. A list of such statements appear on page~\pageref{pageProperties}. 

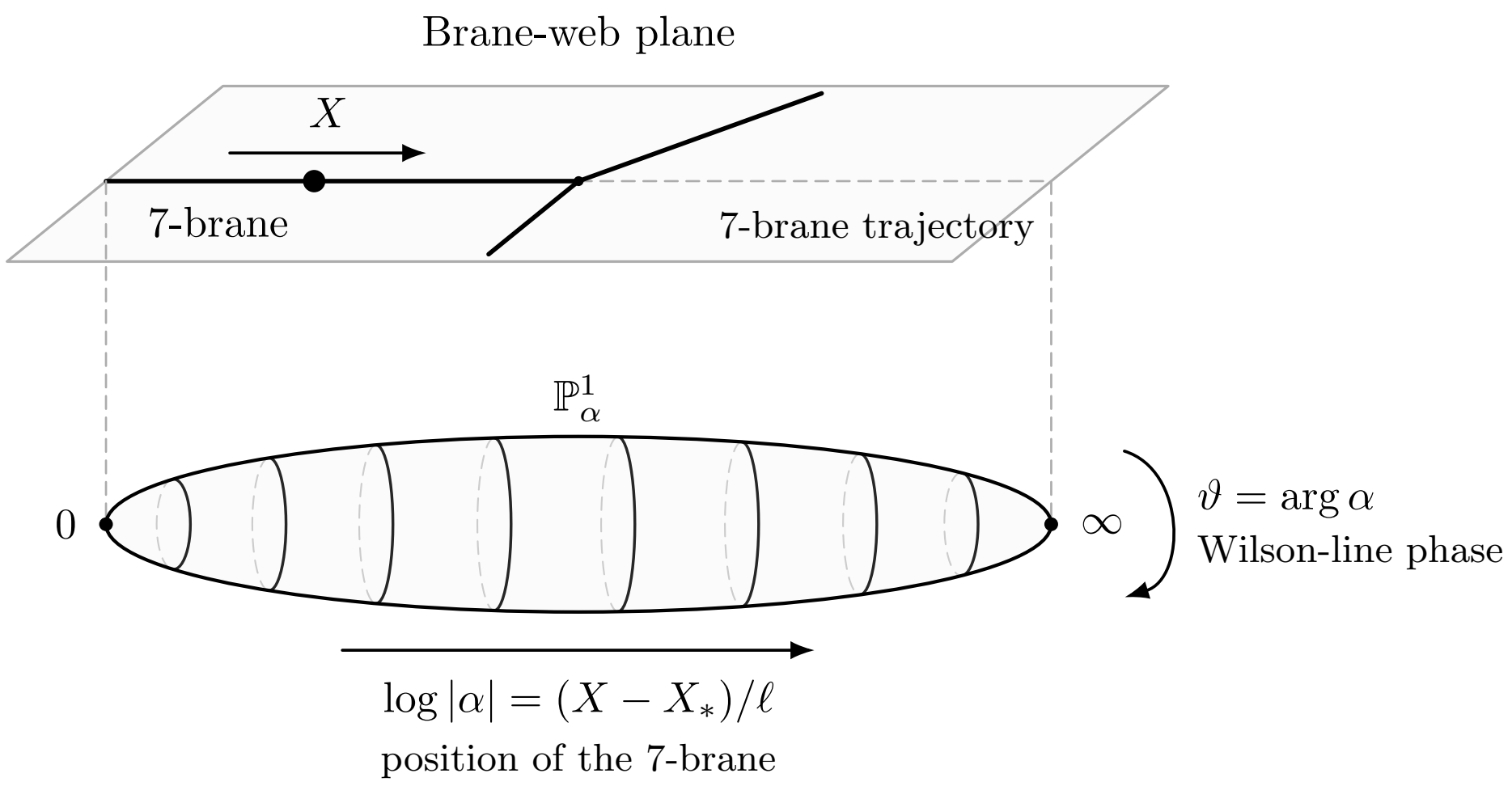
\begin{figure}
    \centering
\begin{tikzpicture}[
    x=.92cm,y=.92cm,
    line cap=round,line join=round,
    every node/.style={font=\small,text=black},
    web/.style={draw=black,line width=1.05pt},
    guide/.style={draw=gray!65,densely dashed,line width=.5pt},
    motion/.style={draw=black,-{Latex[length=2mm]},line width=.7pt}
]
    \filldraw[fill=gray!3,draw=gray!65,line width=.6pt]
        (-5.08,3.335) -- (3.32,3.335) -- (5.24,4.895)
        -- (-3.16,4.895) -- cycle;
    \node at (0,5.35) {Brane-web plane};

    \draw[guide] (-4.2,4.05) -- (4.2,4.05);
    \draw[web] (-4.2,4.05) -- (0,4.05) -- (2.16,4.83);
    \draw[web] (0,4.05) -- (-.8,3.4);

    \fill[black] (0,4.05) circle (1.15pt);
    \fill[black] (-2.35,4.05) circle (2.6pt);
    \node[below left=3pt] at (-2.35,4.05) {$7$-brane};

    \draw[motion] (-3.1,4.30) -- (-1.35,4.30)
        node[midway,above=2pt] {$X$};
    \node[below=3pt,font=\footnotesize] at (2.65,4.05)
        {$7$-brane trajectory};

    \draw[guide] (-4.2,4.05) -- (-4.2,1);
    \draw[guide] (4.2,4.05) -- (4.2,1);

    \fill[gray!3] (0,1) ellipse [x radius=4.2,y radius=.78];

    \begin{scope}
        \clip (0,1) ellipse [x radius=4.2,y radius=.78];

        \foreach \s in {-3.6,-2.75,-1.8,-.75,.35,1.45,2.5,3.4} {
            \pgfmathsetmacro{\phaseheight}{
                .78*sqrt(1-(\s/4.2)^2)
            }

            \draw[gray!40,densely dashed,line width=.4pt]
                plot[domain=90:270,samples=25,variable=\ang]
                ({\s+.15*cos(\ang)},
                 {1+\phaseheight*sin(\ang)});

            \draw[black!85,line width=.65pt]
                plot[domain=-90:90,samples=25,variable=\ang]
                ({\s+.15*cos(\ang)},
                 {1+\phaseheight*sin(\ang)});
        }
    \end{scope}

    \draw[black,line width=.8pt]
        (0,1) ellipse [x radius=4.2,y radius=.78];

    \fill[black] (-4.2,1) circle (1.6pt);
    \fill[black] (4.2,1) circle (1.6pt);
    \node[left=3pt] at (-4.2,1) {$0$};
    \node[right=3pt] at (4.2,1) {$\infty$};
    \node at (0,2.12) {$\mathbb P^1_{\alpha}$};

    \draw[motion] (-2.1,-.12) -- (2.1,-.12);
    \node[below=3pt] at (0,-.12)
        {$\log|\alpha|=(X-X_*)/\ell$};
    \node[below=18pt,font=\footnotesize] at (0,-.12)
        {position of the $7$-brane};

    \draw[motion] (4.85,1.65)
        .. controls (5.4,1.48) and (5.4,.52) .. (4.85,.35);
    \node[right,align=left] at (5.35,1)
        {$\vartheta=\arg\alpha$\\[-1pt]
         \footnotesize Wilson-line phase};
\end{tikzpicture}
    \caption{Cartoon of the interpretation of the deformation parameter $\alpha$. We take for concreteness the deformed $T_3$ theory studied in Section \ref{sec:scattering-basics}, where the threefold equation is $XYZ = W^3 + \alpha X$, see \eqref{eq:T3deformedeq}. The parameter $\alpha$ takes values in $\mathbb{P}^1$, its modulus corresponds to the 7-brane position, and the phase to a Wilson phase (the D6 position on the dual circle after T-duality).  }
    \label{fig:alphainterpretation}
\end{figure}

\paragraph{Plan and summary of results. } As a guide to the paper, we list our main results and the corresponding sections. See also Figure \ref{fig:summary} for a global overview. After reviewing the construction of \cite{Alexeev:2024bko} in Section~\ref{sec:scattering-basics} and examining its relation to violations of the $s$-rule in Section~\ref{sec:s-rule}, we present the following results:
\begin{itemize}
    \item In Section~\ref{sec:hanany-witten}, we study how scattering diagrams transform under Hanany--Witten moves \cite{Hanany:1996ie,DeWolfe:1999hj}. The deformation parameter associated with a single 7-brane is naturally a coordinate on a $\mathbb P^1$, in line with the relation between polytope mutations and deformation families \cite{Ilten_2012,Arias-Tamargo:2024fjt}. Intuitively the radial coordinate on the $\mathbb P^1$ is the position of the 7-brane; the angular part is a Wilson line phase. The two poles recover the singular toric geometries obtained by sending the 7-brane to infinity before or after the move. In the scattering diagram, the 7-brane is delocalized along the direction of its monodromy cut. 
  \item In Section~\ref{sec:ResolvedGeometries}, we extend the formalism to theories on their extended Coulomb branch. Each vertex of the resolved web defines a local scattering diagram, and outgoing walls propagate consistently to neighboring vertices, where they become incoming walls. This gives an explicit description of the geometry from which divisors, curves, triple intersection numbers, resolution maps, and topological data can be extracted.  
  \item The resolved examples in Section~\ref{sec:ResolvedGeometries} show that the geometries constructed in two Hanany--Witten frames describe the same threefold at a generic point of the $\mathbb P^1$ deformation, extending the invariance found in the singular case.
  \item From the explicit description by patches and gluing, one can obtain in an elementary way a basis of divisors for any generalized toric threefold, and then compute their triple intersections. We show on examples that this agrees with \cite{ABBGS}, in which the same data is obtained directly from the combinatoric data of the GTP and Symington polygon. 
  \item In Section~\ref{sec:multiple-seven-branes}, we explore how several 7-branes affect the deformation parameter space. For $n$ 7-branes, this space need not be $(\mathbb P^1)^n$. Its structure depends on the symplectic pairings of their charges and encodes the different 7-brane configurations accessible through Hanany--Witten moves. These pairings define a quiver whose mutations encode Hanany–Witten moves, placing the resulting scattering diagrams in the framework of cluster algebras. 
\end{itemize}

\paragraph{Acknowledgements. } We thank H\"ulya Arg\"uz and Pierrick Bousseau for clarifying certain aspects of scattering diagrams, and Andres Collinucci for discussions about his recent works.  I.C.B. would like to thank Noppadol Mekareeya and Palash Singh for useful conversations. The work of I.C.B. is supported by the Severo Ochoa fellowship  NAC-AT-PUB-ASV-2025 BP24-116 and is partially supported by the Spanish Agencia Estatal de Investigaci\'on and Ministerio de Ciencia, Innovaci\-on y Universidades through the grants PID2021-123021NB-I00 and PID2024-161500NB-I00. I.C.B. would like
to thank IPhT Saclay for the kind hospitality during the start of this work and the University of Milano-Bicocca during the end of this work.  I.C.B. would also like to thank the COST Action CA22113 “Fundamental challenges in
theoretical physics” for financial support during part of this work through an STSM Grant. 
The work of QL is supported by École Normale Supérieure - PSL through a CDSN doctoral grant.

\paragraph{AI statement. } We used AI models (OpenAI ChatGPT-6.1 Sol and Anthropic Claude Opus 5.5) to check the algebraic computations in Section \ref{sec:ResolvedGeometries}, for bibliographic searches, to assist in proofreading the draft and to help write the TikZ code for figures. All of the ideas, computations, coding, writing and figure conception are human. The authors assume full responsibility for the results presented in the present paper.  

\section{Scattering Diagrams, broken lines and theta functions}\label{sec:scattering-basics}

This Section is largely a (hopefully pedagogical and) self-contained review of the construction of affine threefolds dual to GTPs at the SCFT point \cite{Alexeev:2024bko}. Hence we assume the brane web has a unique intersection point, which we take to be the origin of the plane $O$. 
The 5-branes of type $[p,q]$ (with $p,q$ coprime) can either be semi-infinite lines emanating from $O$, or can end on a 7-brane of type $[p,q]$. There can be several 5-branes stacked on each other. When we draw them, we indicate the multiplicity. The \emph{central part} of a brane web is the web obtained by sending all the 7-branes to infinity (equivalently, by zooming in on the origin $O$).  The 7-branes are remembered as boundary conditions for each stack of semi-infinite 5-branes.

\begin{example}[Examples: three running brane webs]
Here are three brane webs, which will serve as running examples to illustrate all the constructions of this section. The black disk represents a 7-brane. The central parts of the first two are identical. 
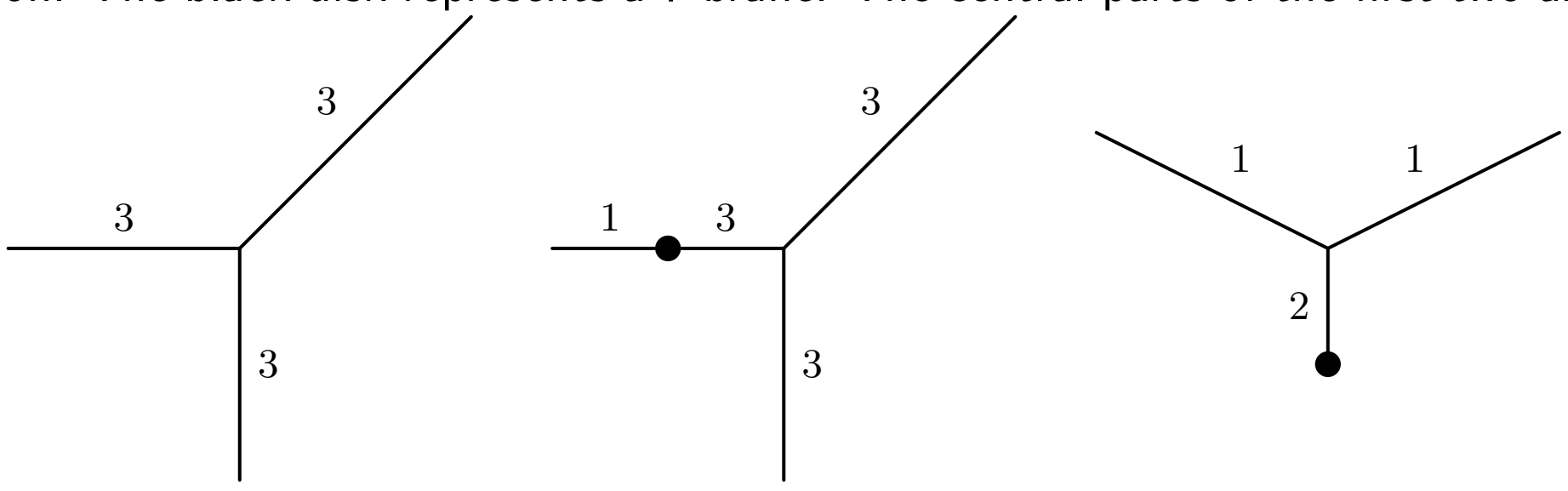
\begin{center}
\begin{tikzpicture}[x=.9cm,y=.9cm]
\begin{scope}[shift={(0,0)}]
\draw[fivebrane] (0,0) -- node[midway,above] {$3$} (-2,0);
    \draw[fivebrane] (0,0) -- node[midway,above left] {$3$} (2,2);
    \draw[fivebrane] (0,0) -- node[midway,right] {$3$} (0,-2);
\end{scope}
\begin{scope}[shift={(4.7,0)}]
\draw[fivebrane] (-1,0) -- node[midway,above] {$1$} (-2,0);
    \draw[fivebrane] (0,0) -- node[midway,above] {$3$} (-1,0);
    \draw[fivebrane] (0,0) -- node[midway,above left] {$3$} (2,2);
    \draw[fivebrane] (0,0) -- node[midway,right] {$3$} (0,-2);
    \fill (-1,0) circle[radius=1mm];
\end{scope}
\begin{scope}[shift={(9.4,0)}]
\draw[fivebrane] (0,0) -- node[midway,left] {$2$} (0,-1);
    \draw[fivebrane] (0,0) -- node[midway,above left] {$1$} (2,1);
    \draw[fivebrane] (0,0) -- node[midway,above right] {$1$} (-2,1);
    \fill (0,-1) circle[radius=1mm];
\end{scope}
\end{tikzpicture} 
\end{center}
\end{example}

\subsection{Scattering Diagrams}\label{sec:scattering_diagrams}

The first step of the procedure is to associate a \emph{scattering diagram} to the brane web. Let $r \geq 0$ be the number of 7-branes; we can label them by an index $i=1 , \dots , r$.

\paragraph{Scattering Diagram. } We first explain what a scattering diagram is. We use the conventions from \cite{Gross_2010,Gross:2011af,Alexeev:2024bko}. See \cite{Gross2011Tropical} for a textbook account. Introduce formal variables $x$, $y$, and $(t_0 , t_1 , \dots , t_r)$. It is convenient to set $I=(t_1,\ldots,t_r)$ and introduce the ring\footnote{Recall that this means an element of $R$ is a possibly infinite sum of monomials in the $I$ variables, with coefficients that are Laurent polynomials in $t_0,x,y$. } 
\begin{equation}\label{eq:formal-wall-ring}
 R=\mathbb Z[t_0^{\pm1},x^{\pm1},y^{\pm1}]
       \llbracket t_1,\ldots,t_r\rrbracket.
\end{equation}
We use the notation $\mathbb{Z}^2_{\rm{pr}}$ for the set of pairs of coprime integers, which correspond to primitive vectors in $\mathbb{Z}^2$. 

A \emph{wall} is a pair $(\mathfrak{d} , f)$ where 
\begin{itemize}
    \item $\mathfrak{d}$ is a half-line from the origin $O$. It can be written uniquely as $\mathfrak{d} = \mathbb{R}_{\geq 0} (p,q)$ with $(p,q)\in\mathbb{Z}^2_{\rm{pr}}$. We call $\mathfrak{d}$ the \emph{support} of the wall, and $(p,q)$ its \emph{direction}.  
    \item $f\in R$ depends on $x,y$ only through $z=x^py^q$, and $f\equiv t_0^n\pmod I$ for some $n\geq0$.\footnote{Thus $f$ is a unit in $R$. } 
\end{itemize}
Here is the vocabulary we use to talk about walls: we say the wall $(\mathfrak{d} , f)$ is 
\begin{itemize}
    \item \emph{incoming} if every monomial in \(x\) and \(y\) occurring in \(f\) is of the form $(x^p y^q)^k$ for $k \in \mathbb{Z}_{\geq 0}$. We draw such a wall with an arrow pointing towards $O$. 
    \item \emph{outgoing} if every monomial in \(x\) and \(y\) occurring in \(f\) is of the form $(x^p y^q)^k$ for $k \in \mathbb{Z}_{\leq 0}$. We draw such a wall with an arrow pointing away from $O$.   
    \item a \emph{kink} if $f = t_0^n$ for some integer $n \geq 1$.\footnote{Hence a kink is both an incoming wall and an outgoing wall. }\footnote{The name \emph{kink} comes from the language of \cite{Gross:2011af}, where they are the places where a piecewise linear function changes slope.} 
    \item a \emph{pure wall} if its wall function is \(1\) plus terms that each involve at least one \(t_i\) and a nonconstant monomial in \(x,y\).
\end{itemize} 
A \emph{scattering diagram} is a (finite or infinite) collection of walls, drawn in the same plane as the brane web plane. If two walls $(\mathfrak{d} , f_1)$ and $(\mathfrak{d} , f_2)$ have the same support, we can construct another diagram by merging them, i.e. replacing them by a single wall $(\mathfrak{d} , f_1 f_2)$. Two scattering diagrams related by a sequence of such mergings (and the addition or removal of walls with function $1$) are called \emph{equivalent}.
For an infinite diagram we require that, for every $N$, only finitely many pure walls have function different from $1$ modulo $I^{N}$, in addition to finitely many kinks. 

\paragraph{Initial Scattering diagram associated to a brane web. } Given a brane web, we associate a scattering diagram, called the \emph{initial scattering diagram}, which consists of the following walls:    
\begin{itemize}
    \item To each stack of $n$ 5-branes in the central part of the brane web, emanating from $O$ in primitive direction $(p,q) \in \mathbb{Z}^2$, associate the kink $(\mathbb{R}_{\geq 0} (p,q) , t_0^n)$. 
    \item To the $i$-th 7-brane, of primitive type $(p,q) \in \mathbb{Z}^2$, associate the incoming wall $(\mathbb{R}_{\geq 0} (p,q) , 1 + t_i x^p y^q)$. 
\end{itemize}

\begin{example}[Examples: initial scattering diagrams]
In our examples, this yields the following initial scattering diagrams: 
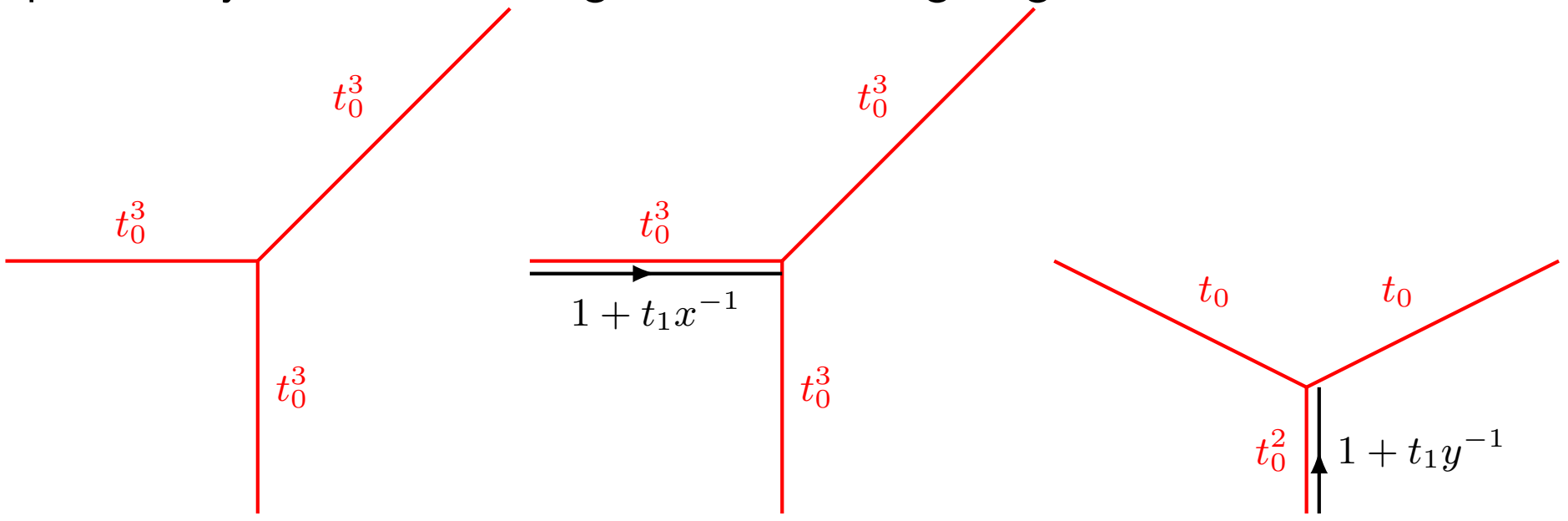
\begin{center}
 \begin{tikzpicture}
\draw[kink,goodred] (0,0)-- node[midway,above left] {$t_0^3$} (2,2);
\draw[kink,goodred] (0,0)-- node[midway,right] {$t_0^3$} (0,-2);
\draw[kink,goodred] (0,0)-- node[midway,above] {$t_0^3$} (-2,0);
\end{tikzpicture}
\begin{tikzpicture}
\draw[kink,goodred] (0,0)-- node[midway,above left] {$t_0^3$} (2,2);
\draw[kink,goodred] (0,0)-- node[midway,right] {$t_0^3$} (0,-2);
\draw[kink,goodred] (0,0)-- node[midway,above] {$t_0^3$} (-2,0);
\draw[wall,black] (-2,-.1)-- node[midway,below] {$1+t_1 x^{-1}$} (0,-.1);
\end{tikzpicture}
\begin{tikzpicture}
\draw[kink,goodred] (0,0)-- node[midway,above left] {$t_0$} (2,1);
\draw[kink,goodred] (0,0)-- node[midway,above right] {$t_0$} (-2,1);
\draw[kink,goodred] (0,0)-- node[midway,left] {$t_0^2$} (0,-1);
\draw[wall,black] (.1,-1)-- node[midway,right] {$1+t_1 y^{-1}$} (.1,0);
\end{tikzpicture}   
\end{center}
\end{example}

\paragraph{Wall-Crossing Automorphism. }  
To a wall $(\mathfrak{d} = \mathbb{R}_{\geq 0} (p,q) , f)$ with $(p,q) \in \mathbb{Z}^2_{\mathrm{pr}}$ one associates the \emph{wall-crossing automorphism} $\Phi_{(\mathfrak{d} , f)} : R \rightarrow R$, defined by its action on monomials:
\begin{equation}
  \Phi_{(\mathfrak{d} , f)} (  x^a y^b ) =  x^a y^b f^{p b - q a} \, . 
\end{equation}
We say that a scattering diagram $\mathfrak{D}$ is \emph{consistent} if the composition $\Phi_{(\mathfrak{d}_1 , f_1)} \circ \dots \circ \Phi_{(\mathfrak{d}_s , f_s)}$ of all its walls, where $\mathfrak{d}_1 , \dots , \mathfrak{d}_s$ are the walls listed in anti-clockwise order from a reference ray, is the identity automorphism $R \rightarrow R$.

\begin{example}[Examples: wall-crossing automorphisms]
The wall-crossing automorphisms associated to the walls in the first example above are: 
\begin{itemize}
    \item For $(\mathbb{R}_{\geq 0} (-1,0) , t_0^3)$, $x^a y^b \mapsto x^a y^b t_0^{-3b}$.
    \item For $(\mathbb{R}_{\geq 0} (0,-1) , t_0^3)$, $x^a y^b \mapsto x^a y^b t_0^{3a}$.
    \item For $(\mathbb{R}_{\geq 0} (1,1) , t_0^3)$, $x^a y^b \mapsto x^a y^b t_0^{3(b-a)}$.
\end{itemize}
Clearly, the composition of all three gives the identity, so the first diagram is consistent. For the second diagram, one should also consider the incoming wall, with associated automorphism $x^a y^b \mapsto x^a y^b (1+ \frac{t_1}{x})^{-b}$. So the total automorphism is the composition (starting from the upper left quadrant): 
\begin{equation}
    x^a y^b \mapsto x^a y^b t_0^{3b-3a} \mapsto x^a y^b t_0^{3b} \mapsto x^a y^b \left(1+ \frac{t_1}{x}\right)^{-b} \, . 
\end{equation}
This is not the identity, so the diagram is not consistent. Similarly, the third example is not a consistent diagram.
\end{example}

\paragraph{Consistent Scattering Diagram $\mathfrak{D}$. }
A crucial fact \cite[Thm.~1.4]{Gross_2010}, \cite[\S3.4]{Gross:2011af} is that for any initial scattering diagram $\mathcal{D}_{\mathrm{in}}$ associated to a brane web as above, there exists a unique (up to equivalence) diagram $\mathcal{D}$ obtained from $\mathcal{D}_{\mathrm{in}}$ by adding only outgoing pure walls, and which is consistent. 

Moreover, $\mathcal{D}$ can be efficiently constructed using an algorithm. Roughly speaking, one starts with the initial diagram and works at fixed order in the variables $t_i$. One then computes the full automorphism; if it is not the identity, this can be fixed by adding a finite number of walls, the walls being prescribed by the difference between the full automorphism and the identity. One then proceeds inductively. We refer to Appendix \ref{app:algoConsistent} for a full and rigorous explanation of this intuition.  

\begin{example}[Examples: consistent completion]
In our examples, this yields the following scattering diagrams:  

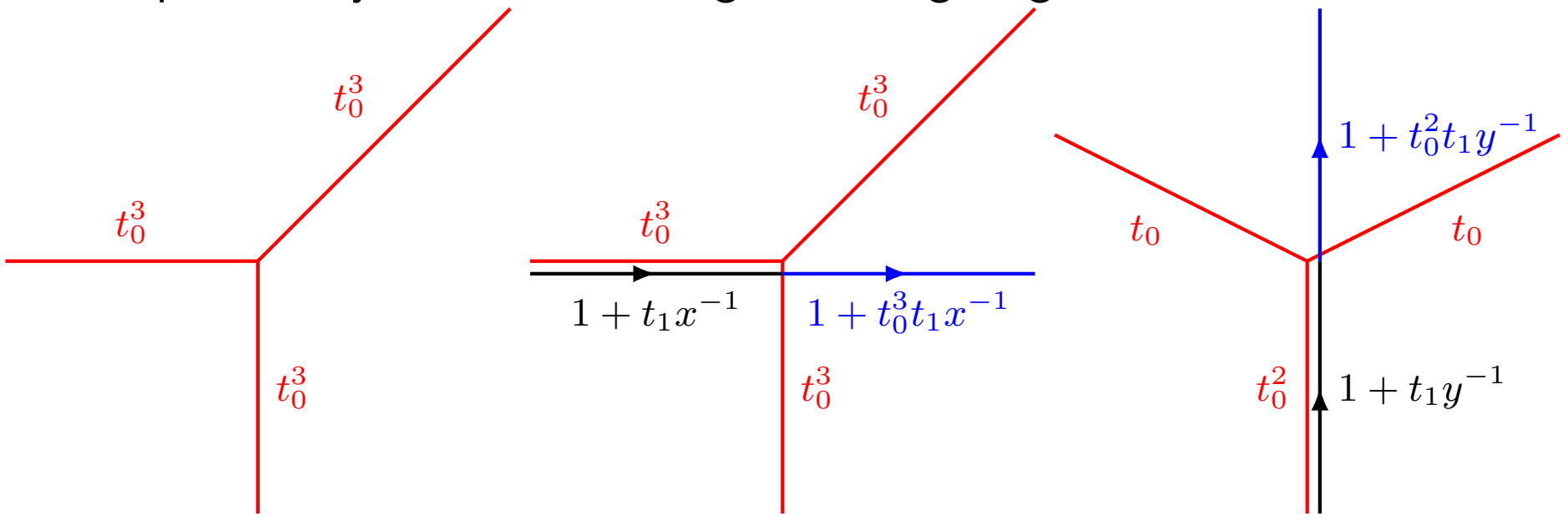
\begin{tikzpicture}
\draw[kink,goodred] (0,0)-- node[midway,above left] {$t_0^3$} (2,2);
\draw[kink,goodred] (0,0)-- node[midway,right] {$t_0^3$} (0,-2);
\draw[kink,goodred] (0,0)-- node[midway,above] {$t_0^3$} (-2,0);
\end{tikzpicture}
\begin{tikzpicture}
\draw[kink,goodred] (0,0)-- node[midway,above left] {$t_0^3$} (2,2);
\draw[kink,goodred] (0,0)-- node[midway,right] {$t_0^3$} (0,-2);
\draw[kink,goodred] (0,0)-- node[midway,above] {$t_0^3$} (-2,0);
\draw[wall,black] (-2,-.1)-- node[midway,below] {$1+t_1 x^{-1}$} (0,-.1);
\draw[wall,blue] (0,-.1)-- node[midway,below] {$1+t_0^3 t_1  x^{-1}$} (2,-.1);
\end{tikzpicture}
\begin{tikzpicture}
\draw[kink,goodred] (0,0)-- node[midway,below right] {$t_0$} (2,1);
\draw[kink,goodred] (0,0)-- node[midway,below left] {$t_0$} (-2,1);
\draw[kink,goodred] (0,0)-- node[midway,left] {$t_0^2$} (0,-2);
\draw[wall,black] (.1,-2)-- node[midway,right] {$1+t_1 y^{-1}$} (.1,0);
\draw[wall,blue] (.1,0)-- node[midway,right] {$1+ t_0^2 t_1 y^{-1}$} (.1,2);
\end{tikzpicture}

At degree one, the second example has obstruction $h_{(1,0)}=-t_1/x$ in the upper-left reference sector. The kinks before the new $(1,0)$ ray give $\kappa_{(1,0)}=3$, so the algorithm adds $1+t_0^3t_1/x$. For the third example, taking the positive horizontal axis as reference gives $h_{(0,1)}=-t_1/y$ and $\kappa_{(0,1)}=2$, hence the new wall function $1+t_0^2t_1/y$. In these single-incoming-wall examples the corrections already give exact consistency, as the following calculation shows.

We can check consistency directly. The first diagram contains only kinks and was already checked above. For the second diagram, choose the reference ray in the upper-left sector and merge the coincident incoming wall and kink. With the \emph{counterclockwise} ordering convention above, we apply the automorphisms in the \emph{clockwise} order, since the rightmost factor ($\Phi_s$) acts first and the leftmost ($\Phi_1$) acts last. So a monomial transforms as
\begin{equation}
 x^a y^b
 \longmapsto t_0^{3(b-a)}x^a y^b 
 \longmapsto t_0^{3(b-a)}x^a y^b
       \left(1+\frac{t_0^3t_1}{x}\right)^b 
 \longmapsto t_0^{3b}x^a y^b
       \left(1+\frac{t_1}{x}\right)^b 
 \longmapsto x^a y^b.
\end{equation}

Similarly for the third diagram, take the positive horizontal axis as the reference ray. The action is
\begin{equation}
 x^a y^b
 \longmapsto t_0^{2a}x^a y^b
       \left(1+\frac{t_1}{y}\right)^a
 \longmapsto t_0^{a-2b}x^a y^b
       \left(1+\frac{t_0^2t_1}{y}\right)^a
 \longmapsto t_0^{a-2b}x^a y^b
 \longmapsto x^a y^b.
\end{equation}
Thus both completed diagrams have total wall-crossing automorphism equal to the identity.
\end{example}

\paragraph{Reduced Consistent Scattering Diagram. } 
Until now, the partition data defining the web has not played any role! Here is how it enters the construction. We operate the substitutions \cite{Alexeev:2024bko} 
\begin{equation}
    t_0 \mapsto t \, , \qquad t_i \mapsto \alpha_i t^{-n_i}
\end{equation}
where $n_i$ is the number of 5-branes which end on the $i$-th 7-brane.
This yields the \emph{reduced consistent scattering diagram} associated to the brane web. 

\begin{example}[Examples: reduced scattering diagrams]
In both nontrivial examples above $n_1=2$: the horizontal outgoing function becomes $1+\alpha t/x$, whereas the vertical outgoing function becomes $1+\alpha/y$, in agreement with the diagrams below.

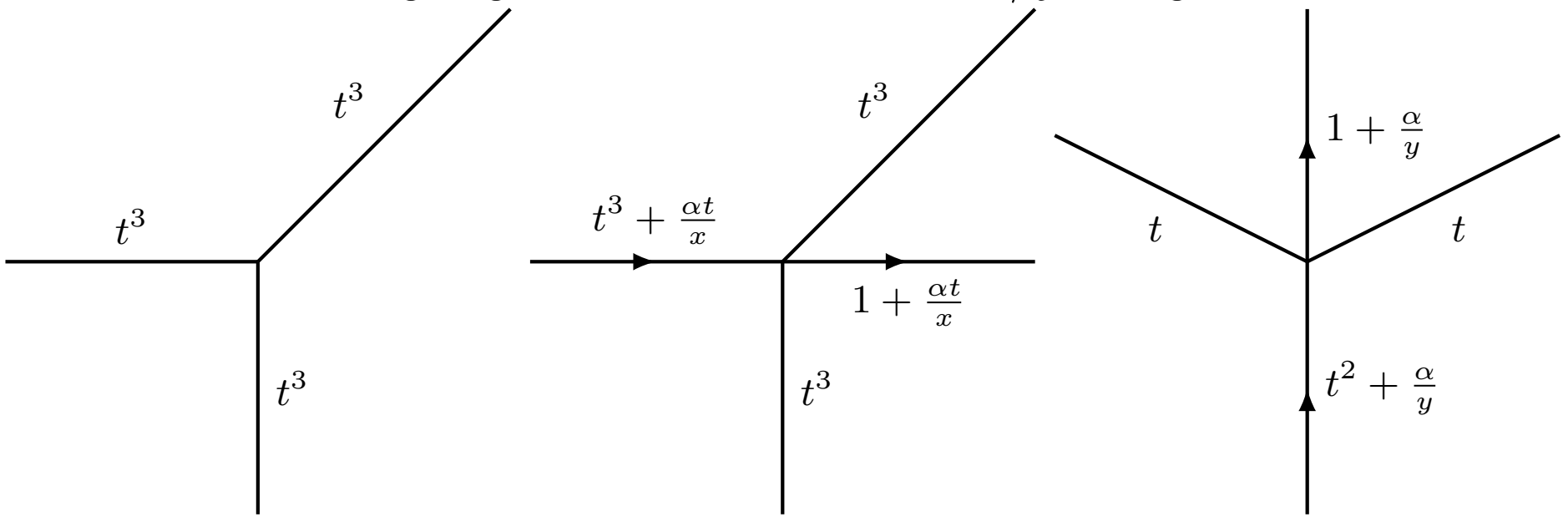
\begin{tikzpicture}
\draw[kink, black] (0,0)-- node[midway,above left] {$t^3$} (2,2);
\draw[kink,black] (0,0)-- node[midway,right] {$t^3$} (0,-2);
\draw[kink,black] (0,0)-- node[midway,above] {$t^3$} (-2,0);
\end{tikzpicture}
\begin{tikzpicture}
\draw[kink,black] (0,0)-- node[midway,above left] {$t^3$} (2,2);
\draw[kink,black] (0,0)-- node[midway,right] {$t^3$} (0,-2);
\draw[wall,black] (0,0)-- node[midway,below] {$1+\frac{\alpha t}{x}$} (2,0);
\draw[wall,black] (-2,0)-- node[midway,above] {$t^3 + \frac{\alpha t}{x}$} (0,0);
\end{tikzpicture}
\begin{tikzpicture}
\draw[kink,black] (0,0)-- node[midway,below right] {$t$} (2,1);
\draw[kink,black] (0,0)-- node[midway,below left] {$t$} (-2,1);
\draw[wall,black] (0,-2)-- node[midway,right] {$t^2 + \frac{\alpha}{y}$} (0,0);
\draw[wall,black] (0,0)-- node[midway,right] {$1 + \frac{\alpha}{y}$} (0,2);
\end{tikzpicture}
\end{example}

\subsection{Broken Lines and Theta Functions}

From now on, we assume that no negative power of $t$ appears in the expansion of any wall function in the reduced scattering diagram $\mathfrak{D}$. We refer to Section \ref{sec:s-rule} for the meaning of negative powers of $t$. 

\paragraph{Broken lines. } Let $\mathfrak{D}$ be a consistent scattering diagram, let $P$ be a point in a generic position\footnote{This means in particular that $P$ should not be on a wall. } in the plane where $\mathfrak{D}$ is drawn, and let $(m,n) \in \mathbb{Z}^2 - \{(0,0)\}$.  A \emph{broken line  ending at $P$  with asymptotic direction $(m,n)$}  is a \emph{continuous} path $\beta: \mathbb{R}_{\leq0} \to \mathbb{R}^2 - \{O\}$ such that 
\begin{itemize}
    \item $\beta (0) = P$. 
    \item The image of $\beta$ is the union of one semi-infinite line $L_1$ and $N(\beta) -1$ compact segments $L_2,\cdots,L_{N(\beta)}$. We call $N(\beta) \in \mathbb{Z}_{\geq 1}$ the \emph{length} of the broken line. 
    \item For all $i=1,\dots,N(\beta)$, $L_i$ is contained in a 2-dimensional cone of the initial diagram $\mathcal{D}_{\mathrm{in}}$. The slope of $L_i$ is required to be rational, i.e. the velocity $(m_i , n_i) := - \beta ' |_{L_i}$ is a vector of (not necessarily coprime) integers. Finally, $L_i \cap L_{i+1}$ (for $i<N(\beta)$) is a point contained in a wall of the consistent diagram $\mathfrak{D}$, with support $\mathbb{R}_{\geq 0} (p_i , q_i)$ with $(p_i , q_i) \in \mathbb{Z}^2_{\mathrm{pr}}$ and function $f_i$. 
    \item We assign to each $L_i$ a monomial $\mu_i = a_i t^{k_i} x^{m_i} y^{n_i}$ where $a_i \in \mathbb{C}$, $k_i \in \mathbb{Z}_{\geq 0}$ and $(m_i , n_i)$ are defined as above. 
    \item We require that $\mu_1 = x^m y^n$, and for each $i=1 , \dots , N(\beta)-1$, $\mu_{i+1}$ is one of the monomials of the formal expansion $\mu_i f_i^{|p_i n_i - q_i m_i|}$. 
\end{itemize}

Call $\mathfrak{L}(P,(m,n))$ the set of broken lines ending at $P$ with asymptotic direction $(m,n)$. 

\paragraph{Theta functions. } Consider a consistent diagram $\mathfrak{D}$ and a generic point $P$. For any $(m,n) \in \mathbb{Z}^2 - \{(0,0)\}$ one associates the \emph{theta function} $\vartheta_{(m,n)}\in\mathbb C[t,x^{\pm1},y^{\pm1}]\llbracket\alpha_1,\dots,\alpha_r\rrbracket$ defined by 
\begin{equation}
    \vartheta_{(m,n)} = \sum\limits_{\beta \in \mathfrak{L}(P,(m,n))} \mu_{N(\beta)} (\beta) \, .  
\end{equation}
So $\vartheta_{(m,n)}$ is the sum of the \emph{final monomials} of all broken lines ending at $P$ with asymptotic direction $(m,n)$. We set $\vartheta_{(0,0)}=1$. 

This gives an algebra over $\mathbb C\llbracket \alpha_i\rrbracket$. In other words, the theta-functions and relations between them are formal power series in the $\alpha_i$. In most of the cases we study, these series are actually polynomials. We can then set the $\alpha_i$ to numerical values and take $\mathcal A$ to be a subalgebra of $\mathbb C[t,x^{\pm 1},y^{\pm 1}]$.

\begin{example}[Example: theta functions for $T_3$ and its deformation]
Consider our three examples in turn. For the first one, the scattering diagram only consists of kinks. In this case, the broken lines are actually not broken. Indeed, the monomials $\mu_i$ along the various segments can only change by powers of $t$, when crossing a wall, so the powers of $x$ and $y$ are constant along the line. Thus the theta functions depend on the position of $P$ in a straightforward way. Consider for instance $P$ to lie in the top left quadrant. The full expression, including the multiplicities acquired at the kinks, is
\begin{equation}
    \vartheta_{(m,n)}=t^{3\max\{0,m-n,-n\}}x^m y^n.
\end{equation}
In particular, it equals $x^m y^n$ when $n\geq0$ and $m\leq n$, since the corresponding line crosses no kink.

The next example is more interesting. Take $P$ in the top left quadrant. The number of broken lines in $\mathfrak{L}(P,(m,n))$ depends strongly on the direction $(m,n)$. If $n \geq 0$, the broken line crosses either no wall or just a kink, and hence is not deflected. One finds 
\begin{equation}
    \vartheta_{(m,n)}=t^{3\max\{m-n,0\}}x^m y^n.
\end{equation}
For $n<0$ on the other hand, the line has to cross a horizontal wall. The starting monomial is $x^m y^n$; when the line hits the horizontal wall, it carries either the monomial $x^m y^n$ (if it first hits the $\mathbb{R}_{\geq 0} (1,0)$ wall) or $t^{3|m|} x^m y^n$ (if it first hits the $\mathbb{R}_{\geq 0} (0,-1)$ wall). We encourage the reader to check that there are exactly $|n|+1$ broken lines, whose final monomials are the monomials obtained by expanding
\begin{equation}\label{eq:broken-line-example-expansion}
 \vartheta_{(m,n)} =t^{3\max\{m,0\}}x^m y^n       \left(t^3+\frac{\alpha t}{x}\right)^{|n|} \, . 
\end{equation}
Here we retain the deformation parameter $\alpha$ of the reduced diagram; the pictures below set $\alpha=1$.  
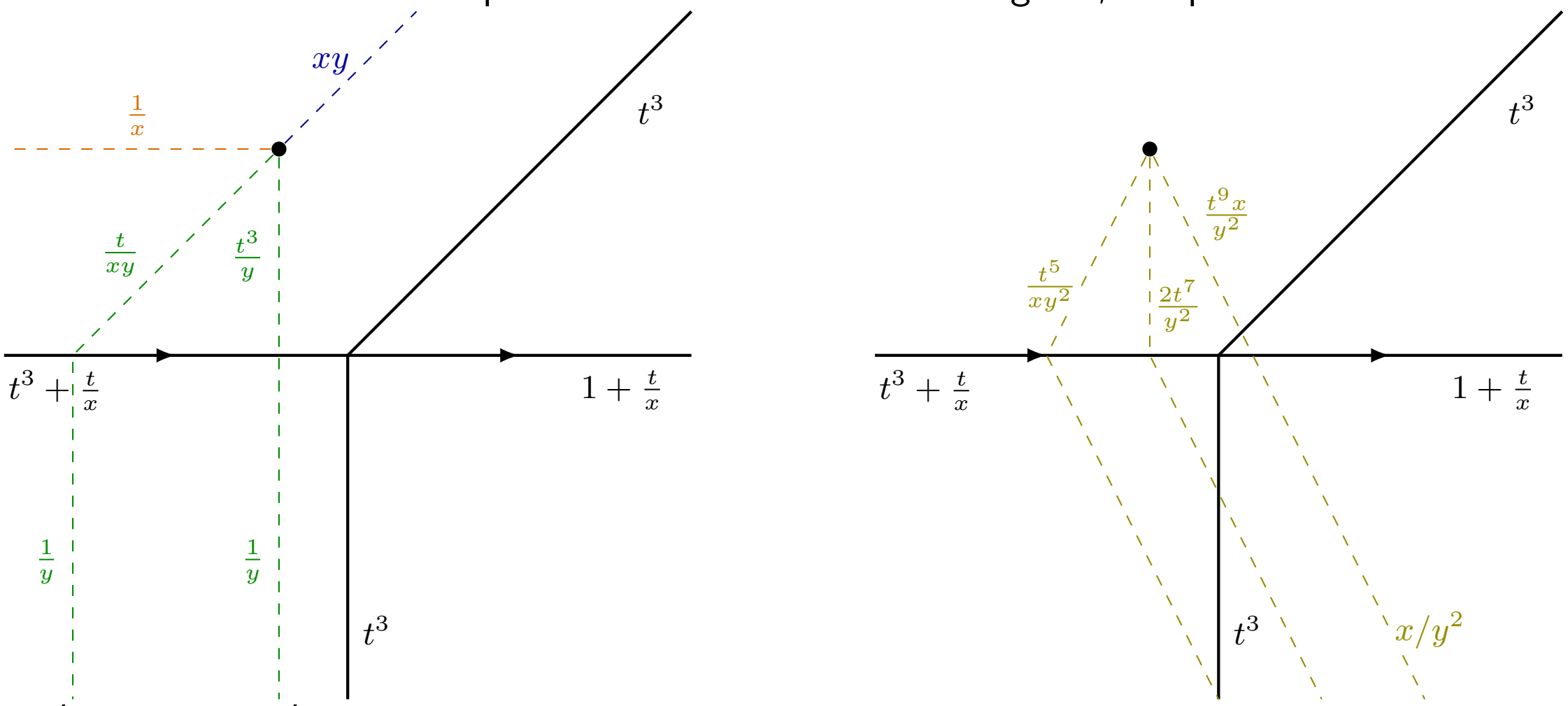
\begin{center}
\begin{minipage}[c]{0.48\linewidth}
    \centering
    \begin{tikzpicture}[x=0.65cm,y=0.65cm]
        \draw[wall,black] (0,0)--(5,0) node[pos=.8,below] {$1+\frac{t}{x}$}; 
        \draw[wall,black] (-5,0)--(0,0) node[pos=.15,below] {$t^3+\frac{t}{x}$}; 
        \draw[kink] (0,0)--(5,5) node[pos=.8,below right] {$t^3$}; 
        \draw[kink] (0,0)--(0,-5) node[pos=.8,right] {$t^3$}; 
        \node[focalPoint,black] (F) at (-1,3) {};
        \draw[broken,goodgreen] (F) -- node[pos=.5,left] {$\frac{t^3}{y}$} (-1,0) -- node[pos=.6,left] {$\frac{1}{y}$} (-1,-5);
        \draw[broken,goodgreen] (F) -- node[pos=.5,left,xshift=-5pt] {$\frac{t}{xy}$} (-4,0) -- node[pos=.6,left] {$\frac{1}{y}$} (-4,-5); 
        \draw[broken,goodorange] (F)--(-5,3) node[pos=.5,above] {$\frac{1}{x}$}; 
        \draw[broken,goodblue] (F)--(1,5) node[pos=.6,left] {$xy$}; 
        \end{tikzpicture}  
        \end{minipage}\hfill%
        \begin{minipage}[c]{0.48\linewidth}
        \centering
           \begin{tikzpicture}[x=0.65cm,y=0.65cm]
        \draw[wall,black] (0,0)--(5,0) node[pos=.8,below] {$1+\frac{t}{x}$}; 
        \draw[wall,black] (-5,0)--(0,0) node[pos=.15,below] {$t^3+\frac{t}{x}$}; 
        \draw[kink] (0,0)--(5,5) node[pos=.8,below right] {$t^3$}; 
        \draw[kink] (0,0)--(0,-5) node[pos=.8,right] {$t^3$}; 
        \node[focalPoint,black] (F) at (-1,3) {}; 
        \draw[broken,olive] (F)-- node[pos=.75,right,fill=white,inner sep=1pt] {$\frac{2t^7}{y^2}$} (-1,0)
         -- (0,-2) -- (1.5,-5);
        \draw[broken,olive] (F)-- node[pos=.65,left,fill=white,inner sep=1pt] {$\frac{t^5}{xy^2}$} (-2.5,0)
         -- (0,-5);
        \draw[broken,olive] (F)-- node[pos=.5,above right,fill=white,inner sep=1pt] {$\frac{t^9x}{y^2}$} (1/3,1/3)
         -- (.5,0) -- node[pos=.8,right,fill=white,inner sep=1pt] {$x/y^2$} (3,-5);
    \end{tikzpicture}  
\end{minipage}
\end{center}
From there, one reads 
\begin{equation}
    \vartheta_{-1,0}=\frac1{x}, \qquad \vartheta_{1,1}=x y\,, \qquad   \vartheta_{(0,-1)}=\frac{t^3}{y}+\frac{\alpha t}{xy},\qquad
 \vartheta_{(1,-2)}=\frac{t^9x}{y^2}
               +\frac{2\alpha t^7}{y^2}
               +\frac{\alpha^2t^5}{xy^2}.
\end{equation}

\end{example}

\subsection{Theta function Algebra and the Threefold}
\label{sec:thetaAlgebra}

We now review some properties of the broken-lines construction of theta-functions. To make the presentation easier to follow, we are deliberately vague about formal details, such as finiteness issues and formal completions that distinguish between formal power series and polynomials. Throughout, one should assume that we work at a finite order $N$ in the $\alpha_i$ so that the scattering diagram is finite and take $N$ to infinity at the end.

All the results here are well known in the literature. We only give a sketch of the proofs to give the reader a sense of how these objects work.

\paragraph{Independence of the point $P$. }
The construction is meaningful only if the algebra of theta functions is independent of the point $P$ chosen to end the broken lines. This is indeed the case, as can be seen as follows. 
 
Take $P$, $Q$ two points in the diagram, $\gamma$ a path from $P$ to $Q$ and $\Phi_{\gamma}$ the wall-crossing automorphism along $\gamma$. It is an algebra automorphism of $\mathbb C[t^{\pm 1},x^{\pm 1},y^{\pm 1}]\llbracket \alpha_i\rrbracket$ that only depends on $P$ and $Q$, as the diagram is consistent. Let $\vartheta_v^{(P)}$ and $\vartheta_v^{(Q)}$ be the theta-functions computed by broken-lines terminating at points $P$ and $Q$ respectively. Then for any $v\in\mathbb Z^2$ 
\begin{equation}\label{eq:thetafunction_transformation}
    \Phi_\gamma (\vartheta_v^{(P)}) = \vartheta_v^{(Q)} \, . 
\end{equation}
This formula is proved in \cite{Gross:2011af,gross:2016,Carl_2024,gross:2019tfv}.   We use it as a black box. The reader is invited to check it by hand on examples until they are convinced of its validity. Equation \eqref{eq:thetafunction_transformation} implies that the algebra generated by theta-functions at $P$ does not depend on $P$, up to isomorphism. Another classical result is that theta-function do not contain any pole in $t$, independently of the point $P$ at which they are computed.

\paragraph{Theta-functions form a linear basis. } Using \eqref{eq:thetafunction_transformation}, we can show that theta-functions form a linear basis (over $\mathbb C[t]\llbracket \alpha_i\rrbracket$) of the algebra they generate.
Assume that there are polynomials $c_w(t)$ such that  
\begin{equation}\label{eq:thetaFunction_linearRelation}
    \sum c_w(t) \vartheta_w = 0 \,.
\end{equation} 
Write $f[w]$ for the coefficient of $x^{w_1}y^{w_2}$ in $f$. Fix $w \in\mathbb Z^2$ and compute the theta-functions near the ray $w\mathbb R_+$. 
The ray $P + w\mathbb R_+$ does not intersect any kink or wall, so there is only one broken line that reaches $P$ with final monomial $x^{w_1}y^{w_2}$ and it appears when computing $\vartheta_{w'}$ with $w'=w$. Therefore:
\begin{equation}
\vartheta_{w'}[w] = \delta_{ww'} \label{eq:product_RHS}
\end{equation}
Applying this to \eqref{eq:thetaFunction_linearRelation}, we get $c_w(t)=0$.

On the other hand, let $F^{(P)} \in \mathbb C[t^{\pm 1},x^{\pm 1},y^{\pm 1}]\llbracket \alpha_i\rrbracket$ be a compatible family, i.e. a family labeled by points in the plane of the scattering diagram that satisfies, for any path $\gamma$ with endpoints $P$ and $Q$ 
\begin{equation}
    \Phi_\gamma (F^{(P)}) = F^{(Q)} \, . 
\end{equation}
We want to show that there are Laurent polynomials $c_w(t)\in \mathbb C[t^{\pm 1}]\llbracket \alpha_i\rrbracket$ such that 
\begin{equation} 
    F = \sum c_w(t) \vartheta_w\,.
\end{equation}
By the same argument as above, we must have $c_w(t) = F^{(P)}[w]$ where $P$ is a point near $w\mathbb R_+$. Use this to define $c_w(t)$ and let $G = F- \sum c_w(t) \vartheta_w$. It satisfies $G^{(P)}[w]=0$ when $P$ is a point near $w\mathbb R_+$.

Assuming that $G\neq 0$, there must be a (non-zero) term in $G^{(P)}$ that has the lowest degree $d$ in the $\alpha_i$. Modulo terms of higher degree, wall-crossing only rescales the terms of order $d$ by powers of $t$. So these terms are non-zero in $G^{(P)}$ for any point $P$. This is a contradiction so $G= 0$. 
In particular, every product of theta-functions forms a compatible family and therefore lies in the linear span of theta-functions.

\paragraph{Product formula. } The theta functions form a linear basis of the algebra, with coefficient polynomials in $t$. Hence the product of two theta functions can be expanded as a linear combination of theta functions. The coefficients in this expansion have an expression in terms of theta functions \cite{Gross:2011af,Gross:2012du,gross:2016} , which we now explain. 

Given $v,v' \in \mathbb Z^2$, there is a unique family of polynomials $c_{vv';w}(t) \in \mathbb C[t]\llbracket \alpha_i\rrbracket$ such that 
\begin{equation}
\vartheta_v \vartheta_{v'} = \sum_w c_{vv';w}(t) \vartheta_w \,.\label{eq:product_over_basis}
\end{equation}
These polynomials can be computed using the same technique as before.   Fix $w$ and compute $\vartheta_v$ and $\vartheta_{v'}$ using theta-functions at a point $P$ near the ray $w \mathbb R_+$. Their product is:
\begin{equation}
\vartheta_v\vartheta_{v'}= \sum_{\beta,\beta'} a_\beta a_{\beta'} t^{k_\beta + k_{\beta'}}x^{u_1 + u'_1} y^{u_2+u'_2} \label{eq:product_LHS}
\end{equation}
where the sum runs over pairs of broken lines $\beta,\beta'$ with asymptotic directions $v$ and $v'$ and ending at $P$  with final monomials $a_\beta t^{k_\beta} x^{u_1} y^{u_2}$ and $a_{\beta'} t^{k_{\beta'}} x^{u'_1} y^{u'_2}$. Selecting the order $ [w] $ terms in \eqref{eq:product_over_basis} and using \eqref{eq:product_RHS} and \eqref{eq:product_LHS}, we get the product formula: 

\begin{equation}\label{eq:product_formula}
 c_{vv';w}(t)  = \sum_{\begin{smallmatrix}\beta,\beta'\\ u+u'=w\end{smallmatrix}} a_\beta a_{\beta'} t^{k_\beta + k_{\beta'}}
\end{equation}
where the sum now runs only over pairs $\beta,\beta'$ of broken lines ending near the ray $w \mathbb R_+$ that satisfy $u+u'=w$.

\paragraph{Theta-function Algebra and the threefold. } 
We denote the algebra of theta-functions by  
\begin{equation}\label{eq:theta_basis}
    \mathcal{A} = \bigoplus\limits_{v \in \mathbb{Z}^2} \mathbb{C}[t]\llbracket \alpha_i\rrbracket \, \vartheta_v \, . 
\end{equation}
This can be seen as an abstract algebra with the product \eqref{eq:product_formula} or as a subalgebra of $\mathbb C[t,x^{\pm 1},y^{\pm 1}]\llbracket \alpha_i\rrbracket$ depending on a point $P$ in the scattering diagram, but independent of it up to isomorphism. 
The M-theory geometry $\mathcal X_{\{\alpha_i\}}$ dual to the brane web with deformation parameters $\{\alpha_i\}$ is then \cite{Alexeev:2024bko}:
\begin{equation}
    \mathcal X_{\{\alpha_i\}} = \operatorname{Spec}(\mathcal A) \, .
\end{equation}
The presentation \eqref{eq:theta_basis} with relations derived from \eqref{eq:product_formula} is the most natural for general considerations and some calculations (see Appendix \ref{sec:Tcones}). In most cases however, we can find simpler presentations with finitely many generators and relations. 

In the cases we study in this section, in particular when there is only one $7$-brane, we have $\vartheta_w \vartheta_{w'} = \vartheta_{w+w'}$ if $w$ and $w'$ lie in the same region of the diagram and are not separated by any wall or kink. 

The algebra given by the theta functions in terms of generators and relations gives an explicit form of the equations of the deformation of the toric geometry of the GTP after adding 5-branes ending on 7-branes. These equations are independent of the choice of $P$, but the theta functions themselves might change.

A natural question is how many theta functions are needed to recover the full equations, since many of them will have trivial relations between them. If there are only kinks (which means we are in the toric case), the theta functions correspond exactly to integral points of the dual cone. In order to recover the full sets of generators of the geometry, we need to find a Hilbert basis of each cone. This can be done using a Hirzebruch-Jung resolution of the original cone (see Appendix \ref{sec:: HJalgo} for details). Even if we turn on deformations, we will only need to add those needed to make the dual lattice saturated.

\begin{example}[Threefold associated to deformed $T_3$]
    We can now investigate the algebra of the theta functions. Since each of the three cones is unimodular, we don't need to add any additional theta functions. One can check this explicitly with the $\vartheta_{(1,-2)}$ computed above. Indeed, a direct calculation shows it can be expressed as\footnote{Equivalently, $\vartheta_{(1,-2)}=\vartheta_{(0,-1)}^2\vartheta_{(1,0)}$ with $\vartheta_{(1,0)}=t^3x=\vartheta_{(0,-1)}\vartheta_{(1,1)}-\alpha t$. } 
\begin{equation}
    \vartheta_{(1,-2)} = \vartheta_{(0,-1)}^3 \vartheta_{1,1} - \alpha t \vartheta_{(0,-1)}^2 = \vartheta_{(0,-1)}^2 (\vartheta_{(0,-1)}  \vartheta_{1,1} - \alpha t) \, . 
\end{equation}

Let us now use the algebra to describe the associated threefold. The relation satisfied by the three generators is 
\begin{equation}
    \vartheta_{-1,0}\vartheta_{0,-1}\vartheta_{1,1}=t^3+\alpha \vartheta_{-1,0} t \, . 
\end{equation}
In other words, the associated threefold can be defined by 
\begin{equation}
\label{eq:T3deformedeq}
    \mathcal{X} = \{ (X,Y,Z,W) \in \mathbb{C}^4 | XYZ = W^3 + \alpha X W \} \, . 
\end{equation}
Setting $\alpha = 0$ recovers the familiar toric case. 

The reader should not deduce from the previous example that it is enough to consider the theta functions corresponding to the asymptotic directions in the brane web. Having a look at our third running example quickly shows that the situation is more complex in general. There, one actually needs eight theta functions, as shown in Figure \ref{fig:Tcone_(2,1)}. 
 
\begin{examplefigure}
\centering
\begin{minipage}[t]{.45\linewidth}
\centering\vspace{0pt}
\begin{tikzpicture}[x=.90cm,y=.90cm]
    \draw[gray,step=1] (-2.4,-1.4) grid (2.4,1.4);
    \foreach \a/\b in {-2/1,2/1,0/-1}
        \draw[very thick] (0,0)--({1.15*\a},{1.15*\b});
    \foreach \a/\b in {-1/1,0/1,1/1,-1/0,1/0}
        \draw[dashed] (0,0)--({1.15*\a},{1.15*\b});
    \foreach \a/\b/\pos in {-2/1/above left,-1/1/above,0/1/above,1/1/above,2/1/above right,-1/0/left,1/0/right,0/-1/below}{
        \fill (\a,\b) circle (1.7pt);
        \node[\pos,font=\scriptsize] at (\a,\b) {$(\a,\b)$};
    }
    \node[below,font=\small] at (0,-1.9) {Eight Hilbert-basis directions};
\end{tikzpicture}
\end{minipage}\hfill
\begin{minipage}[t]{.53\linewidth}
\centering\vspace{0pt}
\begin{tikzpicture}[x=.70cm,y=.70cm,
    every node/.style={font=\scriptsize}]
    \draw[very thick] (-3.2,1.6)--(0,0)--(3.2,1.6);
    \node[below left] at (-2.8,1.4) {$t$};
    \node[below right] at (2.8,1.4) {$t$};
    \draw[wall] (0,-1)--(0,0);
    \draw[wall] (0,0)--(0,4.2);
    \node[above] at (0,4.2) {$1+\alpha/y$};
    \node[below] at (0,-1) {$t^2+\alpha/y$};
    \coordinate (P) at (1,2);
    \draw[broken] (P)--(1,3.7) node[above] {$\vartheta_{0,1}$};
    \draw[broken] (P)--(1,-.5) node[below] {$\vartheta_{0,-1}$};
    \draw[broken] (P)--(3.3,2) node[right] {$\vartheta_{1,0}$};
    \draw[broken] (P)--(2.3,3.3) node[above right] {$\vartheta_{1,1}$};
    \draw[broken] (P)--(3.5,3.25) node[right] {$\vartheta_{2,1}$};
    \draw[broken] (P)--(-3.2,2) node[left] {$\vartheta_{-1,0}$};
    \draw[broken] (P)--(0,1)--(-3.2,1);
    \draw[broken] (P)--(-.8,3.8) node[above left] {$\vartheta_{-1,1}$};
    \draw[broken] (P)--(0,2)--(-1.5,3.5);
    \draw[broken] (P)--(-2.2,3.6) node[left] {$\vartheta_{-2,1}$};
    \draw[broken] (P)--(0,2)--(-2.4,3.2);
    \draw[broken] (P)--(0,1.5)--(-2.8,2.9);
    \fill (P) circle (2pt);
    \node[below right,fill=white,inner sep=1pt] at (P) {$P=(1,2)$};
\end{tikzpicture}
\end{minipage}
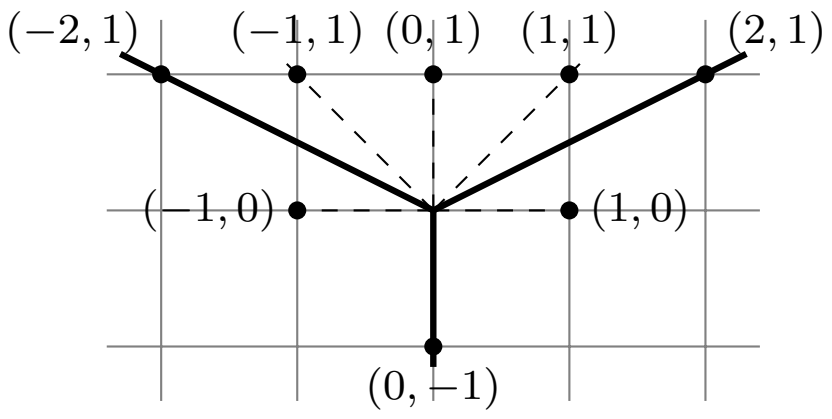
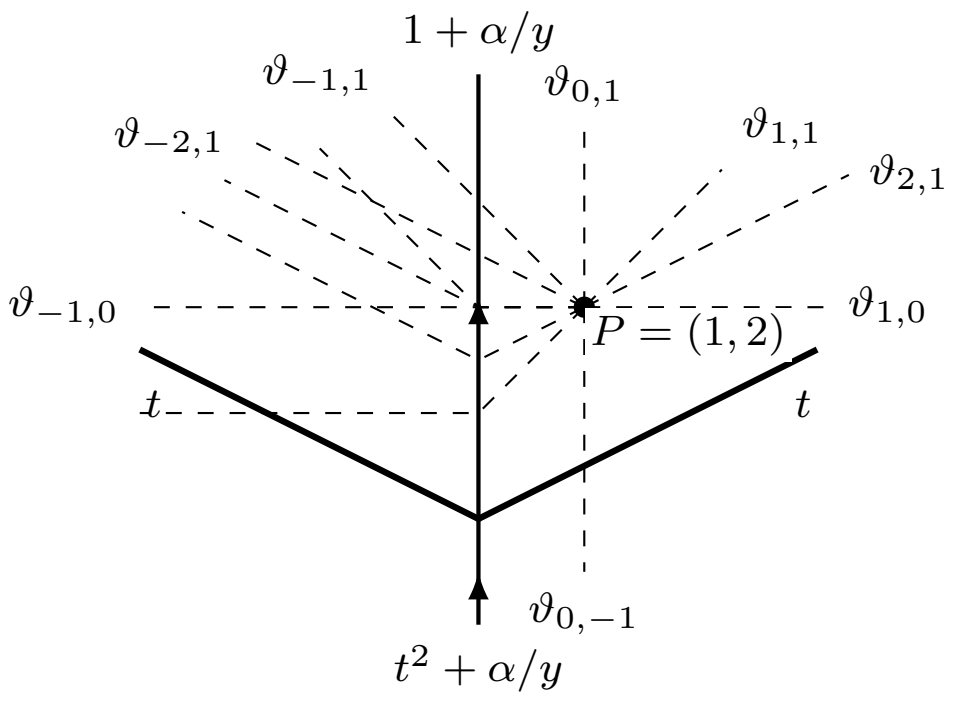
\caption{The $(2,1)$ T-cone. Left: the three kink directions (solid) require five additional Hilbert-basis directions (dashed). Right: broken lines for the eight theta functions at $P=(1,2)$. Crossings of kinks preserve the slope and supply powers of $t$; bends at the upper vertical wall supply powers of $\alpha$.}
\label{fig:Tcone_(2,1)}
\end{examplefigure}

\end{example}
 
\begin{example}[The empty scattering diagram]
    For an empty scattering diagram, we have $\vartheta_{n,m} = x^n y^m$ for any $(n,m)\in\mathbb Z^2$. Therefore $\mathcal A = \mathbb C[x^{\pm 1},y^{\pm 1},t]$ and $\mathcal X = \mathbb C \times (\mathbb C^*)^2$.
\end{example}

\begin{example}[A single wall]
    Consider a scattering diagram with a single horizontal kink and a single horizontal left-bound incoming wall (see Figure \ref{fig:diag_wall}). After making the diagram consistent and merging the walls and kinks, we get a diagram with a single horizontal wall with wall-crossing function $f(x,t) \in\mathbb C[x,t]$. The algebra $\mathcal A$ is then generated by $w_{\pm} = \vartheta_{\pm 1,0}$ and $u = \vartheta_{0,1},v = \vartheta_{0,-1}$, which satisfy:
\begin{subequations}\label{eq:CY3_wall}
\begin{gather}
    w_+ w_- = 1\\
    u v = f(w_+,t)
\end{gather}
\end{subequations}
Geometrically, $\mathcal X$ is a hypersurface in $\mathbb C^* \times \mathbb C^3$.

\begin{figure}[t]
    \centering
    \begin{tikzpicture}[x=1cm,y=1cm,
    every node/.style={font=\scriptsize}]
        \tikzmath{\x=2;\y=1;\z=.9;}

        \draw[wall] (\x,0) -- node[pos=0,right] {$f(x,t)$} (0,0);
        \draw[wall] (0,0) -- node[pos=1,left] {$f(x,t)$}(-\x,0);
        \tikzmath{\px= -.2;\py=.5;}
        \node [focalPoint] at (\px,\py) {};
        \draw[broken] (\px,\py) -- node[pos=1,above] {$u$} (\px,1.5*\y);
        \draw[broken] (\px,\py) -- (\px,-\y);
        \draw[broken] (\px,\py) -- (\px+\py,0) -- (\px+\py, -\y);
        \draw[broken] (\px,\py) -- (\px+2*\py,0) -- (\px+2*\py, -\y);
        \draw[broken] (\px,\py) -- (\px+3*\py,0) -- (\px+3*\py, -\y);
        \draw[broken] (\px,\py) -- node[pos=1,left] {$w_-$} (-\x,\py);
        \draw[broken] (\px,\py) -- node[pos=1,right] {$w_+$}(\x,\py);

        \draw[decorate,decoration={brace, amplitude=6pt},thick] 
            (\px+3*\py+.1,-\y)  -- (\px-.1,-\y)  node[pos=.5,below=6pt] {$v$};
    \end{tikzpicture}
    \caption{Theta functions for a single wall.}
    \label{fig:diag_wall}
\end{figure}
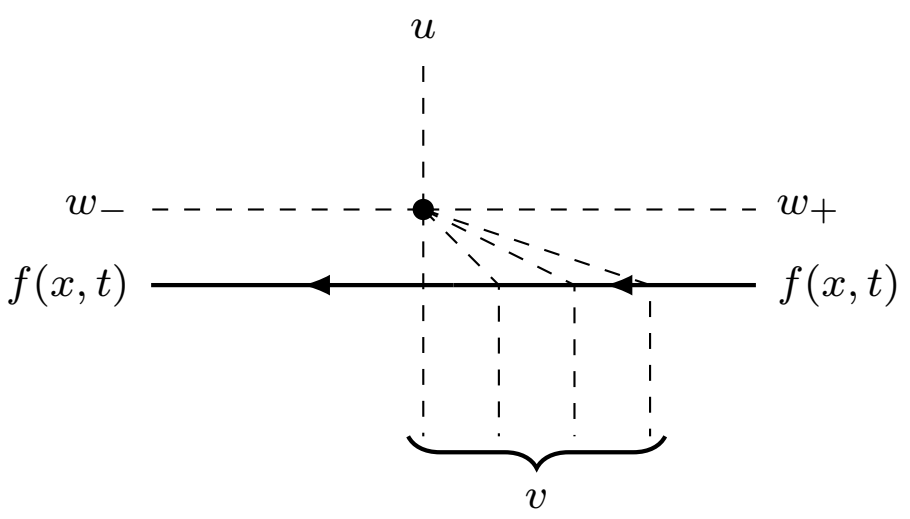

For a stack of $n$ $5$-branes, with $k$ of them ending on the right on $7$-branes according to a partition $\lambda \vdash k$ and $n-k$ going to infinity, we have:
\begin{equation}
f(w_+,t) = t^{n-k}\prod_i (t^{\lambda_i} + \alpha _i w_+)
\end{equation}
This can be derived directly using the method of \cite{Collinucci:2026kom}: choose an $\operatorname{SL}(2,\mathbb Z)$-duality frame where the $5$-branes are D5s and the $7$-branes are D7s. Then,  compactify on a circle a direction transverse to the D5 branes and parallel to the D7 branes, and apply T-duality. The D5 and D7 branes become D6 branes wrapping the curve $f(w_+,t) = 0$. Uplift to M-theory yields the threefold \eqref{eq:CY3_wall}.

\end{example}

\begin{example}[Diagrams for $\mathbb{C}^3$, $T_N$ and affine toric threefolds]
    Consider a scattering diagram consisting only of three kinks constructed from the toric diagram of an area 1/2 triangle (see Figure \ref{fig:GTP_T1}). Without loss of generality, we construct the initial scattering diagram with three kinks in the directions $(-1,0)$,$(0,-1)$ and $(1,1)$ all with the same wall-crossing function $t \in \mathbb{C}[x,t]$. The diagram is already consistent. The algebra $\mathcal{A}$ is generated by $\vartheta_{-1,0}\,, \vartheta_{0,-1},\, \vartheta_{1,1}$ and $t$ with the relation
\begin{equation}
    \vartheta_{-1,0} \vartheta_{0,-1} \vartheta_{1,1}=t\,.
\end{equation}
More generally, the toric diagram consisting of a triangle with sides of lattice length $N$ corresponds to the scattering diagram with three kinks in the directions $(-1,0)$,$(0,-1)$ and $(1,1)$ and with wall-crossing function $t^N$ (see Figure \ref{fig:GTP_TN}). The algebra $\mathcal{A}$ is generated by $\vartheta_{-1,0}\,, \vartheta_{0,-1},\, \vartheta_{1,1}$ and $t$ which satisfy 
\begin{equation}
    \vartheta_{-1,0} \vartheta_{0,-1} \vartheta_{1,1}=t^N\,.
\end{equation}

More generally, one can construct a consistent scattering diagram with only kinks and no walls for an arbitrary toric polygon $\mathcal P$ (not tessellated). The theta-function construction reproduces the affine toric threefold associated with $\mathcal P$ \cite{Gross:2011af,arguz:2023_abc}
\end{example}

\begin{example}[T-cones]
    Another important class of examples are T-cones, which serve as building blocks for resolutions of GTPs \cite{Arias-Tamargo:2024fjt}, \cite{CarrenoBolla:2024fxy}. They are labeled by two integers $(p,q)$ with $0 \leq q \leq p$ and $\gcd{(p,q)}=1$. Their GTPs are triangles with base length and height $p$ (see Figure \ref{fig:Tcone_GTP}). The position of their apex is parameterized by $q$. All $p$ 5-branes of the base end on the same 7-brane. Note that the toric $\mathbb{C}^3$ triangle corresponds to $(p,q)=(1,0)$.

\begin{figure}
\begin{subfigure}{.3\textwidth}
    \centering
    \raisebox{.7\height}{ \scalebox{1}{  
    \begin{tikzpicture}[x=.5cm,y=.5cm]
        \draw[gridline] (-.5,0) -- (1.5,0);
        \draw[gridline] (-.5,1) -- (1.5,1);
        \draw[gridline] (0,-.5) -- (0,1.5);
        \draw[gridline] (1,-.5) -- (1,1.5);
        \draw[ligne] (0,0) -- (0,1) -- (1,0)--(0,0);
        \node[bd] at (0,0) {};
        \node[bd] at (0,1) {};
        \node[bd] at (1,0) {};
    \end{tikzpicture}}}
    \caption{}
    \label{fig:GTP_T1}
\end{subfigure}
\begin{subfigure}{.3\textwidth}
    \centering
    \raisebox{-.5\height}{ \scalebox{1}{  
    \begin{tikzpicture}[x=.5cm,y=.5cm] 
        \draw[gridline] (-1,-4.5)--(-1,0.5); 
        \draw[gridline] (0,-4.5)--(0,0.5); 
        \draw[gridline] (1,-4.5)--(1,0.5); 
        \draw[gridline] (2,-4.5)--(2,0.5); 
        \draw[gridline] (3,-4.5)--(3,0.5); 
        \draw[gridline] (-1.5,-4)--(3.5,-4); 
        \draw[gridline] (-1.5,-3)--(3.5,-3); 
        \draw[gridline] (-1.5,-2)--(3.5,-2); 
        \draw[gridline] (-1.5,-1)--(3.5,-1); 
        \draw[gridline] (-1.5,0)--(3.5,0); 
        \draw[ligne] (3,-4)--(-1,0); 
        \draw[ligne] (-1,0)--(-1,-4); 
        \draw[ligne] (-1,-4)--(3,-4); 
        \node[bd] at (-1,0) {}; 
        \node[bd] at (-1,-1) {}; 
        \node[bd] at (-1,-2) {}; 
        \node[bd] at (-1,-3) {}; 
        \node[bd] at (0,-1) {}; 
        \node[bd] at (1,-2) {}; 
        \node[bd] at (2,-3) {}; 
        \node[bd] at (-1,-4) {}; 
        \node[bd] at (-1,-4) {}; 
        \node[bd] at (3,-4) {}; 
        \node[bd] at (0,-4) {}; 
        \node[bd] at (1,-4) {}; 
        \node[bd] at (2,-4) {}; 
        \draw[
            decorate,
            decoration={brace, amplitude=8pt},
            thick
        ] (-1.5,-4.) -- (-1.5,0)  node[pos=.5,left=8pt] {$N$};
        \draw[
            decorate,
            decoration={brace, amplitude=8pt},
            thick
        ] (3,-4.5) -- (-1,-4.5)  node[pos=.5,below=8pt] {$N$};
    \end{tikzpicture}}}
    \caption{}
    \label{fig:GTP_TN}
\end{subfigure}
\begin{subfigure}{.3\textwidth}
    \centering
    \raisebox{-.5\height}{\scalebox{1}{  
    \begin{tikzpicture}[x=.5cm,y=.5cm] 
        \draw[gridline] (-1,-4.5)--(-1,0.5); 
        \draw[gridline] (0,-4.5)--(0,0.5); 
        \draw[gridline] (1,-4.5)--(1,0.5); 
        \draw[gridline] (2,-4.5)--(2,0.5); 
        \draw[gridline] (3,-4.5)--(3,0.5); 
        \draw[gridline] (-1.5,-4)--(3.5,-4); 
        \draw[gridline] (-1.5,-3)--(3.5,-3); 
        \draw[gridline] (-1.5,-2)--(3.5,-2); 
        \draw[gridline] (-1.5,-1)--(3.5,-1); 
        \draw[gridline] (-1.5,0)--(3.5,0); 
        \draw[ligne] (3,-4)--(0,0); 
        \draw[ligne] (0,0)--(-1,-4); 
        \draw[ligne] (-1,-4)--(3,-4); 
        \node[bd] at (0,0) {}; 
        \node[bd] at (-1,-4) {}; 
        \node[bd] at (3,-4) {}; 
        \node[wd] at (0,-4) {}; 
        \node[wd] at (1,-4) {}; 
        \node[wd] at (2,-4) {}; 
        \draw[
            decorate,
            decoration={brace, amplitude=8pt},
            thick
        ] (-1.5,-4.) -- (-1.5,0)  node[pos=.5,left=8pt] {$p$};
        \draw[
            decorate,
            decoration={brace, amplitude=8pt},
            thick
        ] (3,-4.5) -- (-1,-4.5)  node[pos=.5,below=8pt] {$p$};
        \draw[
            decorate,
            decoration={brace, amplitude=8pt},
            thick
        ] (-1,0.5) -- (0,0.5)  node[pos=.5,above=8pt] {$q$};
    \end{tikzpicture}}}
    \caption{}
    \label{fig:Tcone_GTP}
\end{subfigure}
\caption{GTPs for the toric $\mathbb C^3$ diagram (\subref{fig:GTP_T1}), the toric $T_N$ junction (\subref{fig:GTP_TN}) and a primitive T-cone (\subref{fig:Tcone_GTP})}
\end{figure}
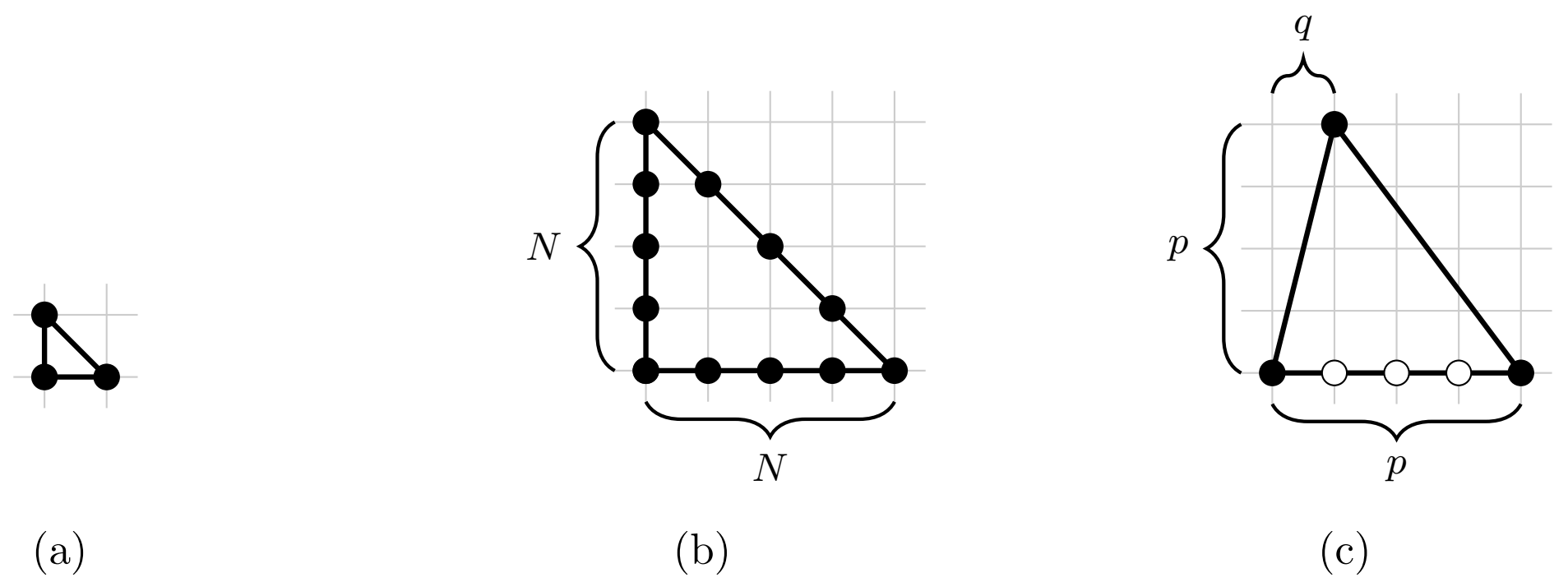

Originally, T-cones were introduced in the mathematical literature \cite{52480b2c6c3e441e8ca0b168fbd5f447},\cite{akhtar2014singularitycontent} . They correspond to quotient surface singularities that admit a one-parameter $\mathbb{Q}-$Gorenstein smoothing. More explicitly, the surfaces are smoothings of $\mathbb{C}^2/ \mathbb{Z}_{p^2}$ constructed as follows: consider the deformation of the $A_{p-1}$ singularity given by
\begin{equation}
    x y = z^p+\alpha\,
\end{equation}
with $\alpha$ the deformation parameter and apply the $\mathbb{Z}_p$ orbifold action $(x,y,z,\alpha)\sim (\omega x, \omega^{-1}y, \omega ^r z, \alpha)$, $\omega^p=1$,  with $0 \leq r \leq p-1$ and $qr \equiv 1 \mod p$. In order to recover a 3-fold we take the total space of the canonical bundle over this smooth surface. If we don't apply the $\mathbb{Z}_p$ action on the fiber coordinate $w$ , $K_S$ will be a non-trivial torsion class. To make the 3-fold  Calabi-Yau, we act on the fiber coordinate $w$ as $w \sim \omega^{-r} w$. The geometry is described by the invariants under the orbifold action and their relations and will generically not be a complete intersection. 

We can construct the same geometry using scattering diagrams. The initial scattering diagram is composed of three kinks at $(0,-1)$,$(-p,q),(p,p-q)$ with wall-crossing functions $t^p, \, t,\, t$ respectively and an incoming wall at $(0,-1)$ with wall-crossing function  $f(x,t)=1+t^{-p}\frac{\alpha}{y}$. It is straightforward to compute the consistent scattering diagram.  

For simplicity, we will focus on the simplest non-trivial case $(p,q)=(2,1)$. The theta functions satisfy polynomial relations equivalent to: 

\begin{equation}\label{eq:Tcone_rank_relation}
   \text{rank} \left(
\begin{array}{cccc}
 \vartheta_{0,1} & t & \vartheta_{1,1} & \vartheta_{-1,1} \\
 t & \vartheta_{0,-1} & \vartheta_{1,0} & \vartheta_{-1,0} \\
 \vartheta_{1,1} & \vartheta_{1,0} & \vartheta_{2,1} & \alpha +\vartheta_{0,1} \\
 \vartheta_{-1,1} & \vartheta_{-1,0} & \alpha +\vartheta_{0,1} & \vartheta_{-2,1} \\
\end{array}
\right) \leq 1
\end{equation}

In Appendix \ref{sec:Tcones}, we show explicitly that the broken line procedure reproduces the geometric quotient construction outlined above, for the general T-cone. The explicit change of variables between theta-functions and invariant monomials in $(x,y,z,w)$ also identifies which open subsets of the $T$-cones correspond to the external legs of the scattering diagram and, in turn, how the $T$-cones can be glued together to form larger threefolds (using the method of Section \ref{sec:ResolvedGeometries}).

\end{example}

\subsection{Consistency and the $s$-rule}
\label{sec:s-rule}

In this subsection, we address certain subtle issues related to negative powers of $t$ that can appear in the reduced consistent scattering diagram associated to a brane web.\footnote{We always assume the diagram has been maximally simplified by merging all walls sharing the same support. } Roughly speaking, $t$ keeps track of the number of 5-branes located on a wall. When several distinct powers of $t$ cohabit on the same wall, this means that the presence of 7-branes makes the number of 5-branes on the wall ambiguous. This is particularly clear in the examples considered above: in the deformed $T_3$ case, the $(-1,0)$ wall carries the wall-crossing function $t^3 + \frac{\alpha t}{x}$, corresponding to the fact that there are respectively one and three 5-branes on each side of the 7-brane. 

Given this, one can expect that negative powers of $t$ are associated to the presence of anti-5-branes, leading to supersymmetry violation. However, the story appears to be a bit more subtle. Before making general statements, it is enlightening to have a look at two examples. 

\begin{example}[Examples of diagrams with poles in $t$. ]
We describe two extremely simple examples which display poles in $t$. 
\begin{itemize}
    \item The simplest example of s-rule violation by a GTP is (see Figure 6 of \cite{Benini:2009gi})
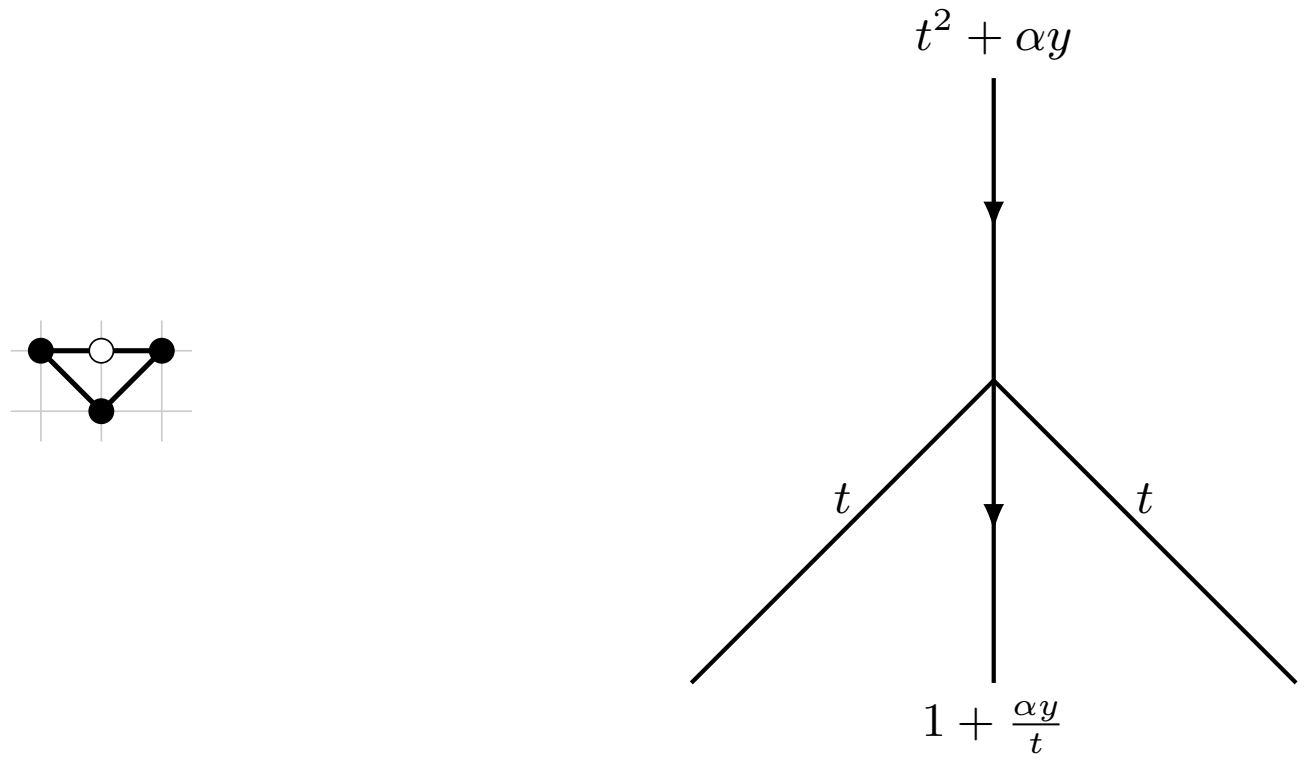
\begin{center}
\begin{tikzpicture}
    \node at (-4,0) {\raisebox{-.5\height}{ \scalebox{.8}{  
        \begin{tikzpicture}[x=.5cm,y=.5cm] 
            \draw[gridline] (-1,-0.5)--(-1,1.5); 
            \draw[gridline] (0,-0.5)--(0,1.5); 
            \draw[gridline] (1,-0.5)--(1,1.5); 
            \draw[gridline] (-1.5,0)--(1.5,0); 
            \draw[gridline] (-1.5,1)--(1.5,1); 
            \draw[ligne] (-1,1)--(0,0); 
            \draw[ligne] (0,0)--(1,1); 
            \draw[ligne] (1,1)--(-1,1); 
            \node[bd] at (0,0) {}; 
            \node[bd] at (1,1) {}; 
            \node[bd] at (-1,1) {}; 
            \node[wd] at (0,1) {}; 
        \end{tikzpicture}}}};
    \node at (2,0) { 
        \begin{tikzpicture}
            \tikzmath{\x=2;\y=1.5;}
            \draw[kink]  (0,0) -- node[above]{$t$} (\x,-\x) ;
            \draw[kink]  (0,0) -- node[above]{$t$}  (-\x,-\x) ;
            \draw[wall]  (0,0) --     (0,-\x)  node[below]{$1 +\frac{\alpha y}{ t}$} ;
            \draw[wall]  (0,\x)  node[above]{$t^2 +\alpha y$} --   (0,0)  ;
        \end{tikzpicture} };
\end{tikzpicture}
\end{center}
The reduced consistent scattering diagram obtained following the algorithm above is shown on the right. When $\alpha \neq 0$, one has a pole $\frac{\alpha y}{t}$ in the wall-crossing function, signalling that attempting to move the 7-brane across the brane junction would create an anti-brane. 
\item Another example is given by a simple intersection of branes: 
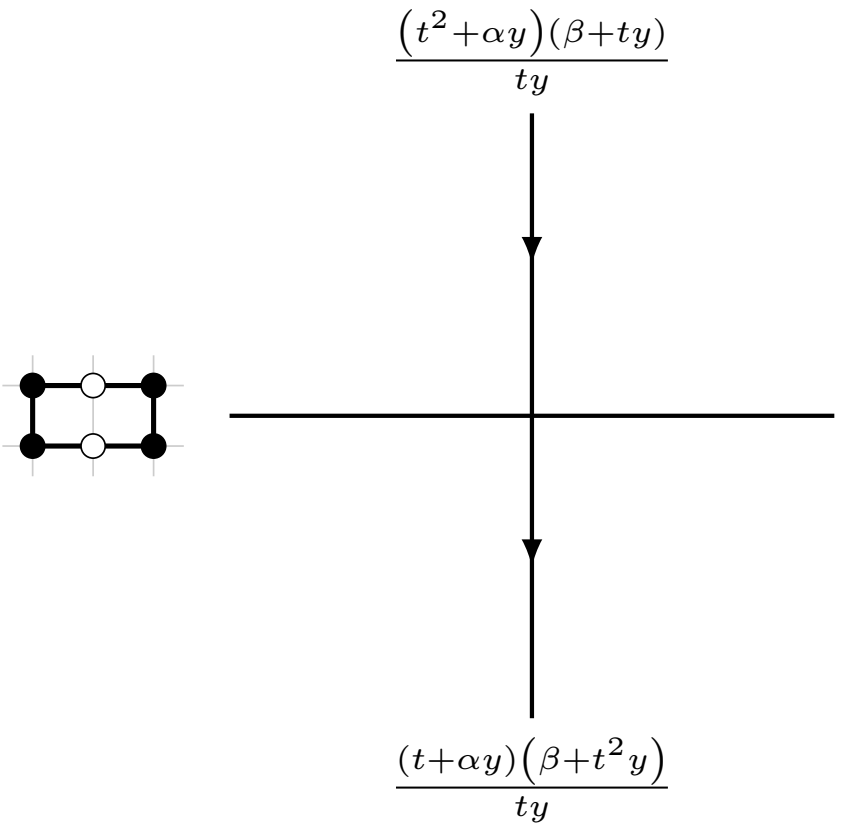
\begin{center}
\begin{tikzpicture}
    \node at (-1,0) {\raisebox{-.5\height}{ \scalebox{.8}{  
        \begin{tikzpicture}[x=.5cm,y=.5cm] 
            \draw[gridline] (0,-1.5)--(0,0.5); 
            \draw[gridline] (1,-1.5)--(1,0.5); 
            \draw[gridline] (2,-1.5)--(2,0.5); 
            \draw[gridline] (-0.5,-1)--(2.5,-1); 
            \draw[gridline] (-0.5,0)--(2.5,0); 
            \draw[ligne] (0,-1)--(0,0); 
            \draw[ligne] (0,0)--(2,0); 
            \draw[ligne] (2,0)--(2,-1); 
            \draw[ligne] (2,-1)--(0,-1); 
            \node[bd] at (0,0) {}; 
            \node[bd] at (2,0) {}; 
            \node[bd] at (2,-1) {}; 
            \node[bd] at (0,-1) {}; 
            \node[wd] at (1,0) {}; 
            \node[wd] at (1,-1) {}; 
        \end{tikzpicture}}}};
    \node at (2,0) { 
        \begin{tikzpicture}
            \tikzmath{\x=2;\y=1.5;}
            \draw[kink]  (0,0) -- (\x,0) ;
            \draw[kink]  (0,0) -- (-\x,0) ;
            \draw[wall]  (0,0) --     (0,-\x)  node[below]{$\frac{(t+\alpha  y) \left(\beta +t^2 y\right)}{t y}$} ;
            \draw[wall]  (0,\x)  node[above]{$\frac{\left(t^2+\alpha  y\right) (\beta +t y)}{t y}$} --   (0,0)  ;
        \end{tikzpicture} };
\end{tikzpicture}
\end{center}
For this GTP, there is a stack of two 5-branes suspended between two 7-branes with deformation parameters $\alpha,\beta$. This web is supersymmetric, but the consistent scattering diagram associated to it has negative powers of $t$ if both $\alpha\neq 0 \neq \beta$. This type of web is dubbed inconsistent in the language of \cite{Alexeev:2024bko}. 
\end{itemize}
\end{example}

In order to make general statements, we first set up some notations. 
\begin{itemize}
    \item For a given brane web, we call $\mathcal{C}$ the extended Coulomb Branch. By definition, $\mathcal{C}$ is the space of parameters that completely defines the shape of the web; in a gauge theory picture, these are Coulomb branch parameters, and masses of hypermultiplets and instantons. On $\mathcal{C}$ the web is always supersymmetric by definition. In particular, certain parameters are constrained by supersymmetry: in the second example above, the transverse positions of the two 7-branes are forced to be equal. 
    \item Despite always being supersymmetric, the web can be \emph{inconsistent} in the sense of \cite{Alexeev:2024bko}. We introduce the \emph{inconsistent locus} $\mathcal{I} \subset \mathcal{C}$ as the locus where the web is inconsistent. Note in particular that if Hanany-Witten moves can be performed and yield two disconnected webs (this includes a single disconnected 7-brane), then the web is inconsistent. 
    \item A third notion, which is the one we are really interested in for this paper, corresponds to the threefold being indeed given as the spectrum of the theta function algebra. For lack of a better word, we say that a point in $\cal C$ is \emph{scatterable} when this is the case. The \emph{non-scatterable locus} $\mathcal{N} \subset \cal C$ is the subset of the extended Coulomb branch where the web is not scatterable.  
\end{itemize}

Based on the above examples, we now propose the following conjectures: 
\begin{enumerate}
    \item[$(i)$] For any brane web, $\mathcal{N} = \mathcal{C}$ $\Longleftrightarrow$ there is a pole in $t$ in $\mathfrak{D}$.
    \item[$(ii)$] $\mathcal{N} \subset \mathcal{I}$. 
\end{enumerate}
In particular, $(i)$ says that $\cal{X}$ is given by $\mathrm{Spec}(\cal{A})$ only when no poles in $t$ are present, since a pole implies $\mathcal{C} \backslash \mathcal{N} = \emptyset$. Also, $(ii)$ says that if the web is consistent, then it is scatterable, as one could hope. However, we see in the example below that the converse is not true: there are scatterable webs which are inconsistent. 
Note also that $(i)$ implies, in the particular case $\mathcal{N} = \mathcal{C}=\emptyset$, 
\begin{enumerate}
    \item[$(i')$] If the s-rule is violated, in the sense that one can move\footnote{The allowed moves should keep the 7-branes in generic positions with respect to each other, while preserving supersymmetry, which implies not colliding two 7-branes. } 7-branes and create a negative number of 5-branes, then a pole in $t$ develops.  
\end{enumerate}
 On $\mathcal{N}$, the method a priori doesn't work, and one might have to resort to other techniques, like those of \cite{Collinucci:2026kom}.  

Examples of scatterable but inconsistent webs are T-cones and the $(1,-3)$-curve studied below.\footnote{Despite the terminology, "inconsistent" webs are perfectly fine objects to study, being of physical interest, supersymmetric, and accessible through scattering diagram techniques. Physical relevance of the non scatterable notion is confirmed by the observation in \cite{Garavaglia:2026_1,Garavaglia:2026_2} that it is related to the appearance of distributions in the superconformal index. } A more extensive study of this topic is left for future work. \emph{From now on, until the end of this paper, we study scatterable webs, which can be either consistent or inconsistent. }

\begin{example}
The last example can be generalized to a 2-parameter family as follows:
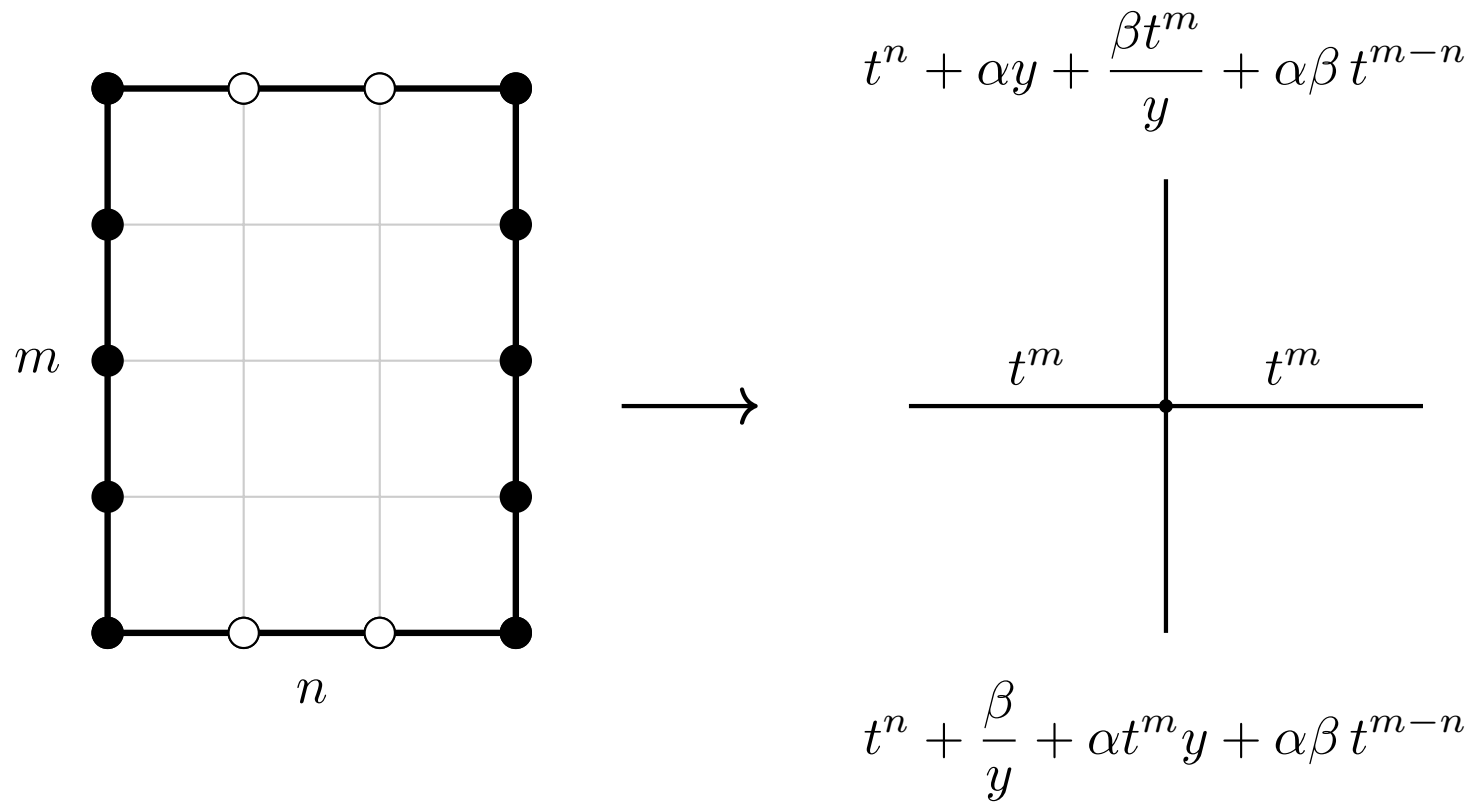
\begin{center}
    \begin{tikzpicture}[
    every node/.style={font=\small},
    gtpblack/.style={
        circle,draw=black,fill=black,
        inner sep=0pt,minimum size=2mm
    },
    gtpwhite/.style={
        circle,draw=black,fill=white,
        inner sep=0pt,minimum size=2mm
    }
]
\def\GTPn{3}
\def\GTPm{4}
\begin{scope}[shift={(-6,-1.5)},x=.9cm,y=.9cm]
    \draw[step=1,black!20]
        (0,0) grid (\GTPn,\GTPm);
    \draw[very thick]
        (0,0) rectangle (\GTPn,\GTPm);

    \foreach \i in {0,...,\GTPn}{
        \node[gtpblack] at (\i,0) {};
        \node[gtpblack] at (\i,\GTPm) {};
    }
    \foreach \j in {0,...,\GTPm}{
        \node[gtpblack] at (0,\j) {};
        \node[gtpblack] at (\GTPn,\j) {};
    }
    \ifnum\GTPn>1
        \pgfmathtruncatemacro{\GTPinterior}{\GTPn-1}
        \foreach \i in {1,...,\GTPinterior}{
            \node[gtpwhite] at (\i,0) {};
            \node[gtpwhite] at (\i,\GTPm) {};
        }
    \fi

    \node[below=5pt] at ({\GTPn/2},0) {$n$};
    \node[left=5pt] at (0,{\GTPm/2}) {$m$};
\end{scope}

\draw[->,thick] (-2.6,0) -- (-1.7,0);

\begin{scope}[shift={(1,0)}]
    \draw[thick] (0,0) -- (1.7,0)
        node[midway,above] {$t^m$};
    \draw[thick] (0,0) -- (-1.7,0)
        node[midway,above] {$t^m$};

    \draw[thick] (0,0) -- (0,1.5)
        node[above=5pt] {$\displaystyle
            t^n+\alpha y+\frac{\beta t^m}{y}
            +\alpha\beta\,t^{m-n}
        $};
    \draw[thick] (0,0) -- (0,-1.5)
        node[below=5pt] {$\displaystyle
            t^n+\frac{\beta}{y}+\alpha t^m y
            +\alpha\beta\,t^{m-n}
        $};

    \fill (0,0) circle (1.3pt);
\end{scope}
\end{tikzpicture}
\end{center}
One checks directly on the scattering diagram that there is a pole in $t$ if and only if $m<n$. 

This is in agreement with the analysis above. Indeed, for any value of $m$ and $n$, the web is at least supersymmetric when the brane web forms a simple cross. So $\mathcal{C} \neq \emptyset$ in all cases, as it contains at least $\{0\}$. 
\begin{itemize}
    \item When $m<n$, we actually have $\mathcal{C} = \mathcal{I} = \mathcal{N} = \mathbb{R}^{m-1}$, parametrized only by the positions of the D5 branes. The configuration is always inconsistent and non-scatterable, although supersymmetric. Physically, the two 7-branes of charge $[0,1]$ are forced by the s-rule to be aligned in the vertical direction. 
    \item When $m = n$, the s-rules does not require the 7-branes to be aligned. Now $\mathcal{N} \subsetneq \mathcal{I} = \mathcal{C}$. The last equality translates the fact that a Hanany-Witten move can detach a 7-brane. However the first strict inclusion means we can use scattering diagrams to find the threefold, and correspondingly, there is no pole in $t$. 
    \item For $m > n$, not only are the 7-branes not forced to be aligned, but the web is now large enough so that 7-branes can not be detached. Hence $\mathcal{N} \subseteq \mathcal{I} \subsetneq \mathcal{C}$. 
\end{itemize}
\end{example}

\section{Hanany Witten moves and the $\mathbb{P}^1$ deformation}\label{sec:hanany-witten}

In this section we will see how Hanany-Witten moves affect scattering diagrams. The 5d SCFTs engineered on two brane webs related by HW moves should be the same, there are no physical moduli associated with the 7-brane motion across its monodromy cut. We expect that the geometric avatar of this fact is that the geometries defined by two GTPs related by Hanany-Witten moves should be the same when the 7-branes are at finite distance. 

This can be seen explicitly using the theta functions. The setup is as follows: we start with a toric diagram which defines a toric variety, then we introduce a deformation by ending a set of external 5-branes in a  7-brane. This deformation is parameterized by a parameter $\alpha \in \mathbb{P}^1$. We can HW cross the 7-brane, which may result in a GTP. After sending the 7-branes to infinity this gives another toric geometry \footnote{It is worth remarking that this is not the geometry of the GTP, as the latter should have some moduli/divisors frozen in some sense.}. The original toric geometry is recovered when $\alpha=0$ and the toric geometry corresponding to the crossed diagram is found when $\alpha=\infty$ \cite{Ilten_2012}.
We will illustrate this explicitly with the HW move that transforms local $\mathbb{F}_0$ into local $\mathbb{F}_2$.  

\begin{figure}
\centering
\begin{tikzpicture}
    \node at (-2,0)     {\raisebox{-.5\height}{ \scalebox{.8}{  
        \begin{tikzpicture}[x=.5cm,y=.5cm] 
            \draw[gridline] (-1,-1.5)--(-1,1.5); 
            \draw[gridline] (0,-1.5)--(0,1.5); 
            \draw[gridline] (1,-1.5)--(1,1.5); 
            \draw[gridline] (-1.5,-1)--(1.5,-1); 
            \draw[gridline] (-1.5,0)--(1.5,0); 
            \draw[gridline] (-1.5,1)--(1.5,1); 
            \draw[ligne] (-1,-1)--(0,1); 
            \draw[ligne,goodred] (0,1)--(1,1); 
            \draw[ligne] (1,1)--(0,-1); 
            \draw[ligne] (0,-1)--(-1,-1); 
            \node[bd] at (0,1) {}; 
            \node[bd] at (1,1) {}; 
            \node[bd] at (0,-1) {}; 
            \node[bd] at (-1,-1) {}; 
        \end{tikzpicture}}}} ;
    
    \node at (2,0) {\raisebox{-.5\height}{ \scalebox{.8}{  
        \begin{tikzpicture}[x=.5cm,y=.5cm] 
            \draw[gridline] (-1,-1.5)--(-1,1.5); 
            \draw[gridline] (0,-1.5)--(0,1.5); 
            \draw[gridline] (1,-1.5)--(1,1.5); 
            \draw[gridline] (-1.5,-1)--(1.5,-1); 
            \draw[gridline] (-1.5,0)--(1.5,0); 
            \draw[gridline] (-1.5,1)--(1.5,1); 
            \draw[ligne] (-1,-1)--(0,1); 
            \draw[ligne] (0,1)--(1,-1); 
            \draw[ligne,goodred] (1,-1)--(0,-1); 
            \draw[ligne] (0,-1)--(-1,-1); 
            \node[bd] at (0,1) {}; 
            \node[bd] at (1,-1) {}; 
            \node[bd] at (0,-1) {}; 
            \node[bd] at (-1,-1) {}; 
        \end{tikzpicture}}}};
    
    \node at (-1.9,1) {\large $\mathbb{F}_0$};
    \node at (2.1,1) {\large  $\mathbb{F}_2$};
    
    \node at (-2,-2)  {\scalebox{.8}{
        \begin{tikzpicture}[x=.55cm,y=.55cm] 
            \draw[gridline] (-3.5,1) -- (3.5,1);
            \draw[gridline] (-3.5,0) -- (3.5,0);
            \draw[gridline] (-3.5,-1) -- (3.5,-1);
            \foreach \n in {0,...,6} 
            \draw[gridline] (-3+\n,-1.5) -- (-3+\n,1.5);
            
            \draw[black] (0,0)--(0,2);
            \draw[black] (0,0)--(0,-2);
            \draw[black] (0,0)--(-3,1.5);
            \draw[black] (0,0)--(3,-1.5);
            
            \fill[goodred](0,0) circle (2.4pt);
            \fill[blue](0,1) circle (2.4pt);
            \fill[blue](0,-1) circle (2.4pt);
            \fill[blue](-2,1) circle (2.4pt);
            \fill[blue](2,-1) circle (2.4pt);
            \fill[black](-1,1) circle (2.4pt);
            \fill[black](-1,0) circle (2.4pt);
            \fill[black](1,0) circle (2.4pt);
            \fill[black](1,-1) circle (2.4pt);
        \end{tikzpicture}}};

    \node at (2,-2)  {\scalebox{.8}{
        \begin{tikzpicture}[x=.55cm,y=.55cm] 
            \draw[gridline] (-3.5,1) -- (3.5,1);
            \draw[gridline] (-3.5,0) -- (3.5,0);
            \draw[gridline] (-3.5,-1) -- (3.5,-1);
            \foreach \n in {0,...,6} 
            \draw[gridline] (-3+\n,-1.5) -- (-3+\n,1.5);
            
            \draw[black] (0,0)--(0,2);
            \draw[black] (0,0)--(0,-2);
            \draw[black] (0,0)--(-3,1.5);
            \draw[black] (0,0)--(3,1.5);
            
            \fill[goodred](0,0) circle (2.4pt);
            \fill[blue](0,1) circle (2.4pt);
            \fill[blue](0,-1) circle (2.4pt);
            \fill[blue](-2,1) circle (2.4pt);
            \fill[blue](2,1) circle (2.4pt);
            
            \fill[black](-1,1) circle (2.4pt);
            \fill[black](-1,0) circle (2.4pt);
            \fill[black](1,0) circle (2.4pt);
            \fill[black](1,1) circle (2.4pt);
        \end{tikzpicture}}};
\end{tikzpicture}
\caption{Above, the polygons defining each geometry. We paint the edges corresponding to the 5-branes attached to the 7-brane to be crossed in red. Below, the generators of the dual cone. The origin is painted red, the generators corresponding to the kinks are in blue and in black are the generators added by the HJ algorithm. Both lattices are saturated.}
\label{fig:f0tof2}
\end{figure}
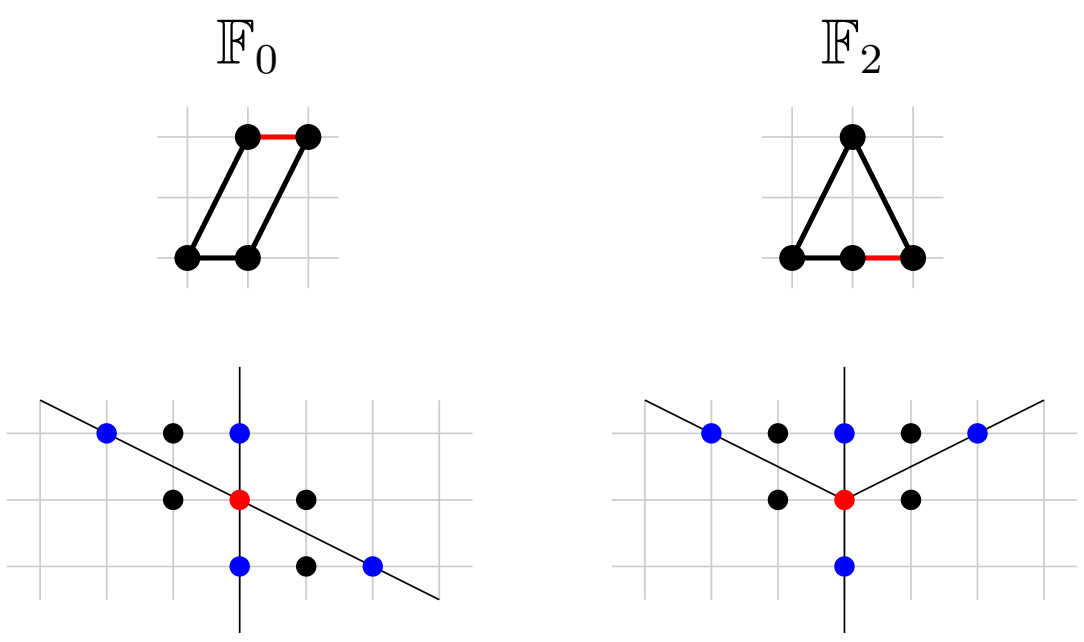

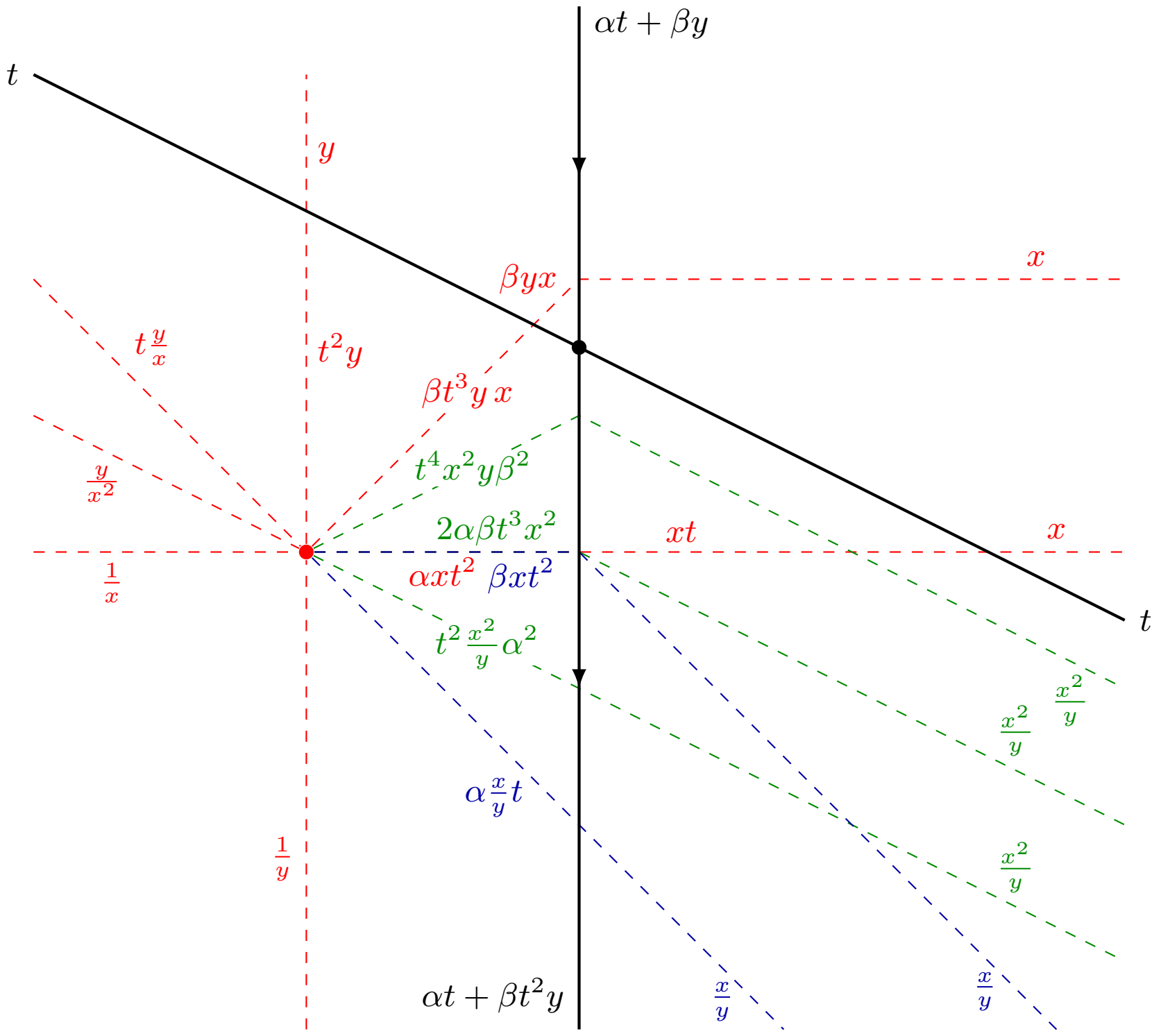
\begin{figure}
    \centering
    \begin{tikzpicture}[
        x=0.65cm,
        y=0.65cm]
    
        \node[focalPoint,goodred] (F) at  (-4,0) {};
        
        \draw[broken,goodred] (-8,0) -- (F) -- (0,0) -- (6,0) -- (8,0);
        \draw[broken,goodred] (-4,-7) -- (F) -- (-4,5) -- (-4,7);
        \draw[broken,goodred] (-8,4) -- (F);
        \draw[broken,goodred] (-8,2) -- (F);
        \draw[broken,goodred] (F) -- node[pos=.7,fill=white,inner sep=0pt] {$\beta t^3 y\,x$} (-.66666,3.33333) -- (0,4) -- (8,4);

        \node[goodred] at (-3.7,5.85) {$y$};
        \node[goodred] at (-3.5,2.95) {$t^2 y$};
        \node[goodred] at (1.5,0.3) {$x t$};
        \node[goodred] at (-0.75,4) {$\beta y x  $};
        \node[goodred] at (7,0.3) {$x $};
        \node[goodred] at (6.7,4.3) {$x$};
        \node[goodred] at (-6.85,-0.45) {$\frac{1}{x}$};
        \node[goodred] at (-4.35,-4.5) {$\frac{1}{y}$};
        \node[goodred] at (-6.25,3.0) {$t \frac{y}{x}$};
        \node[goodred] at (-7,1) {$\frac{y}{x^2}$};
        \node[goodred] at (-2.,-0.29) {$\alpha x t^2$};
        \node[goodred] at (-2.15,2.75) {};
        
        \draw[broken,goodgreen] (F) -- node[pos=.6,fill=white,inner sep=0pt] {$t^4 x^2 y\beta^2$}(0,2) --node[pos=.9,below] {$\frac{x^2}{y}$} (8,-2);
        \draw[broken,goodgreen] (F) -- (0,0) -- node[pos=.8,above] {$\frac{x^2}{y}$}(8,-4);
        \draw[broken,goodgreen] (F) -- (0,-2) -- node[pos=.8,above] {$\frac{x^2}{y}$} (8,-6);
        
        \node[goodgreen] at (-1.2,0.35) {$2\alpha\beta t^3 x^2$};
        \node[goodgreen,fill=white,inner sep=-1pt] at (-1.35,-1.35) {$t^2\frac{x^2}{y}\alpha^2$};
        
        \draw[broken,goodblue] (F) -- (0,0) -- node[pos=.85,below] {$\frac{x}{y}$} (7,-7);
        \draw[broken,goodblue] (F) -- (0,-4) -- node[pos=.7,below] {$\frac{x}{y}$} (3,-7);
        
        \node[goodblue] at (-.85,-0.35) {$\beta x t^2$};
        \node[goodblue] at (-1.25,-3.6) {$\alpha \frac{x}{y}t$};
        \node[goodblue] at (6.15,-6.8) {};
        \node[goodblue] at (2.2,-6.8) {};
        
        \draw[wall] (0,8) -- node[pos=.05,right] {$\alpha t+\beta y$} (0,3);
        \draw[wall] (0,3) -- node[pos=.95,left] {$\alpha t+\beta t^2 y$} (0,-7);
        \draw[kink] (-8,7) -- node[pos=0,left] {$t$} (0,3) -- node[pos=1,right] {$t$} (8,-1);
        \node[focalPoint] at (0,3) {};
    \end{tikzpicture} 
    \caption{Broken lines for $\mathbb{F}_0$. This is consistent for all $\alpha \neq 0$ and all $\beta$, corresponding to one patch of the $\mathbb{P}^1$}
    \label{fig:brokenline_f0}
\end{figure}

We start with the brane web of $\mathbb{F}_0$ and add a single 7-brane ending on the external $(0,1)$ brane, then HW cross it to the web of $\mathbb{F}_2$, so this crossing does not generate any white dot. The dual cone is not saturated in either case and additional generators have to be added using the Hirzebruch-Jung algorithm (see Appendix \ref{sec:: HJalgo} for details) until we have a total of 9 for each diagram. We depict this in Figure \ref{fig:f0tof2}.
We now start computing the initial scattering diagram. As usual, we add a kink for each side of the toric diagram, and add the following ingoing wall
\begin{equation}
     ( \left(\begin{smallmatrix}0\\1\end{smallmatrix}\right)\mathbb R_+,\alpha \,  +\beta\, t_1 y)\,.
\end{equation}
So $[1:0]$ corresponds to turning off the deformation. We apply the consistency algorithm, set $t_1 \to t^{-1}$ and compute the broken lines to find the following theta functions for the choice of $p$ in Figure \ref{fig:brokenline_f0}
\begin{equation}
    \begin{aligned}
        \vartheta_{0,1}&=y t^2,\; \vartheta_{-1,1}=t \frac{y}{x},\; \vartheta_{-2,1}=\frac{y}{x^2},\\\; \vartheta_{1,0}&=t^2(\alpha+\beta t y)\,x,\,\; \vartheta_{-1,0}=\frac1x,\\
        \vartheta_{2,-1}&= t^2 (\alpha+\beta t y)^2\,\frac{x^2}{y},\; \vartheta_{1,-1}=t (\alpha+\beta t y)\,\frac{x}{y},\; \vartheta_{0,-1}=\frac1{y}\;.
           \end{aligned}
\end{equation}
At the north pole $[1:0]$ we recover the generators for the toric geometry of $\mathbb{F}_0$ as expected. We can depict these theta functions in a grid whose coordinates are given by their powers in $x,y$ and find the bottom left diagram of Figure \ref{fig:f0tof2}. We can do the same for the point [0:1] and recover the bottom right grid of Figure \ref{fig:f0tof2}. These functions can be obtained from the initial scattering diagram given by the $\mathbb{F}_2$ toric diagram and the following ingoing wall
\begin{equation}
    \left( \left(\begin{smallmatrix}0\\-1\end{smallmatrix}\right)\mathbb R_+,\frac{\alpha}{y} \,  +\beta\,\right)\,,
\end{equation}
and the same argument follows by swapping $\alpha$ and $\beta$.

We managed to find a family of geometries fibered by a $\mathbb{P}^1$ that give toric geometries at the north and south poles and coincide for generic points.

\begin{figure}
    \centering
    \begin{tikzpicture}[
        x=0.65cm,
        y=0.65cm]
    
        \node[focalPoint,goodred] (F) at  (-4,0) {};
        
        \draw[broken,goodred] (-8,0) -- (F) -- (0,0) -- (8,8);
        \draw[broken,goodred] (-4,-6) -- (F) -- (-4,5) -- (-4,7);
        \draw[broken,goodred] (-8,4) -- (F);
        \draw[broken,goodred] (-8,2) -- (F);
        \draw[broken,goodred] (F) -- node[pos=.7,fill=white,inner sep=0pt] {$\beta t^3 y\,x$} (-.66666,3.33333) -- (0,4) --node[pos=.7,left]{$x y$} (4,8);

        \node[goodred] at (-3.7,5.85) {$y$};
        \node[goodred] at (-3.5,2.95) {$t^2 y$};
        \node[goodred] at (3.55,3) {$x y t$};
        \node[goodred] at (-0.75,4) {$\beta y x  $};
        \node[goodred] at (6.8,7.3) {$x y $};
        \node[goodred] at (-6.85,-0.45) {$\frac{1}{x}$};
        \node[goodred] at (-4.35,-4.5) {$\frac{1}{y}$};
        \node[goodred] at (-6.25,3.0) {$t \frac{y}{x}$};
        \node[goodred] at (-7,1) {$\frac{y}{x^2}$};
        \node[goodred] at (-2.,-0.29) {$\alpha x t^2$};
        \node[goodred] at (-2.15,2.75) {};
        
        \draw[broken,goodgreen] (F) -- node[pos=.6,fill=white,inner sep=0pt] {$t^4 x^2 y\beta^2$}(0,2) --node[pos=.9,below] {$x^2 y$} (8,6);
        \draw[broken,goodgreen] (F) -- (0,0) -- node[pos=.8,above] {$x^2y$}(8,4);
        \draw[broken,goodgreen] (F) -- (0,-2) -- node[pos=.8,above] {$x^2y$} (8,2);
        
        \node[goodgreen] at (-1.2,0.35) {$2\alpha\beta t^3 x^2$};
        \node[goodgreen,fill=white,inner sep=-1pt] at (-1.35,-1.35) {$t^2\frac{x^2}{y}\alpha^2$};
        
        \draw[broken,goodblue] (F) -- (0,0) -- node[pos=.85,below] {$x$} (8,0);
        \draw[broken,goodblue] (F) -- (0,-4) -- node[pos=.7,below] {$x $} (8,-4);
        
        \node[goodblue] at (-.85,-0.35) {$\beta x t^2$};
        \node[goodblue] at (-1.25,-3.6) {$\alpha \frac{x}{y}t$};
        \node[goodblue] at (6.15,-6.8) {};
        \node[goodblue] at (2.2,-6.8) {};

        \draw[wall] (0,3) -- node[pos=.95,right] {$\beta +\alpha t/y$} (0,8);
        \draw[wall] (0,-6) -- node[pos=.05,left] {$\beta t^2 +\alpha t/y$} (0,3);
        \draw[kink] (-8,7) -- node[pos=0,left] {$t$} (0,3) -- node[pos=1,right] {$t$} (8,7);
        \node[focalPoint] at (0,3) {};
    
    \end{tikzpicture} 
    \caption{Broken lines for $\mathbb{F}_2$. This is consistent for all $\alpha$ and all $\beta \neq 0$, corresponding to the other patch of the $\mathbb{P}^1$.}
    \label{fig:brokenline_f2}
\end{figure}
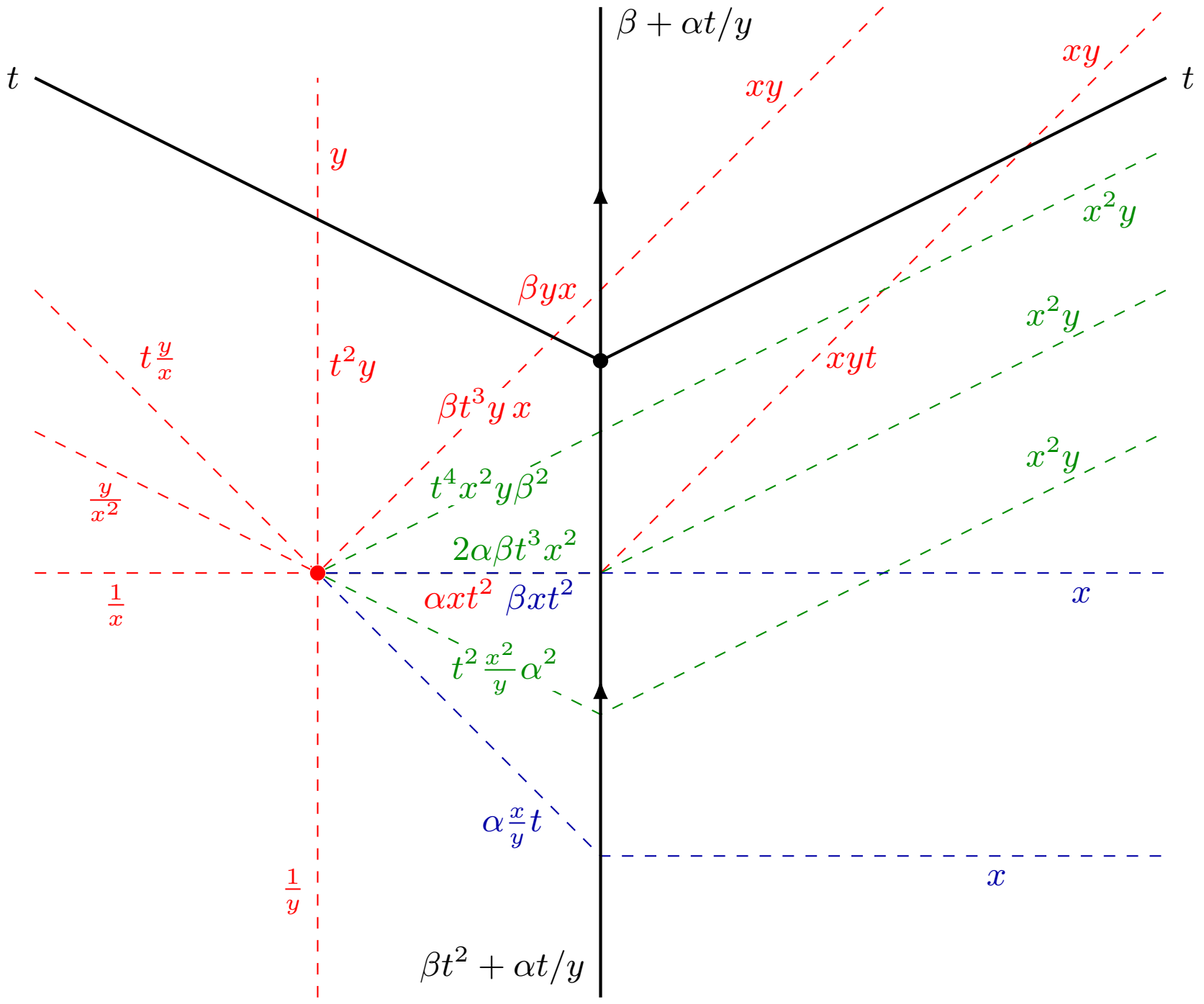

We can change the Hanany-Witten frame as follows. We start in Figure \ref{fig:brokenline_f0}, with a 7-brane at finite distance, but far away in the vertical direction, above the web. We take this 7-brane and move it down through the web, sending it far in the other direction. Then, we take the monodromy cut of the 7-brane, now lying on the vertical axis, and rotate it by 180° thus applying the $\operatorname{SL}(2,\mathbb Z)$ monodromy transformation to half of the diagram (here we choose to apply it on the right). This is the natural transformation of the 5-branes as the monodromy cut of the crossed 7-brane sweeps them. The $SL(2,\mathbb{Z})$ matrix associated to the monodromy cut of a $[p,q]$ 7-brane is
\begin{equation}
    M= \left(\begin{array}{cc}
       1-p q  &  p^2\\
        -q^2 &1+ pq 
    \end{array}\right)\,.
\end{equation}
If the 7-brane branch cut sweeps a 5-brane counterclockwise, its charge $(p',q')^T$ is transformed into $(p'',q'')^T=M\cdot(p',q')^T $; in the other direction, its action is given by $M^{-1}$.

All the kinks, walls and broken lines in this half of the diagram see their direction change, and are subject to an  $\operatorname{SL}(2,\mathbb Z)$  change of variables
\begin{equation}
    \left(\begin{array}{cc}
       a  & b\\
        c &d 
    \end{array}\right) \in\operatorname{SL}(2,\mathbb Z) \,, \quad x \to x^a y^c, \; y \to x^b y^d\,,
\end{equation}
which in this case is $x,y\to xy,y$. To keep the diagram consistent and the rules computing the theta-functions unchanged, we also need to multiply the vertical wall by $1/y$. This way, the incoming wall $( \left(\begin{smallmatrix}0\\1\end{smallmatrix}\right)\mathbb R_+,\alpha t  +\beta y)$ in Figure \ref{fig:brokenline_f0} becomes the outgoing wall $(\left(\begin{smallmatrix}0\\1\end{smallmatrix}\right)\mathbb R_+,\beta  +\alpha t/y)$ in Figure \ref{fig:brokenline_f2}. As long as the parameters $[\alpha:\beta]$ take a generic value, the two diagrams are equivalent: they have the same theta-functions\footnote{Up to relabeling.} and construct the same threefold. The diagram in Figure \ref{fig:brokenline_f0} is in the frame appropriate to take the limit $[\alpha:\beta] \to [1:0]$, where we recover the toric local $\mathbb F_0$.  The diagram in Figure \ref{fig:brokenline_f2} is in the frame appropriate to take the limit $[\alpha:\beta] \to [0:1]$, where we get a singular toric threefold.

In the following subsection we will extend this case to a large class of examples.
\subsection{A class of examples}
Consider the brane web given by the following external 5-branes
\begin{equation}\label{eq::W0}
   W_0:= \{\,n(0,-1),\;(-m,-s),\; n(0,1),\; (m,s)\}\,
\end{equation}
where the prefactor counts the multiplicity of 5-branes with coprime charge, with $m \geq n>0$ and $s=0,1$. We end $l\leq n$ (0,-1) 5-branes on a [0,-1] 7-brane and HW cross it. We end with the following web  
   \begin{equation}
   W_1:= \{\,(n-l)(0,-1),\;(-m,-s),\; (n+m-l)(0,1),\; (m,s-m)\}\,
\end{equation}
and the [0,1] 7-brane now has multiplicity $m-l$. For this GTP to be supersymmetric we need $l\leq m$. We depict the GTPs and their saturated dual cones in Figure \ref{fig:sfam} and Figure \ref{fig:dualcone_s1} .
\begin{figure}[h]
    \centering
    \begin{tikzpicture}
        \node at (-2,0) {\raisebox{-.5\height}{ \scalebox{.8}{  
            \begin{tikzpicture}[x=.5cm,y=.5cm] 
                \draw[gridline] (-2,-2.5)--(-2,3.5); 
                \draw[gridline] (-1,-2.5)--(-1,3.5); 
                \draw[gridline] (0,-2.5)--(0,3.5); 
                \draw[gridline] (1,-2.5)--(1,3.5); 
                \draw[gridline] (2,-2.5)--(2,3.5); 
                \draw[gridline] (3,-2.5)--(3,3.5); 
                \draw[gridline] (-2.5,-2)--(3.5,-2); 
                \draw[gridline] (-2.5,-1)--(3.5,-1); 
                \draw[gridline] (-2.5,0)--(3.5,0); 
                \draw[gridline] (-2.5,1)--(3.5,1); 
                \draw[gridline] (-2.5,2)--(3.5,2); 
                \draw[gridline] (-2.5,3)--(3.5,3); 
                \draw[ligne] (3,-2)--(2,3); 
                \draw[ligne] (2,3)--(1,3); 
                \draw[ligne] (1,3)--(0,3); 
                \draw[ligne] (0,3)--(-1,3); 
                \draw[ligne] (-1,3)--(-2,3); 
                \draw[ligne] (-2,3)--(-1,-2); 
                \draw[ligne] (-1,-2)--(2,-2); 
                \draw[ligne] (2,-2)--(3,-2); 
                \node[bd] at (2,3) {}; 
                \node[bd] at (1,3) {}; 
                \node[bd] at (0,3) {}; 
                \node[bd] at (-1,3) {}; 
                \node[bd] at (-2,3) {}; 
                \node[bd] at (-1,-2) {}; 
                \node[bd] at (2,-2) {}; 
                \node[bd] at (3,-2) {}; 
                \node[wd] at (0,-2) {}; 
                \node[wd] at (1,-2) {}; 
                
                \draw[
                    decorate,
                    decoration={brace, amplitude=8pt},
                    thick
                ] (2,-2.5) -- (-1,-2.5)  node[pos=.5,below=8pt] {$l$};
                \draw[
                    decorate,
                    decoration={brace, amplitude=8pt},
                    thick
                ] (3,-3.75) -- (-1,-3.75)  node[pos=.5,below=8pt] {$n$};
                
                \draw[
                    decorate,
                    decoration={brace, amplitude=8pt},
                    thick
                ] (-2.5,-2) -- (-2.5,3)  node[pos=.5,left=8pt] {$m$};
                \draw[
                    decorate,
                    decoration={brace, amplitude=8pt},
                    thick
                ] (2,3.5) -- (3,3.5)  node[pos=.5,above=8pt] {$s$};
            \end{tikzpicture}}}};

    \node at (2,0.5) {\raisebox{-.5\height}{ \scalebox{.8}{  
        \begin{tikzpicture}[x=.5cm,y=.5cm] 
            \draw[gridline] (-4,-2.5)--(-4,3.5); 
            \draw[gridline] (-3,-2.5)--(-3,3.5); 
            \draw[gridline] (-2,-2.5)--(-2,3.5); 
            \draw[gridline] (-1,-2.5)--(-1,3.5); 
            \draw[gridline] (0,-2.5)--(0,3.5); 
            \draw[gridline] (1,-2.5)--(1,3.5); 
            \draw[gridline] (2,-2.5)--(2,3.5); 
            \draw[gridline] (3,-2.5)--(3,3.5); 
            \draw[gridline] (-4.5,-2)--(3.5,-2); 
            \draw[gridline] (-4.5,-1)--(3.5,-1); 
            \draw[gridline] (-4.5,0)--(3.5,0); 
            \draw[gridline] (-4.5,1)--(3.5,1); 
            \draw[gridline] (-4.5,2)--(3.5,2); 
            \draw[gridline] (-4.5,3)--(3.5,3); 
            \draw[ligne] (3,-2)--(2,3); 
            \draw[ligne] (2,3)--(1,3); 
            \draw[ligne] (1,3)--(0,3); 
            \draw[ligne] (0,3)--(-1,3); 
            \draw[ligne] (-1,3)--(-2,3); 
            \draw[ligne] (-2,3)--(-4,3); 
            \draw[ligne] (-4,3)--(2,-2); 
            \draw[ligne] (2,-2)--(3,-2); 
            \node[bd] at (2,3) {}; 
            \node[bd] at (1,3) {}; 
            \node[bd] at (0,3) {}; 
            \node[bd] at (-1,3) {}; 
            \node[bd] at (-2,3) {}; 
            \node[bd] at (-4,3) {}; 
            \node[bd] at (2,-2) {}; 
            \node[bd] at (3,-2) {}; 
            \node[wd] at (-3,3) {}; 
            \draw[
                decorate,
                decoration={brace, amplitude=8pt},
                thick
            ] (-4,3.5) -- (-2,3.5)  node[pos=.5,above=8pt] {$m-l$};
            \draw[
                decorate,
                decoration={brace, amplitude=8pt},
                thick
            ] (-4,4.75) -- (2,4.75)  node[pos=.5,above=8pt] {$n+m-l$};
            \draw[
                decorate,
                decoration={brace, amplitude=8pt},
                thick
            ] (2,3.5)  -- (3,3.5)  node[pos=.5,above=8pt] {$s$};          
            \draw[
                decorate,
                decoration={brace, amplitude=8pt},
                thick
            ] (3,-2.5)  -- (2,-2.5)  node[pos=.5,below=8pt] {$n-l$};           
            \draw[
                decorate,
                decoration={brace, amplitude=8pt},
                thick
            ] (3.5,3)  -- (3.5,-2)  node[pos=.5,right=8pt] {$m$};
        \end{tikzpicture}}}};
    \end{tikzpicture}
    \caption{}
    \label{fig:sfam}
\end{figure}

We will study the geometry of $W_0$ with the 7-branes. We first consider the case $s=1$. The lattice generated by \eqref{eq::W0} is not saturated. We add the necessary generators using the HJ algorithm and we end up with the following set of generators of the dual cone
\begin{equation}
    \{(1,0)\, (-1,0),\, (i,1),\,(-i,-1),\quad 0\leq i \leq m\,\}.
\end{equation}
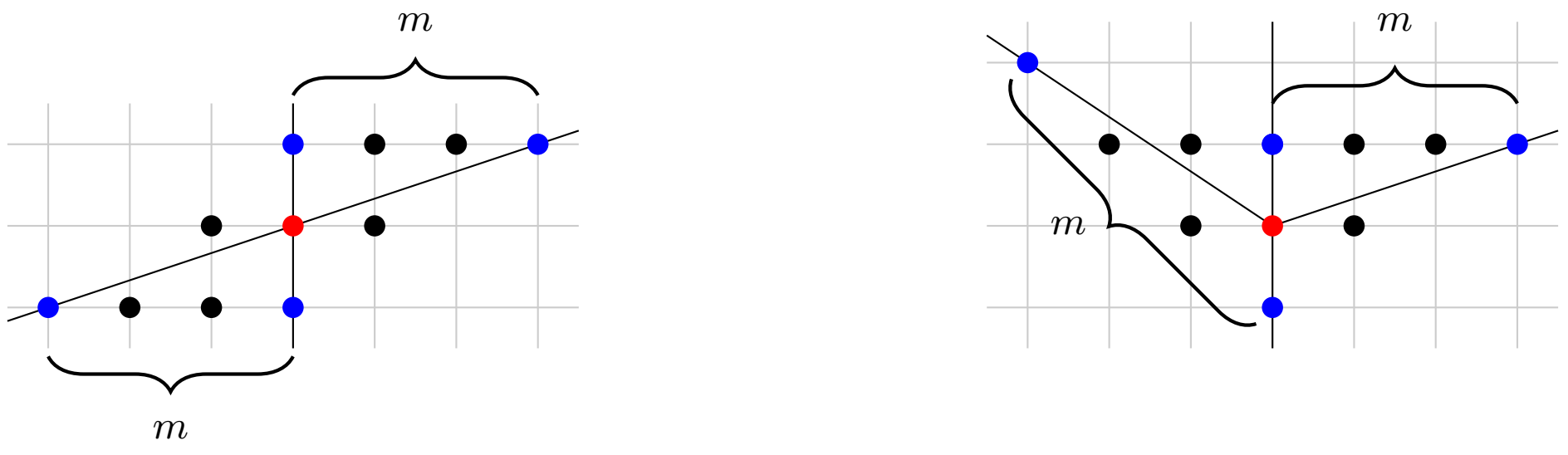
\begin{figure}[h!]
    \centering
    \begin{tikzpicture}[x=0.65cm,y=0.65cm]
        \foreach \n in {3,...,9}
            \draw[gridline] (\n,-1.5) -- (\n,2.5);
        \foreach \n in {-1,...,2}
            \draw[gridline] (2.5,\n) -- (9.5,\n);
        \draw (6,0) -- (2.5,2.33333);
        \draw (6,-1.5) -- (6,2.5);
        \draw (6,0) -- (9.5,1.16666);
        
        \fill[goodred] (6,0) circle (2.4pt);
        \fill[blue] (6,1) circle (2.4pt);
        \fill[blue] (6,1) circle (2.4pt);
        \fill[black] (7,1) circle (2.4pt);
        \fill[black] (8,1) circle (2.4pt);
        \fill[blue] (9,1) circle (2.4pt);
        \fill[black] (5,1) circle (2.4pt);
        \fill[black] (4,1) circle (2.4pt);
        \fill[black] (5,0) circle (2.4pt);
        \fill[black] (7,0) circle (2.4pt);
        \fill[blue] (3,2) circle (2.4pt);
        \fill[blue] (6,-1) circle (2.4pt);
        
        \draw[
            decorate,
            decoration={brace, amplitude=8pt},
            thick
        ] (5.8,-1.2)--(2.8,1.8);
        \node[black,rotate=0] at (3.5,0) {$m$};
        \draw[
            decorate,
            decoration={brace, amplitude=8pt},
            thick
        ] (6,1.5) -- (9,1.5);
        \node[black,rotate=0] at (7.5,2.5) {$m$};
          
        \draw[gridline] (-9.5,1) -- (-2.5,1);
        \draw[gridline] (-9.5,0) -- (-2.5,0);
        \draw[gridline] (-9.5,-1) -- (-2.5,-1);
        \foreach \n in {3,...,9} 
            \draw[gridline] (-\n,-1.5) -- (-\n,1.5);
        \draw (-6,-1.5) -- (-6,1.5);
        \draw (-9.5,-1.16666) -- (-2.5,1.166666);
        
        \fill[goodred] (-6,0) circle (2.4pt);
        \fill[blue] (-6,1) circle (2.4pt);
        \fill[black] (-5,1) circle (2.4pt);
        \fill[black] (-4,1) circle (2.4pt);
        \fill[blue] (-3,1) circle (2.4pt);
        \fill[black] (-5,0) circle (2.4pt);
        \fill[black] (-7,0) circle (2.4pt);
        \fill[blue] (-6,-1) circle (2.4pt);
        \fill[black] (-7,-1) circle (2.4pt);
        \fill[black] (-8,-1) circle (2.4pt);
        \fill[blue] (-9,-1) circle (2.4pt);

        \draw[
            decorate,
            decoration={brace, amplitude=8pt},
            thick
        ] (-6,-1.6) -- (-9,-1.6);
        \node[black,rotate=0] at (-4.5,2.5) {$m$};
        \draw[
            decorate,
            decoration={brace, amplitude=8pt},
            thick
        ] (-6,1.6) -- (-3,1.6);
        \node[black,rotate=0] at (-7.5,-2.5) {$m$};
    \end{tikzpicture} 
    \caption{Saturated dual cones for $W_0$ on the left and for $W_1$ on the right for $s=1$.}
    \label{fig:dualcone_s1}
\end{figure}

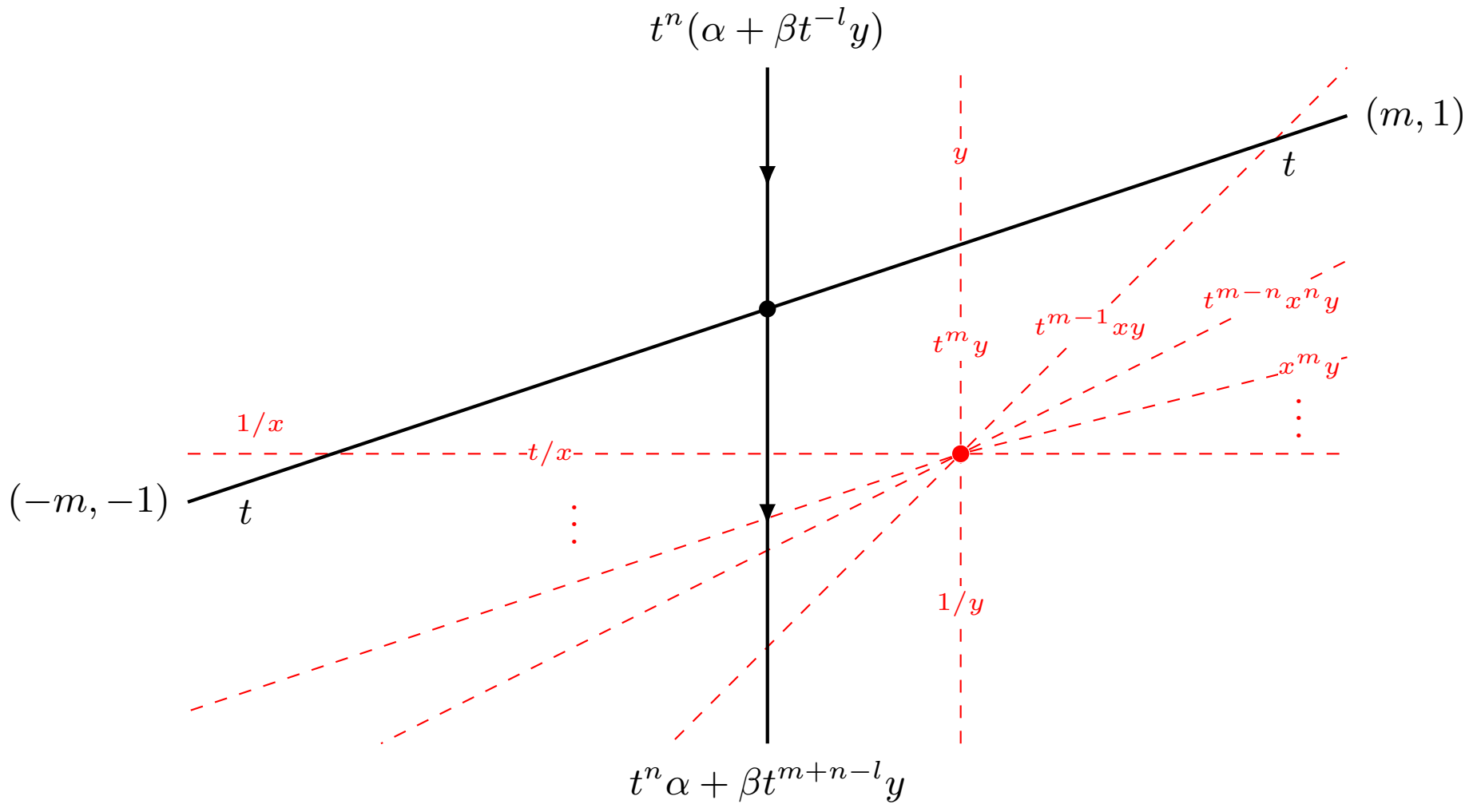
\begin{figure}
    \centering
    \begin{tikzpicture}[x=.4cm,y=.4cm]
    
        \node[focalPoint,goodred] (F) at (4,0) {};               
        \draw[broken,goodred] (-12,0) -- node[pos=.5,above] {$\scriptstyle 1/x$} (-9,0) -- node[pos=.5,fill=white,inner sep=0pt] (p1) {$\scriptstyle t/x$} (0,0) -- (F) -- (12,0);
        \draw[broken,goodred] (4,-6) --node[pos=.5,fill=white,inner sep=1pt] {$\scriptstyle 1/y$} (F) --node[pos=.5,fill=white,inner sep=1pt] {$\scriptstyle t^my$} (4,4.3333) -- node[pos=.5,fill=white,inner sep=1pt] {$\scriptstyle y$} (4,8);
        \draw[broken,goodred] (F) -- node[pos=.4,fill=white,inner sep =1pt] {$\scriptstyle t^{m-1}x y$} (10.5,6.5) -- (12,8);
        \draw[broken,goodred] (F) -- node[pos=.8,fill=white,inner sep=0pt] {$\scriptstyle t^{m-n}x^{n} y$} (12,4);
        \draw[broken,goodred] (F) -- node[pos=.9,fill=white,inner sep=0pt] {$\scriptstyle x^m y$} (12,2);
        \draw[broken,goodred] (F) -- (0,-4) -- (-2,-6);
        \draw[broken,goodred] (F) -- (0,-2) -- (-8,-6);
        \draw[broken,goodred] (F) -- (0,-1.333) -- (-12,-5.333);

        \node[goodred] at (-4,-1.2) {$\vdots$};
        \node[goodred] at (11,1) {$\vdots$};
        
        \draw[wall] (0,8) -- node[pos=0,above] {$t^n(\alpha+\beta t^{-l}y)$} (0,3);
        \draw[wall] (0,3) -- node[pos=1,below] {$t^n \alpha+ \beta t^{m+n-l} y$}(0,-6);
        \draw[kink] (-12,-1) -- node[pos=.1,below] {$t$} node[pos=0,left]{$(-m,-1)$} (0,3) -- node[pos=.9,below] {$t$} node[pos=1,right] {$(m,1)$} (12,7);
        
        \node[focalPoint] at (0,3) {};
    \end{tikzpicture} 
    \caption{Broken lines for $W_0$ if $s=1$.}
    \label{fig:brokenlines_w0}
\end{figure}

We compute the broken lines in Figure \ref{fig:brokenlines_w0}. We find the following theta functions for the choice of $p$: 
\begin{equation}
\begin{aligned}
    \vartheta_{1,0}&=x,\;\vartheta_{-1,0}=\frac{t}{x}(t^n \alpha+\beta t^{n-l+m}y),\, \\ \vartheta_{i,1}&=t^{m-i}x^i y,\;\vartheta_{-i,-1}=\frac{1}{x^i y}(t^n \alpha+\beta t^{n-l+m}y)^i,\quad 0\leq i \leq m
\end{aligned}   
\end{equation}
For $[1:0]$ we recover the grid to the left of Figure \ref{fig:dualcone_s1} and for $[0:1]$ we recover the grid to the right of  Figure \ref{fig:dualcone_s1}.

\begin{figure}[h]
    \centering
    \begin{tikzpicture}
    
       \node at (0,0){ \begin{tikzpicture}
           \draw[wall] (0,1.5) -- node[pos=.05,right] {$\alpha t^n + \beta y t^{n-\ell}$} (0,0);
        \draw[wall] (0,0) -- node[pos=.95,right] {$\alpha t^n + \beta yt^{n-\ell+m}$} (0,-3);
        \draw[kink] (-4,0) -- node[pos=0,left] {$t^m$} (0,0) -- node[pos=1,right] {$t^m$} (4,0);
        \node[focalPoint,goodred] (F) at (-1.5,-2) {};
        \draw[broken, red] (F) -- node[pos=.5,below] {$\alpha t^n x$} (0,-2) -- node[pos=.5,below] {$x$} (4,-2);
        \draw[broken,goodred] (F) -- node[pos=.5,inner sep=0pt] (p1) {} (0,-.5) -- node[pos=.5,below] {$x$} (4,-.5);
        \node[goodred] (l1) at (1.3,-1.5) {$\beta xyt^{n-\ell+m}$};
        \draw[red,-stealth,bend left=30,inner sep=0pt] (l1.190) to (p1);
        \draw[broken, red] (F) -- node[pos=.7,left] {$\frac{1}{y}$} (-1.5,-3); 
        \draw[broken,goodred] (F) -- node[pos=.7,above]{$\frac{1}{x}$} (-4,-2);
        \draw[broken,goodred] (F) -- node[pos=.5,left] {$t^my$} (-1.5,0) -- node[pos=.5,left] {$y$} (-1.5,1.5);
        \end{tikzpicture}};

        \node at (6,0){
            \begin{tikzpicture} [x=0.65cm,y=0.65cm]
                \draw[gridline] (-1.5,1) -- (1.5,1);
                \draw[gridline] (-1.5,0) -- (1.5,0);
                \draw[gridline] (-1.5,-1) -- (1.5,-1);
                \foreach \n in {1,...,3} 
                    \draw[gridline] (-2+\n,-1.5) -- (-2+\n,1.5);
                \draw (-1.5,0) -- (1.5,0);
                \draw (0,-1.5) -- (0,1.5);
                
                \fill[goodred] (0,0) circle (2.4pt);
                \fill[blue] (0,1) circle (2.4pt);
                \fill[blue] (-1,0) circle (2.4pt);
                \fill[blue] (0,-1) circle (2.4pt);
                \fill[blue] (1,0) circle (2.4pt);
            \end{tikzpicture}};
    
        \node at (9,0){
            \begin{tikzpicture} [x=0.65cm,y=0.65cm]
                \draw[gridline] (-1.5,1) -- (1.5,1);
                \draw[gridline] (-1.5,0) -- (1.5,0);
                \draw[gridline] (-1.5,-1) -- (1.5,-1);
                \foreach \n in {1,...,3} 
                \draw[gridline] (-2+\n,-1.5) -- (-2+\n,1.5);
                \draw (-1.5,1.5)--(0,0) -- (1.5,0);
                \draw (0,-1.5) -- (0,1.5);
                
                \fill[goodred] (0,0) circle (2.4pt);
                \fill[blue] (0,1) circle (2.4pt);
                \fill[blue] (-1,1) circle (2.4pt);
                \fill[blue] (0,-1) circle (2.4pt);
                \fill[blue] (1,0) circle (2.4pt);
            \end{tikzpicture}};
    \end{tikzpicture}
    \caption{Broken lines and dual lattices for $s=0$.}
    \label{fig:scattdiag_s0}
\end{figure}
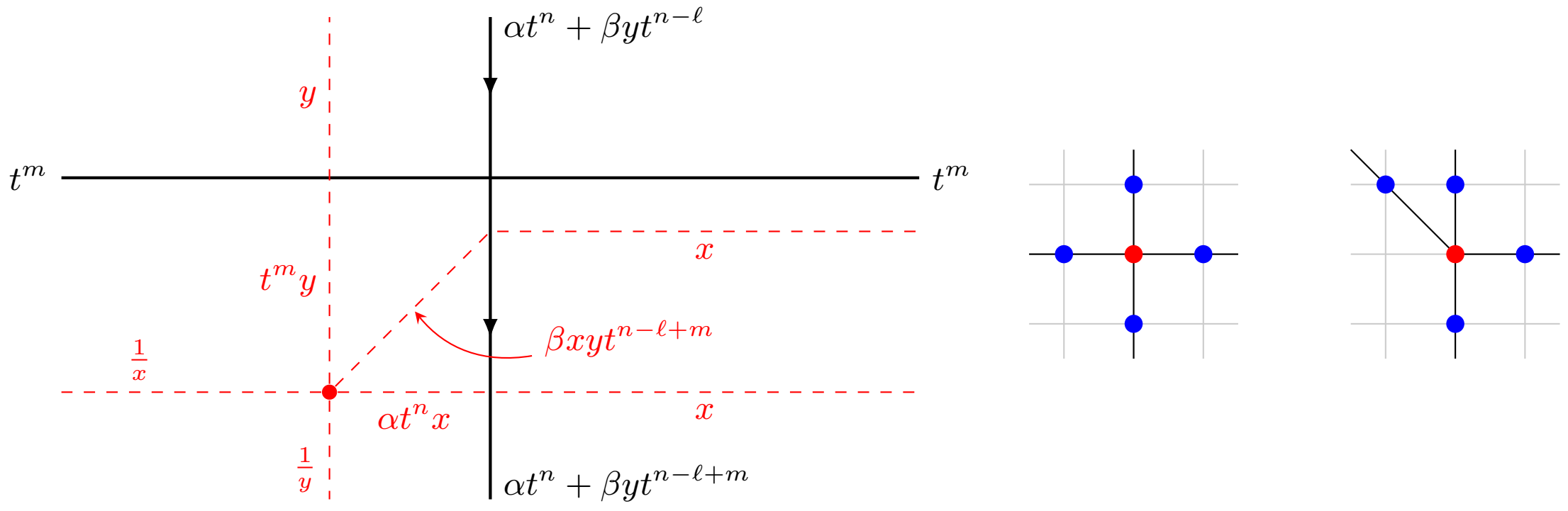

We will now consider the case $s=0$. Here, the lattice generated by the external 5-branes is already saturated, so we do not need to add additional asymptotic directions for the theta functions. We summarize this case in Figure \ref{fig:scattdiag_s0} . We find the following theta functions
\begin{equation}
     \vartheta_{1,0}=\alpha t^n x+ \beta \,t^{n-l+m} x y ,\; \vartheta_{0,1}=t^m y ,\;\vartheta_{-1,0}=\frac{1}{x},\, \vartheta_{0,-1}=\frac{1}{ y}\,.
\end{equation}
We see again that for $[1:0]$ we recover the dual lattice of $W_0$ while for $[0:1]$ we recover that of $W_1$ as expected.

\subsection{Physical interpretation}

We may interpret the $\mathbb{P}^1$ fibration as a proxy for the 7-brane position. At a generic point, the 7-brane is located at any finite position along its monodromy cut and imposes boundary conditions in the 5d theory geometrized by the algebraic deformation. The 7-brane is effectively delocalized along the wall in the scattering diagram. Its position does not affect the physics. On the other hand, going to the poles of the $\mathbb{P}^1$ corresponds to sending the 7-branes to infinity and restoring toricity: only 5-branes are present. Setting $[\alpha:\beta] \to [1:0],[0:1]$ recovers the web that engineers each toric geometry with only kinks. It is crucial to remark that the toric geometries at the poles need not be equivalent: some moduli have to be frozen in order for them to engineer the same SCFT. Thus, we identify the generic point of $\mathbb{P}^1$ with the genuine GTP geometry. We will extend this discussion to resolved geometries in the next section.

\section{Resolved geometries} \label{sec:ResolvedGeometries}

In this section, we extend the results that we have obtained for webs engineering SCFTs to their extended Coulomb branches. This involves opening up the web into several junctions, and as a consequence the scattering diagram has several vertices. We present a general prescription for constructing consistent scattering diagrams in this case with and without 7-branes. We obtain a very explicit description of the resolved singularity, from which geometric and physical data can be extracted.

\subsection{Toric threefolds} \label{sec::resolvedconif}

\paragraph{A simple example in detail: the conifold.}
Start with the small resolution of the conifold singularity. The toric polygon of the conifold singularity $\mathcal X$ is a square of area $1$. The corresponding scattering diagram is composed of only four kinks $(\left(\begin{smallmatrix} \pm 1 \\ 0\end{smallmatrix}\right) ,t)$ and $(\left(\begin{smallmatrix} 0\\ \pm 1 \end{smallmatrix}\right) ,t)$. 
$\mathcal X$ is then the affine hypersurface $Z_1Z_2=Z_3Z_4$, with the four coordinates equal to the theta-functions $\vartheta_{-1,0},\vartheta_{1,0},\vartheta_{0,1},\vartheta_{0,-1}$ respectively.

In toric geometry, one can construct the small resolution $\widetilde{\mathcal X}$ of the conifold by adding the edge $(0,1)-(1,0)$ in the toric polygon. The corresponding threefold is the total space of the rank $2$ bundle $\mathcal O(-1)\oplus\mathcal O(-1)$ over $\mathbb P^1$. Let us review in detail how we can construct it with scattering diagrams.

\begin{figure}[t]
\centering
\begin{tikzpicture}
    \tikzmath{\x = 2.2;\y =.5;\z = 1.2;}
    \draw[kink] (-\x,0) -- (0,0) -- (0,-\x);
    \draw[kink] (0,0) -- (\x,\x) -- (2*\x,\x);
    \draw[kink] (\x,\x) -- (\x,2*\x);

    \node[focalPoint,goodred] at (-\y,2*\y) {};
    \draw[broken,goodred] (-\y,2*\y) -- node[pos=.5,fill=white,inner sep=2pt] {$y$} (-\y,2*\x);
    \draw[broken,goodred] (-\y,2*\y) -- node[pos=.5,fill=white,inner sep=2pt] {$1/x$} (-\x,2*\y);
    \draw[broken,goodred] (-\y,2*\y) -- node[pos=.5,fill=white,inner sep=2pt] {$t/y$} (-\y,0) -- node[pos=.5,fill=white,inner sep=2pt] {$1/y$} (-\y,-\x);
    \draw[broken,goodred] (-\y,2*\y) -- node[pos=.5,fill=white,inner sep=2pt] {$tx$} (2*\y,2*\y) -- node[pos=.5,fill=white,inner sep=2pt] {$x$} (2*\x,2*\y);
    \draw[broken,goodred] (-\y,2*\y) --node[pos=.5,fill=white,inner sep=2pt] {$txy$} (\x,\x+3*\y) -- node[pos=.5,fill=white,inner sep=1pt] {$xy$} (2*\x -3*\y,2*\x);
    \draw[broken,goodred] (-\y,2*\y) -- node[pos=.5,fill=white,inner sep = 2pt,xshift=-5pt] {$t/xy$} (-3*\y,0) -- node[pos=.8,xshift=15pt,inner sep = 2pt] {$1/xy$} (-\x,-\x+3*\y);

    \node[draw,rectangle,dashed,inner sep=3.3cm] (p0) at (.5*\x,.5*\x) {};
    \node[anchor=north west] at (p0.north west) {(a)};
    
    \node[draw,rectangle,densely dotted,inner sep = .3cm] (p1) at (0,0) {};
    \node[inner sep = 2cm] (p2) at (3,-2.8) {};
    \draw[dotted] (p1.north east) -- (p2.north east);
    \draw[dotted] (p1.south west) -- (p2.south west);
    \draw[dotted] (p1.north west) -- (p2.north west);
    \draw[dotted] (p1.south east) -- (p2.south east);
    \node[draw,rectangle,fill=white,dashed,inner sep = 2cm] at (p2) {};
    \node (pp) at (p2) {
        \begin{tikzpicture}
            \tikzmath{\x = 2;\y =.7;\z = 1.2;}
            \draw[kink] (-\x,0) -- (0,0) -- (0,-\x);
            \draw[kink] (0,0) -- (\x,\x);

            \node[focalPoint,goodblue] at (-\y,\y) {};
            \draw[broken,goodblue] (-\y,\y) -- node[pos=.5,fill=white,inner sep=2pt] {$1/x$} (-\x,\y);
            \draw[broken,goodblue] (-\y,\y) -- node[pos=.5,xshift=2pt,fill=white,inner sep=2pt] {$t/y$} (-\y,0) --node[pos=.5,fill=white,inner sep=2pt] {$1/y$} (-\y,-\x);
            \draw[broken,goodblue] (-\y,\y) --node[pos=.5,fill=white,inner sep=2pt] {$tx$} (\y,\y) -- node[pos=.5,fill=white,inner sep=2pt] {$x$}(\x,\y);
            \draw[broken,goodblue] (-\y,\y) -- node[pos=.5,fill=white,inner sep=2pt] {$xy$} (\x-2*\y,\x);
            \draw[broken,goodblue] (-\y,\y) -- node[pos=.5,fill=white,inner sep=2pt] {$y$} (-\y,\x);
            \draw[broken,goodblue] (-\y,\y) -- node[pos=.5,xshift=-6pt,fill=white,inner sep=0pt] {$t/xy$} (-2*\y,0)-- node[pos=.55,xshift=5pt,fill=white,inner sep=0pt]{$1/xy$} (-\x,-\x +2*\y);
            
        \end{tikzpicture}
    };
    \node[anchor=south west] at (p2.south west) {(c)};

    \node[draw,rectangle,densely dotted,inner sep = .3cm] (p3) at (\x,\x) {};
    \node[inner sep = 2cm] (p4) at (7.1,4.4) {};
    \draw[dotted] (p3.north west) -- (p4.north west);
    \draw[dotted] (p3.south west) -- (p4.south west);
    \draw[dotted] (p3.north east) -- (p4.north east);
    \draw[dotted] (p3.south east) -- (p4.south east);
    \node[draw,rectangle,fill=white,dashed,inner sep = 2cm] at (p4)  {};
    \node (pp1) at (p4) {
        \begin{tikzpicture}
            \tikzmath{\x = 2;\y =.7;\z = 1.2;}
            \draw[kink] (\x,0) -- (0,0) -- (0,\x);
            \draw[kink] (0,0) -- (-\x,-\x);

            \node[focalPoint,goodblue] at (-1.5*\y, .5*\y) {};
            \draw[broken,goodblue] (-1.5*\y, .5*\y) --node[pos=.5,fill=white,inner sep=2pt] {$y$} (-1.5*\y, \x);
            \draw[broken,goodblue] (-1.5*\y, .5*\y) --node[pos=.5,fill=white,inner sep=2pt,xshift=3pt] {$t/y$} (-1.5*\y,-1.5*\y) --node[pos=.3,fill=white,right,inner sep=2pt] {$1/y$} (-1.5*\y, -\x);
            \draw[broken,goodblue] (-1.5*\y, .5*\y) -- node[pos=.5,fill=white,inner sep=2pt] {$tx$} (0, .5*\y)-- node[pos=.5,fill=white,inner sep=1pt] {$x$} (\x, .5*\y);
            \draw[broken,goodblue] (-1.5*\y, .5*\y) --node[pos=.6,fill=white,inner sep=2pt] {$1/x$} (-\x, .5*\y);
            \draw[broken,goodblue] (-1.5*\y, .5*\y) -- node[pos=.5,fill=white,inner sep=1pt] {$txy$} (0, 2*\y) -- node[pos=.5,fill=white,inner sep=1pt,xshift=3pt] {$xy$}(\x-2*\y,\x);
            \draw[broken,goodblue] (-1.5*\y, .5*\y) --node[pos=.6,fill=white,inner sep=0pt] {$1/xy$} (-\x,2*\y -\x);
        \end{tikzpicture}
    };
    \node[anchor=north east] at (p4.north east) {(b)};

    \node[draw,rectangle,densely dotted,inner sep = .2cm] (p5) at ($(pp) + (.15*\x,.15*\x)$) {};
    \node[draw,rectangle,densely dotted,inner sep = .2cm] (p6) at ($(pp1) + (-.7*\x,-.7*\x)$) {};

    \node[inner sep = 2cm] (p7) at (7.8,-.2) {};
    \draw[dotted] (p5.north west) -- (p7.north west);
    \draw[dotted] (p5.south west) -- (p7.south west);
    \draw[dotted] (p6.north east) -- (p7.north east);
    \draw[dotted] (p6.south west) -- (p7.south west);
    \draw[dotted] (p5.north east) -- (p7.north east);
    \draw[dotted] (p5.south east) -- (p7.south east);
    \draw[dotted] (p6.north west) -- (p7.north west);
    \draw[dotted] (p6.south east) -- (p7.south east);
    \node[draw,rectangle,fill=white,dashed,inner sep = 2cm] at (p7) {};
    \node (pp2) at (p7){
        \begin{tikzpicture}
            \tikzmath{\x = 2;\y =.7;\z = 1.2;}
            \draw[kink] (\x,\x) -- (-\x,-\x);

            \node[focalPoint,goodgreen] at (-\y,\y) {};
            \draw[broken, goodgreen] (-\y,\y) -- node[pos=.5,fill=white,inner sep=1pt] {$y$} (-\y,\x);
            \draw[broken, goodgreen] (-\y,\y) -- node[pos=.5,fill=white,inner sep=1pt] {$1/x$} (-\x,\y);
            \draw[broken, goodgreen] (-\y,\y) --node[pos=.6,fill=white,inner sep=1pt] {$1/xy$}  (-\x,-\x+2*\y);
            \draw[broken, goodgreen] (-\y,\y) -- node[pos=.5,fill=white,inner sep=1pt] {$xy$} (\x-2*\y,\x);
            \draw[broken, goodgreen] (-\y,\y) --node[pos=.5,fill=white,inner sep=1pt] {$t/y$}  (-\y,-\y) --node[pos=.5,fill=white,inner sep=1pt] {$1/y$}  (-\y,-\x);
            \draw[broken, goodgreen] (-\y,\y) --node[pos=.5,fill=white,inner sep=1pt] {$tx$}  (\y,\y) --node[pos=.5,fill=white,inner sep=1pt] {$x$}  (\x,\y);
            
        \end{tikzpicture}
    };

    \node[anchor=south east] at (p7.south east) {(d)};
\end{tikzpicture}
\caption{Scattering diagram construction of the small resolution of the conifold singularity. The broken lines on the diagrams $\mathfrak D$ (a), $\mathfrak D_1$ (b), $\mathfrak D_2$ (c) and $\mathfrak D_{12}$ (d) construct the algebra of holomorphic functions on the conifold, on its two affine charts and their intersection respectively. We show broken lines with six asymptotic directions: they generate the algebra, though in each case a subset of three or four is enough.}
\label{fig:conifoldResolution}
\end{figure}
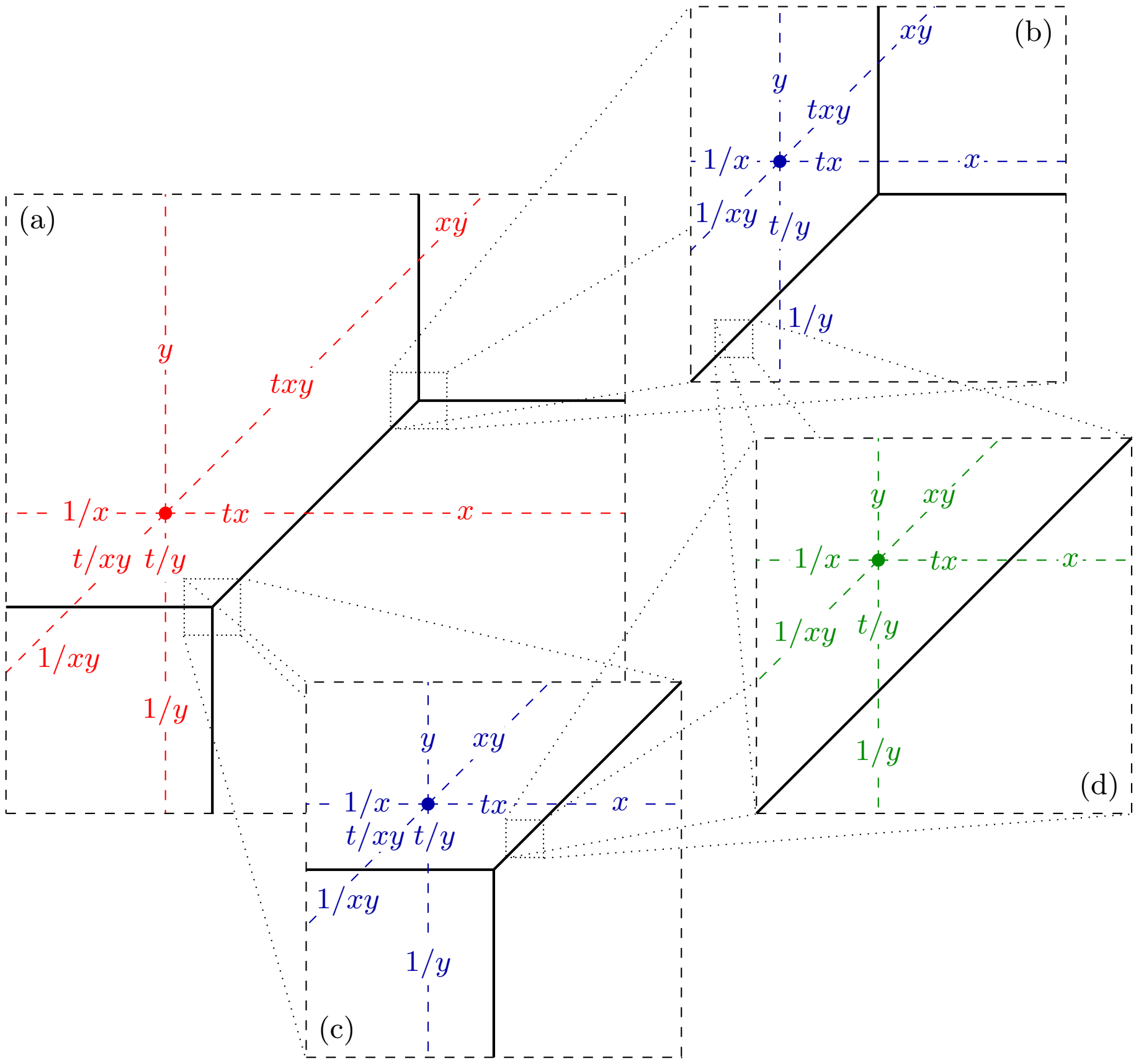

The corresponding scattering diagram $\mathfrak D$ is shown in box (a) of Figure \ref{fig:conifoldResolution}. Its theta-functions construct the algebra of global algebraic functions on $\widetilde{\mathcal X}$, equal to that of $\mathcal X$. To see the whole geometry, we consider the scattering diagrams $\mathfrak D_1,\mathfrak D_2$ obtained by restriction to a formal neighborhood of the two points $p_1,p_2$ where kinks meet, shown in boxes (b) and (c). They construct two threefolds $\mathcal U_1,\mathcal U_2$ both isomorphic to $\mathbb C^3$, and we want to show that these can be thought of as two affine open subsets  of $\widetilde{\mathcal X}$. For this, we need to use the scattering diagram to construct the natural restriction maps $\mathcal O(\widetilde{\mathcal X})\rightrightarrows \mathcal O(\mathcal U_i)\rightrightarrows \mathcal O(\mathcal U_{1}\cap \mathcal U_2)$.

Choose a point $q$ in $\mathfrak D_1$ or, equivalently, a point in $\mathfrak D$ close enough to $p_1$. The theta-function construction gives embeddings of $\mathcal O(\widetilde{\mathcal X})$ and $\mathcal O(\mathcal U_1)$ in $\mathbb C[t,x^{\pm 1},y^{\pm 1}]$. By inspecting the theta-functions shown in Figure \ref{fig:conifoldResolution}, it is easy to see that there is an inclusion\footnote{For instance, a function in $\mathcal O(\widetilde{\mathcal X})$ is $txy$, which is the product of $t$ and $xy$ which are in $\mathcal O(\mathcal U_1)$. The reverse inclusion does not hold, as $xy \notin \mathcal O(\widetilde{\mathcal X})$. } $\mathcal O(\widetilde{\mathcal X}) \hookrightarrow \mathcal O(\mathcal U_1)$. This map is independent of $q$, even if the embedding in $\mathbb C[t,x^{\pm 1},y^{\pm 1}]$ is not.
Likewise, choosing a point $p$ near $p_2$ gives us a natural map $\mathcal O(\widetilde{\mathcal X}) \hookrightarrow \mathcal O(\mathcal U_2)$.

To see how the two affine varieties $\mathcal U_1$, $\mathcal U_2$ are glued together to form $\mathcal X$, we need to consider the scattering diagram $\mathfrak D_{12}$ obtained by restriction to a formal neighborhood of a point on the edge between $p_1$ and $p_2$. It is shown in box (d) of Figure \ref{fig:conifoldResolution} and realizes the threefold $\mathcal V\simeq \mathbb C^2\times \mathbb C^*$ (recall the example in Section \ref{sec:thetaAlgebra}). By choosing a point $q\in\mathfrak D_{12}$, we get natural inclusions of $\mathcal O(\widetilde{\mathcal X}),\mathcal O(\mathcal U_1)$ and $\mathcal O(\mathcal U_2)$ in $\mathcal O(\mathcal V)$. There is a region of $\mathfrak D$ where we can move $p$ so that it gets close to $p_1,p_2$ and the edge $[p_1,p_2]$ without changing the embedding. Therefore, the maps of algebras we constructed are compatible: all the compositions $\mathcal O(\widetilde{\mathcal X}) \to \mathcal O(\mathcal V)$ are equal.

Let us see show that these maps allow us to glue $\mathcal U_1$ and $\mathcal U_2$ together into a smooth threefold, and that this is the same as the standard construction of $\widetilde{\mathcal X}$. With the embeddings chosen as in Figure \ref{fig:conifoldResolution}, the generators of our algebras are summarized in Table \ref{tab:thetafunctions_resolved_conifold}.

\begin{table}[h!]
    \centering
    \begin{tabular}{c|c|c|c}
    $\mathcal O(\widetilde{\mathcal X})$   & $\mathcal O(\mathcal U_1) $& $\mathcal O(\mathcal U_2)$ &$\mathcal O(\mathcal V)$ \\ \hline
    $\begin{array}{l}
    Z_1 = \vartheta_{-1,0} = 1/x \\
    Z_2= \vartheta_{1,0} = tx\\
    Z_3 = \vartheta_{0,-1} =t/ y\\
    Z_4 = \vartheta_{0,1}= y\end{array}$  & $\begin{array}{l}
    X_1 = \vartheta_{1,1}= xy\\
    X_2 = \vartheta_{-1,0} = 1/x\\
    X_3 = \vartheta_{0,-1} = t/y\end{array}$
    & $\begin{array}{l}
    Y_1 = \vartheta_{-1,-1} = 1/xy\\
    Y_2 = \vartheta_{1,0} = tx\\
    Y_3 = \vartheta_{0,1} = y 
    \end{array}$ & $\begin{array}{l} \vartheta_{-1,-1} = 1/xy \\
    \vartheta_{1,1}=xy \\
    \vartheta_{0,1} = y\\
    \vartheta_{0,-1} = t/y\end{array}$
    \end{tabular}
    \caption{Theta-functions generating the different algebras in Figure \ref{fig:conifoldResolution}.}
    \label{tab:thetafunctions_resolved_conifold}
\end{table}

From there, we see that $\mathcal O(\mathcal V)$ is a localization:
\begin{equation}
\mathcal O(\mathcal V) \simeq \mathcal O(\mathcal U_1)[X_1^{-1}] \simeq \mathcal O(\mathcal U_2)[Y_1^{-1}]
\end{equation}
This gives an isomorphism between $\mathcal V$ and an open subset of $\mathcal U_1$ (resp. $\mathcal U_2$), namely the locus where the theta-function along the leg, $X_1$ (resp. $Y_1$), is non-zero. We can use it to glue $\mathcal U_1$ and $\mathcal U_2$ together. 

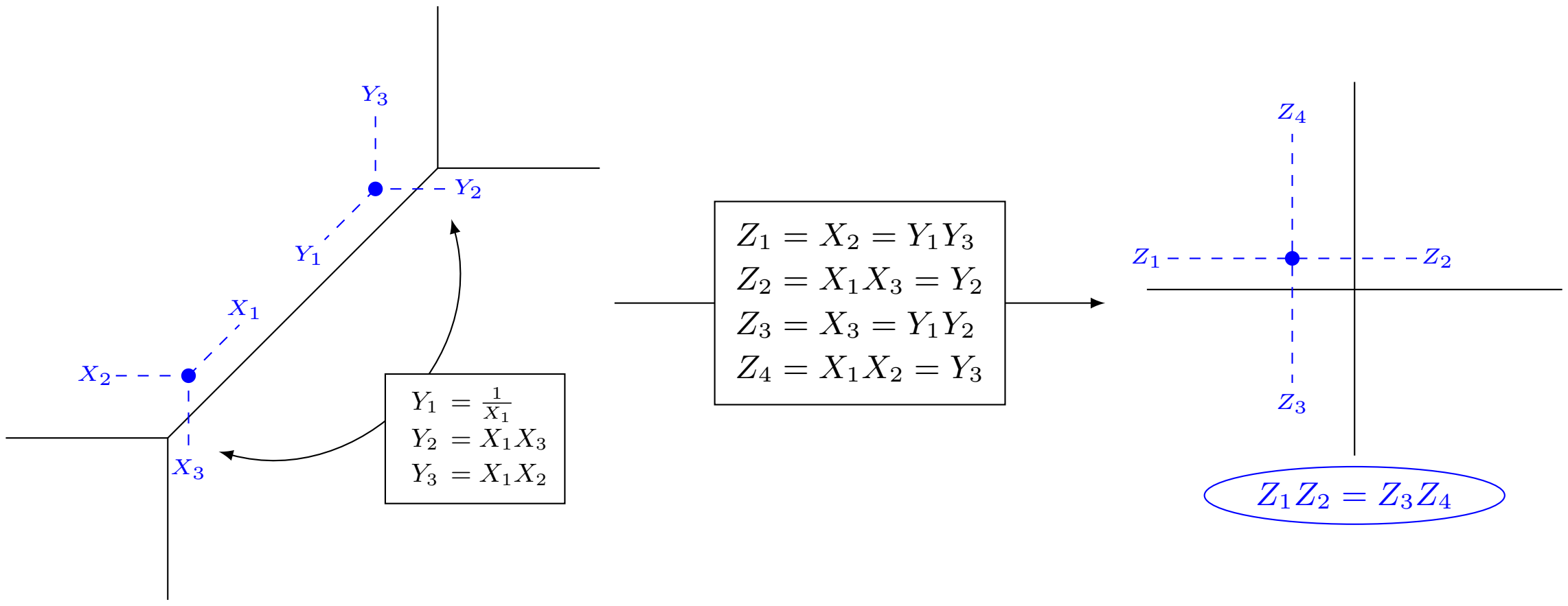
\begin{figure}[h]
\begin{center}
\begin{tikzpicture}
    \node (A) at (0,0) {
        \begin{tikzpicture}
            \tikzmath{\x = 2.6;\y =.2;\z = .7;} 
            \draw (-.6*\x,0) -- (0,0) -- (\x,\x)-- (1.6*\x,\x);
            \draw (\x,\x) -- (\x,1.6*\x);
            \draw (0,-.6*\x) -- (0,0);
        
            \node[focalPoint,blue]  at (\y,3*\y) {};
            \node[circle,minimum size = 1cm] (X) at (0,0) {};
            \draw[broken,blue] (\y-\z,3*\y) -- (\y,3*\y) -- (\y,3*\y-\z);
            \draw[broken,blue] (\y,3*\y) -- (\y+.7*\z,3*\y+.7*\z);
            \node[blue] at (\y+.7*\z+.05,3*\y+.7*\z+.15) {$\scriptstyle X_1$};
            \node[blue] at (\y-\z-.2,3*\y) {$\scriptstyle X_2$};
            \node[blue] at (\y,3*\y-\z-.2) {$\scriptstyle X_3$};
        
            \node[focalPoint,blue]  at (\x - 3*\y,\x - \y) {};
            \node[circle,minimum size = 1cm] (Y) at (\x,\x) {};
            \draw[broken,blue] (\x - 3*\y,\x - \y+\z) -- (\x - 3*\y,\x - \y) -- (\x - 3*\y+\z,\x - \y);
            \draw[broken,blue] (\x - 3*\y,\x - \y) -- (\x - 3*\y-.7*\z,\x - \y-.7*\z);
            \node[blue] at (\x - 3*\y-.7*\z-.15,\x - \y-.7*\z-.15) {$\scriptstyle Y_1$};
            \node[blue] at (\x - 3*\y,\x - \y+\z + .2) {$\scriptstyle Y_3$};
            \node[blue] at  (\x - 3*\y+\z+.2,\x - \y) {$\scriptstyle Y_2$};
        
            \draw[latex-latex,bend right=60,min distance = 40pt] (X) to  node[pos=.5,xshift=20pt,yshift=-10pt,draw,fill=white,inner sep = 3pt] {\scalebox{.8}{
            $\begin{array}{rl}
                Y_1 &= \frac{1}{X_1}\\
                Y_2 &= X_1 X_3 \\
                Y_3 &= X_1 X_2
            \end{array}$}} (Y) ;    
        
        \end{tikzpicture}
    };
\node (B) at (10,0) {
    \begin{tikzpicture}[decoration={markings,mark=at position 0.5 with {\arrow{latex}}}]
        \tikzmath{\x = 2;\y =.3;\z = 1.2;} 
        \draw (-\x,0) -- (0,0) -- (\x,0);
        \draw (0,-.8*\x) -- (0,0) -- (0,\x);
    
        \node[focalPoint,blue] at (-2*\y,\y) {};
        \draw[broken,blue] (-2*\y-\z,\y) -- (-2*\y,\y) -- (-2*\y,\y-\z);
        \draw[broken,blue] (-2*\y+\z ,\y) -- (-2*\y ,\y) -- (-2*\y,\y+\z);
        \node[blue] at (-2*\y-\z-.2,\y) {$\scriptstyle Z_1$};
        \node[blue] at (-2*\y,\y-\z-.2) {$\scriptstyle Z_3$};
        \node[blue] at (-2*\y+ \z+.2,\y) {$\scriptstyle Z_2$};
        \node[blue] at (-2*\y,\y+\z+.2) {$\scriptstyle Z_4$};

        \node[draw,ellipse,blue,inner sep=2pt,anchor=north] at (0,-.8*\x-.1) {$Z_1Z_2 = Z_3Z_4$};
    \end{tikzpicture}
    };
\draw[-latex] (A) --node[pos=.5,draw,fill=white] {$\begin{array}{l}
Z_1 = X_2 = Y_1Y_3\\
Z_2 = X_1 X_3 = Y_2\\
Z_3 = X_3 = Y_1 Y_2\\
Z_4 = X_1 X_2 = Y_3
\end{array}$} (B);
\end{tikzpicture}
\end{center}
\caption{Transition map between charts of the resolution and blow-down map for the conifold}
\label{fig:conifoldMaps}
\end{figure}

Using the generators in Table \ref{tab:thetafunctions_resolved_conifold}, we can write down explicitly the transition function and the blow-down map. We summarize them in Figure \ref{fig:conifoldMaps}. It is easy to see that they indeed reproduce the toric construction of the conifold and its small resolution.

In the rest of the paper, we show transition functions and affinization maps as presented in Figure \ref{fig:conifoldMaps}.  

\paragraph{Toric threefolds.} This construction can be generalized to construct any toric Calabi-Yau threefold. Starting with the toric polygon, the scattering diagram is simply the dual graph, with edges labelled by the appropriate power of $t$. It is automatically consistent. 

Theta-functions near the vertices of this diagram construct affine charts that are glued together using theta-functions near the edges. It is clear that this gives the same threefold as the more usual methods of toric geometry. It should not be hard to provide a rigorous proof of this fact, though we do not attempt to do it here.

\paragraph{Intersections of charts and faces of the diagram.}

For a smooth toric web, whose toric polygon is fully tessellated by triangles of area $1/2$, brane junctions correspond to $\mathbb C^3$ charts, edges to $\mathbb C^*\times \mathbb C^2$ charts and faces to $(\mathbb C^*)^2\times \mathbb C$ charts. There is a unique $(\mathbb C^*)^3$ chart, corresponding to the apex of the toric fan, which does not appear naturally in the web. 

The intersection of charts corresponding to two vertices is: 
\begin{compactitem}[$\qquad\cdot$]
    \item if they share an edge, the chart corresponding to that edge,
    \item if they share a face, but not an edge, the chart corresponding to that face,
    \item if they do not share a face, the global $(\mathbb C^*)^3$ chart.
\end{compactitem}
We can determine in the same way an arbitrary intersection of toric charts.

We expect that these rules extend to the non-toric case. Charts corresponding to vertices and edges are built using theta-functions and can be non-toric. Charts corresponding to faces of the diagram can be built using theta-functions on the empty scattering diagram\footnote{Induced on a formal neighborhood of a point of the full diagram, not contained in any edge.} and are therefore $\mathbb C\times (\mathbb C^*)^2 = \operatorname{Spec}\mathbb C[t,x^{\pm 1},y^{\pm 1}]$. There is also a global chart isomorphic to $(\mathbb C^*)^3 =\operatorname{Spec}\mathbb C[t^{\pm 1},x^{\pm 1},y^{\pm 1}]$.

One example where charts corresponding to faces of the diagram are directly needed is a stack of  $n$ separated $5$-branes. We can resolve $\mathcal X^{\rm can} = \mathbb C^* \times (\mathbb C^2 / \mathbb Z_n)$ by moving the branes away from each other. In a configuration where all the branes are separated, the scattering diagram consists of $n$ parallel lines (say horizontal). From each of the lines, we construct a copy of $\mathbb C^2 \times \mathbb  C^*$ with coordinates $u_i,v_i,t,w_{i,\pm}$ and relations $w_+ w_- = 1, u_i v_i =t$. Using the embedding in $\mathbb C[x^{\pm 1},y^{\pm 1},t]$ obtained from a point between the $i$-th and $(i+1)$-th branes, we have:
\begin{equation}
\begin{array}{ccc}
    u_i = ty, & \qquad v_i =1/y,\qquad & w_{i,\pm }= x^{\pm 1}\\
    u_{i+1} = y, & v_{i+1} =t/y, & w_{i+1,\pm }= x^{\pm 1}
\end{array}
\end{equation}
and therefore the transition function is 
\begin{equation}
\begin{array}{cl}
    v_i &= 1/u_{i+1} \\
    u_i &= u_{i+1}^2 v_{i+1} \\
    w_{i,\pm} &= w_{i+1,\pm}
\end{array}
\end{equation}
This is a description of the crepant resolution $\widetilde{\mathcal X}=\widetilde{\mathbb C^2 / \mathbb Z_n} \times \mathbb C^*$, with the resolution map $\widetilde{\mathcal X}\to \mathcal X^{\rm{can}}$ given in chart $i$ by $u = t^{i-1} u_i, \,v = t^{n-i}v_i \,.$

\subsection{General prescription}\label{sec:general_prescription}

We can generalize the method described in the previous section to construct the crepant resolution of the non-toric Calabi-Yau threefold dual to webs of 5-branes and 7-branes on their extended Coulomb branch. For such a web, we construct a consistent scattering diagram $\mathfrak D$ where walls can be placed along arbitrary rays (not necessarily ending at the origin) and line segments. Applying the theta-function construction on the whole diagram gives the algebra of global holomorphic functions on the threefold: this gives us the affine threefold $\mathcal X$. Theta-functions on the scattering diagrams $\mathfrak D_P$ in the formal neighborhood of points $P$ in $\mathfrak D$ construct affine charts of the crepant resolution $\widetilde{\mathcal X}$. Except in a degenerate case where all walls are parallel, the charts associated with the vertices of the scattering diagram form an open covering of $\widetilde{\mathcal X}$. The diagrams associated with the edges of $\mathfrak D$ tell us how to glue these charts together.

\paragraph{Scattering diagrams with multiple centers.} 

\begin{figure}[h]
    \centering
    \scalebox{.8}{
    \begin{tikzpicture}
        \tikzmath{\x = 1.6;}
        \draw[wall] (0,2*\x) -- (0,1*\x);
        \draw[wall] (0,1*\x) -- (0,0);
        \draw[wall] (0,0) -- (0,-1*\x);
        \draw[wall] (0,0) -- (-1*\x,-1*\x);
        \draw[wall] (2*\x,0) -- (1*\x,0);
        \draw[wall] (1*\x,0) -- (0,0);
        \draw[wall] (0,-1*\x) -- (0,-2*\x);
        \draw[wall] (0,0) -- (-2.5*\x,0);
        \draw[wall] (-1*\x,-1*\x) -- (-2*\x,-2*\x);
        \draw[wall] (-2.5*\x,0) -- (-5.5*\x,0);
        \draw[kink] (-3.5*\x,-1*\x) -- (1*\x,-1*\x) -- (1*\x,0) -- (0,1*\x) -- (-1.5*\x,1*\x)--(-3.5*\x,-1*\x);
        \draw[kink] (-3.5*\x,-1*\x) -- (-5.5*\x,-2*\x);
        \draw[kink] (1*\x,-1*\x) -- (2*\x,-2*\x);
        \draw[kink] (-1.5*\x,1*\x) -- (-1.5*\x,2*\x);
        \node at (.8,2*\x) {$c e t + d y$};
        \node at (2*\x + 0.8,0) {$a e t + b x$};
        \node at (.8,-2*\x) {$c + t \, ady  $};
        \node at (-2*\x -.8,-2*\x) {$e + tbd x y$};
        \node at (-5.5 *\x+.5, 0.2) {$a + b c tx $};
        \node at (-2*\x+.8,.2) {$a + b c x$};
        \node at (.5*\x , .2) {$ae+bx$};
        \node at (.7,-.5*\x) {$c+ady$};
        \node at (-.5*\x - .9,-.5*\x) {$e + bd xy$};
        \node at (-.7,.5*\x) {$c e + dy$};
    \end{tikzpicture}}
    
    \caption{Scattering diagram with two $7$-branes crossing through a web of $5$-branes. Lines without arrows are kinks and carry a factor of $t$, not shown in the figure. Letters $a,b,c,d,e$ denote complex parameters. See Section \ref{sec:multiple-seven-branes} for a detailed study of this kind of diagram.   }
    \label{fig:diag_resolved}
\end{figure}
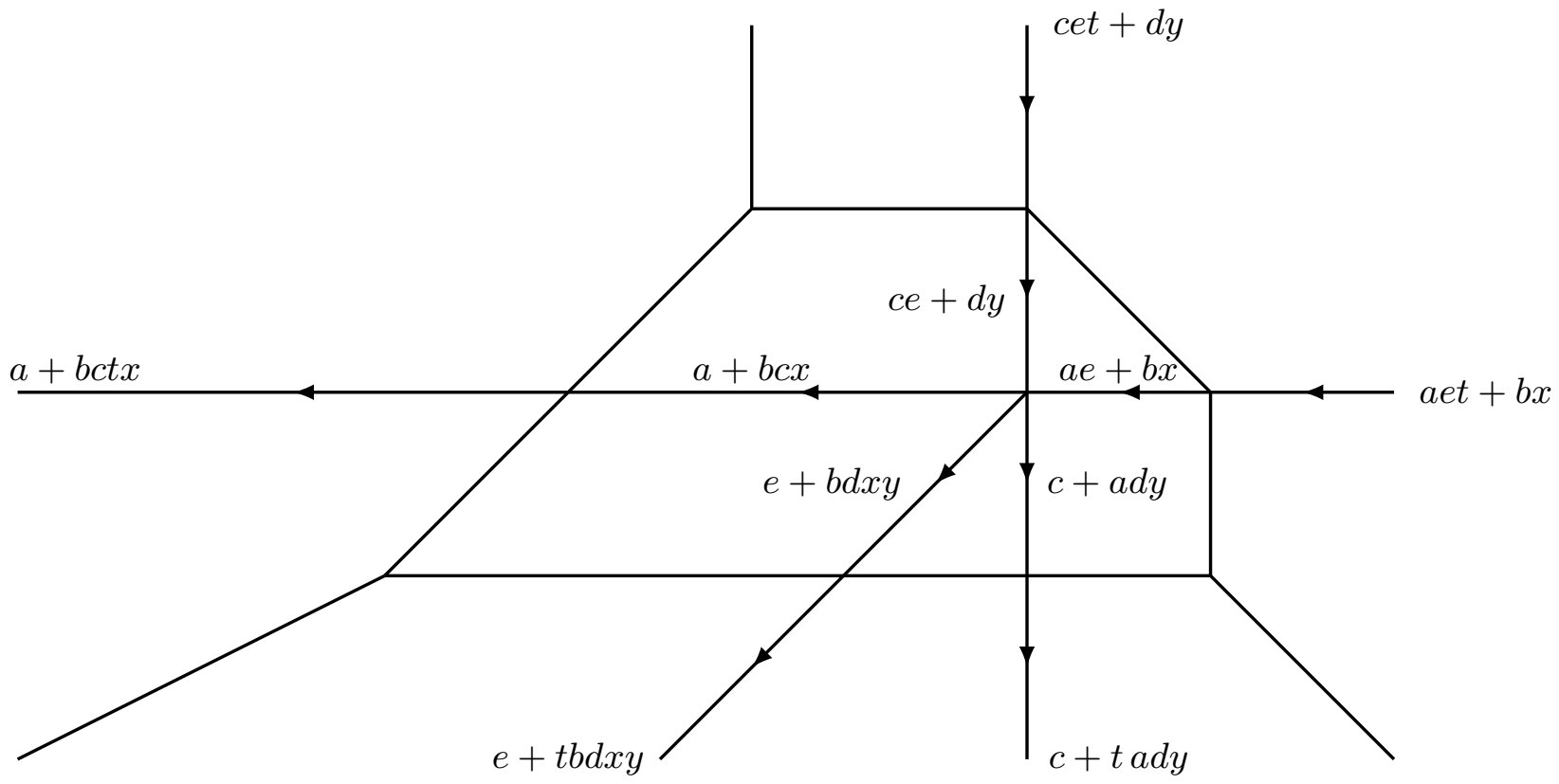

We now allow a more general class of scattering diagrams in the plane, like the one in Figure \ref{fig:diag_resolved}, where walls and kinks are allowed to be segments and lines in the affine plane. Monodromies along curves are computed as before. Such a diagram is \emph{coherent} if the monodromy around any loop is trivial, or equivalently if the monodromies along any two paths with the same endpoints are equal.

To construct the consistent scattering diagram associated to a tessellated GTP, the procedure is as follows: 
\begin{itemize}
    \item construct the kinks: they correspond exactly to the brane web, with powers of $t$ equal to the number of $5$-branes stacked together in a given segment;
    \item add incoming walls: for a given $(p,q)$ 7-brane with $k_i$ 5-branes ending on it, the wall is $(P + \mathbb R_+ \left(\begin{smallmatrix} p\\ q \end{smallmatrix}\right),1+\alpha_i x^p y^q t^{-k_i})$. The position of the infinite line $P + \mathbb R\left(\begin{smallmatrix} p\\ q \end{smallmatrix}\right)$ corresponds to the position of the $7$-brane. $P$ is the first point where this line intersects the rest of the web. 
    \item make the diagram consistent: at each point $P$ where the incoming walls intersect other walls or kinks, we run the algorithm described in Section \ref{sec:scattering_diagrams} and Appendix \ref{app:algoConsistent}. This gives a list of germs of outgoing walls necessary to make the diagram locally consistent around $P$. We extend these outgoing walls until they intersect another wall or kink in the diagram. At a finite order in the $\alpha_i$, this procedure ends after a finite number of steps.
\end{itemize}

\paragraph{Theta-functions and the threefolds.}

If the consistent scattering diagram $\mathfrak D$ is finite, we construct the threefold $\widetilde{\mathcal X}$ by gluing the affine charts built from the vertices of the diagram.
For each vertex $P_i$ in the diagram, where kinks and/or walls intersect, we construct the algebra $\mathcal A_i$ of theta functions associated with the local diagram $\mathfrak D_{P_i}$, consisting only of the walls and kinks that meet at $P_i$. Write $\mathcal U_i = \operatorname{Spec}(\mathcal A_i)$ for the corresponding affine variety.

Consider a segment $[P_iP_j]$ of the diagram  (wall or kink) between two vertices, with primitive tangent vector $w\in\mathbb Z^2$ (oriented from $P_i$ to $P_j$). Let $\mathcal A_i,\mathcal A_j$ be the algebras of theta-functions around the two vertices and $\mathcal A_{ij}$ the algebra of theta-functions around the edge. We claim that as in the toric case, there are natural embeddings 
\begin{equation}\label{eq:thetafunction_embedding}
\mathcal A_i\hookrightarrow \mathcal A_{ij}\hookleftarrow \mathcal{A}_j \,.    
\end{equation}
Moreover, with $\vartheta^{(i)}_w\in \mathcal A_i$ and $\vartheta^{(j)}_{-w} \in \mathcal A_j$  the theta-functions associated with the direction of the edge, we see that, in $\mathcal A_{ij}$, we have $\vartheta^{(i)}_w\vartheta^{(j)}_{-w} = 1$. We claim that the embeddings \eqref{eq:thetafunction_embedding} identify $\mathcal A_{ij}$ as the localization: 
\begin{equation}\label{eq:thetafunction_localization}
\mathcal A_{ij}\simeq \mathcal{A}_i[(\vartheta^{(i)}_w)^{-1}] \simeq \mathcal{A}_j[(\vartheta^{(j)}_{-w})^{-1}] \,.    
\end{equation}
Geometrically, we get an isomorphism between the open subsets $\{x\in\mathcal U_i|\vartheta^{(i)}_w(x) \neq 0\}$ and $\{y\in\mathcal U_j|\vartheta^{(j)}_{-w}(y)\neq 0\}$, which we can use to glue $\mathcal U_i$ and $\mathcal U_j$ together. We obtain $\mathcal X$ by repeating this process for all vertices and edges in the diagram.
Intersections between the different charts behave as in the toric case. The only difference is that the charts associated with vertices and edges are constructed from diagrams that may contain walls. In particular, they may be non-toric.

As everything lives in the large scattering diagram $\mathfrak D$, we get natural maps $\mathcal A\hookrightarrow \mathcal A_i$ where $\mathcal A$ is the algebra of theta-functions on $\mathfrak D$. These maps are compatible with the gluing maps \eqref{eq:thetafunction_embedding}. $\mathcal A$ is also the algebra of theta-functions on the diagram $\mathfrak D^{\rm{asymp}}$ where all the walls and kinks of finite length have been removed and the semi-infinite ones have been moved to end at the origin. In other words, $\mathfrak D^{\rm{asymp}}$ is the consistent scattering diagram for the GTP with its tessellation removed. Geometrically, this gives an affinization map $\varphi:\widetilde{\mathcal X}\to \mathcal{X}$.
We claim that $\varphi$ is a crepant resolution.

For an infinite diagram (with accumulation points or dense regions): at order $N$ the diagram is finite. We can apply this method to get a scheme over $\mathbb C\llbracket \alpha_i\rrbracket /I^N$. There may then be a way to get a Calabi-Yau variety in the limit $N\to \infty$. However, this seems unlikely to be of practical use, at least in the simple physical settings, so we do not attempt to do it here. 

Just like in the affine case, if there are white dots / several $5$-branes ending on the same $7$-brane, then $\mathcal X$ is a partial smoothing of the singular toric variety obtained by removing the $7$-branes. $7$-branes with a single $5$-brane ending on them simply induce a non-toric deformation.
As in the affine case, the diagrams transform in a simple way under Hanany-Witten transitions and the threefold is invariant.

Although the construction mimics that of toric varieties, thus justifying the naming of generalized toric varieties, one should be careful about hasty extensions of known properties of toric varieties. Here is a short list of such properties: \label{pageProperties}
\begin{itemize}
    \item In toric varieties, finite edges and faces in the brane web imply the existence of associated compact holomorphic curves and divisors. This is not the case in generalized toric geometry. 
    \item In toric geometry, opening up a face in the web or changing the triangulation implies a topological change (blow-up, flop). This is not systematically the case in generalized toric geometry. 
    \item In generalized toric geometry, some open charts associated with vertices may not add any new point to $\mathcal X$ and instead be fully contained in the union of the other charts. 
    \item In toric geometry, triple intersections of divisors can be directly read from the toric polygon / brane web. In the generalized toric setting, this becomes more intricate due to the appearance of exuberant divisors \cite{ABBGS}. Our methods can exhibit the exuberant divisors and reproduce these results in an elementary fashion. 
    \item In the toric case, the Euler characteristic is simply the number of vertices of the web \cite{cox2011toric}. This can be computed by an inclusion-exclusion generalization of the Mayer-Vietoris formula: the vertices provide an open cover and each chart has $\chi = 1$. All non-empty intersections of charts contain a factor of $\mathbb C^*$ and therefore have $\chi =0$. In the presence of $7$-branes, the situation is more complicated. In particular, some vertices may have $\chi =0$ and some edges $\chi =1$.
\end{itemize} 
Rather than attempting to give formal proofs of these statements (which we assume present no particular conceptual difficulty, only careful bookkeeping), we now explore each of these phenomena in detail through examples in the next subsections.

\subsection{Revisiting Hanany-Witten moves}

In this section, we will study how Hanany-Witten moves are realized in scattering diagrams corresponding to resolved brane webs, using as an illustratory example the transition $\mathbb{F}_0 \leftrightarrow \mathbb{F}_2$ induced by ending an external 5-brane on a 7-brane and crossing it.

\subsubsection{$\mathbb{F}_0 \leftrightarrow \mathbb{F}_2$}

We start with the brane web of the resolved local $ \mathbb{F}_2$ at a concrete point of the extended Coulomb branch. This web has 4 trivalent junctions, so in principle its underlying variety will be covered by 4 patches.  

\begin{figure}[h]
\centering
\scalebox{.9}{
\begin{tikzpicture}[
    x=0.65cm,
    y=0.65cm,
    >=Stealth,
    every node/.style={font=\small},
    redline/.style={red, very thick,dashed},
    greenline/.style={green!55!black, very thick,dashed},
    orangeline/.style={orange, very thick,dashed},
    magentaline/.style={magenta, very thick,dashed},
    blueline/.style={goodblue, very thick,dashed},
    blackline/.style={black, very thick}
]

\coordinate (A) at (-9,-2);
\coordinate (a) at (-6,0);
\coordinate (B) at (-3,4);
\coordinate (b) at (-4,2);
 \coordinate (C) at (3,4);
  \coordinate (c) at (2,2);
\coordinate (D) at (9,-2);
\coordinate (d) at (6,0);
\coordinate (E) at (3,-2);
\coordinate(e) at (2,-0.5);

\draw[kink] (A) -- (E) node[pos=.5,below]{$t$} --(D) node[pos=.5,below]{$t$};
\draw[kink] (A) -- (B)  node[pos=.5, above left]{$t$};
\draw[kink] (B) -- (C)  node[pos=.5,above]{$t$};
\draw[kink] (D) -- (C)  node[pos=.5,above right]{$t$};
\draw[wall] (C) -- (E)  node[pos=.5,right]{$\alpha +\beta y$};
\draw[wall] (E) -- ++(0,-3)  node[pos=.5,right]{$\alpha +\beta t y$};
\draw[wall]  ($(C)+(0,3)$)--(C)  node[pos=.5,right]{$\alpha t +\beta y $};
\draw[kink] (B) -- ++(0,3) node[pos=.5,left]{$t$};
\draw[kink] (A) -- ++(-4,-2) node[pos=.5,above left]{$t$};
\draw[kink] (D) -- ++(4,-2) node[pos=.5,below left]{$t$};

\draw[broken, goodgreen] (a)--($(a)+(1,1)$) node[pos=0.5, right]{$A_3$};
\draw[broken, goodgreen] (a)--++(1,0)  node[pos=0.5, below]{$A_2$};
\draw[broken, goodgreen] (a)--++(-2,-1) node[pos=0.5, below]{$A_1$};
\node[focalPoint,goodgreen] at (a) {};

\draw[broken, goodblue] ($(a)+(1,1)$)--(b) node[pos=0.5, right]{$B_3$};
\draw[broken, goodblue] (b)--++(0,1.5) node[pos=0.5, above left]{$B_1$};
\draw[broken, goodblue] (b)-- ($(b)+(1.5,0)$)  node[pos=0.5, above]{$B_2$};
\node[focalPoint,goodblue] at (b) {};

\draw[broken,goodred] ($(c)+(-1.5,0)$)--(c) node[pos=0.5, above]{$C_2$};
\draw[broken,goodred] (c)--($(c)+(1.5,-1.5)$) ;
\draw[broken,goodred] (c)--($(c)+(1.0,0)$) node[pos=0.5,above]{$C_3$}--($(c)+(2.0,-1.0)$);
\draw[broken,goodred] ($(c)+(0,-1)$)--(c) node[pos=0.5, left]{$C_4$}--($(c)+(0,1.5)$) node[pos=1, right]{$C_1$};
\node[focalPoint,goodred] at (c) {};

\draw[broken, orange] ($(d)+(-1.5,0)$)--(d)  node[pos=0.5, above]{$D_2$};
\draw[broken, orange] ($(d)+(-1.5,1.5)$)--(d)  node[pos=0.5, above]{$D_3$};
\draw[broken, orange](d)--++(3,-1.5) node[pos=0.5, above right]{$D_1$};
\node[focalPoint,orange] at (d) {};

\draw[broken,magenta] ($(e)+(-1.5,0)$)--(e) node[pos=0.5, above]{$E_3$};
\draw[broken,magenta] (e)--($(e)+(1.5,-1.5)$) ;
\draw[broken,magenta] (e)--($(e)+(1.0,0)$) node[pos=0.5,above]{$E_2$}--($(e)+(2.0,-1.0)$);
\draw[broken,magenta] ($(e)+(0,-1)$)--(e) node[pos=0.5, left]{$E_1$}--($(e)+(0,1.5)$) node[pos=0.75, right]{$E_4$};
\node[focalPoint,magenta] at (e) {};

\fill[green!55!black] (A) circle (2.4pt);
\fill[goodblue] (B) circle (2.4pt);
\fill[goodred] (C) circle (2.4pt);
\fill[orange] (D) circle (2.4pt);
\fill[magenta] (E) circle (2.4pt);

\node[black] at ($(A)+(0,-0.5)$) {$A$};
\node[black] at ($(B)+(0.5,0.5)$) {$B$};
\node[black] at ($(C)+(0.5,0.5)$) {$C$};
\node[black] at ($(D)+(0.5,0.5)$) {$D$};
\node[black] at ($(E)+(0.5,-0.5)$) {$E$};

\end{tikzpicture} 
}
    \caption{ Resolved local $\mathbb F_2$ with a 7-brane on the $(0,1)$ leg at $C$, in the frame adapted to $[\alpha:\beta]\to[1:0]$.  }
    \label{fig:resolvedF2}
\end{figure}
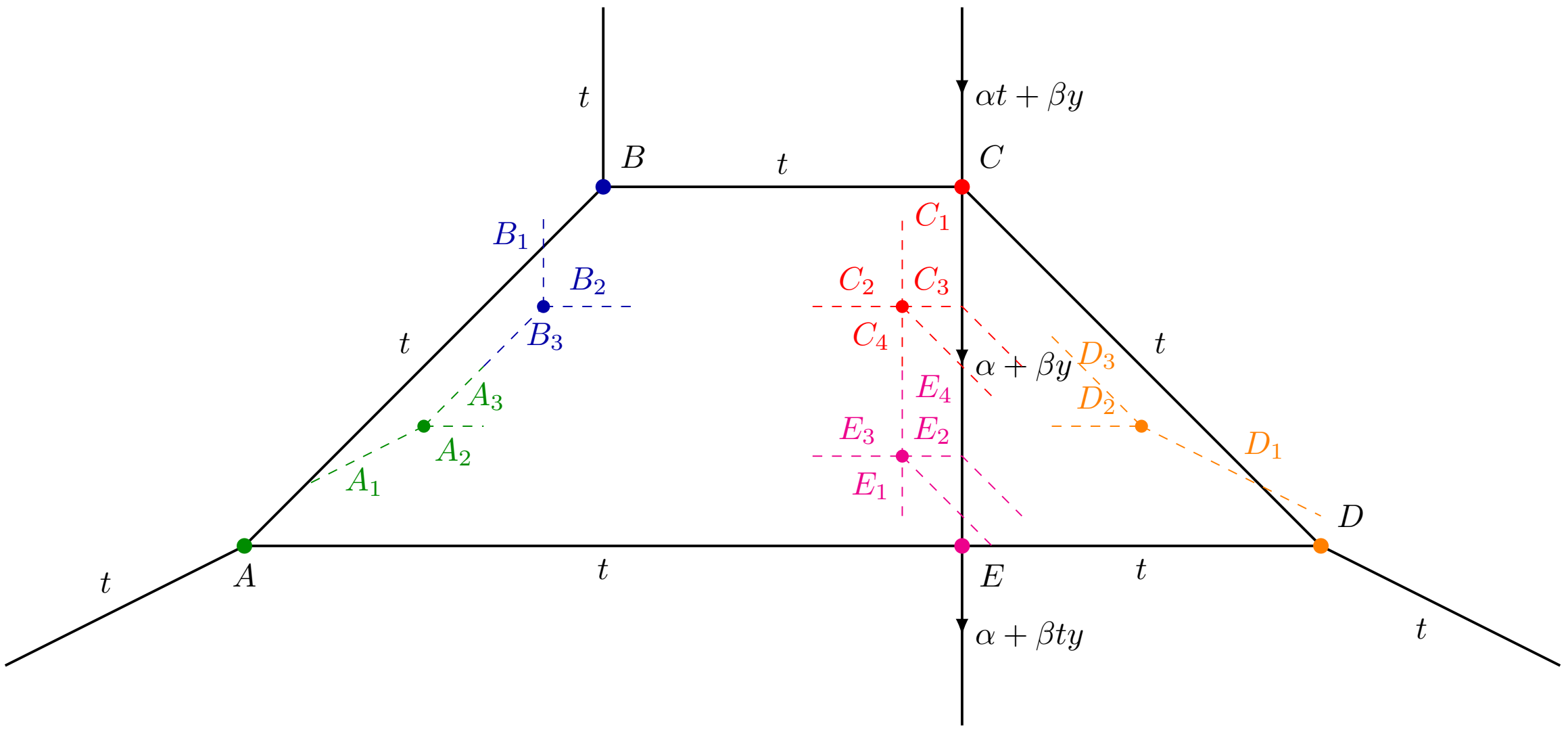

We attach a 7-brane to one of the external $(0,-1)$ 5-branes, so we add an incoming wall at that junction and run the consistency algorithm. As a result, we find an outgoing wall in the $(0,1)$ direction, which intersects a $(1,0)$ brane and creates a new junction. This results in the addition of a fifth patch. We demand consistency in this new patch, which results in the full scattering diagram in Figure \ref{fig:resolvedF2}.

The 5 patches are read from the theta-functions (computed at the appropriate points) given by
\begin{equation}
\begin{aligned}
     \mathcal U_A&=\mathbb{C}[A_1,A_2,A_3] \cong \mathbb{C}^3, \quad  \mathcal U_B=\mathbb{C}[B_1,B_2,B_3] \cong \mathbb{C}^3, \quad \mathcal U_C=\mathbb{C}[C_1,C_2,C_3,C_4]/(C_2 C_3-\alpha C_4- \beta) ,\\
     \mathcal U_D&=\mathbb{C}[D_1,D_2,D_3],\quad \mathcal U_E=\mathbb{C}[E_1,E_2,E_3,E_4]/(E_2 E_3-\alpha-\beta E_4)\,,
\end{aligned}
   \end{equation}
   with transition functions 
   \begin{equation}
   \begin{aligned}
       AB&: B_1= A_1 A_3^2,\quad B_2=A_2,\quad B_3=A_3^{-1},\\
       BC&: C_1= B_1, \quad C_2=B_2^{-1},\quad C_3= \alpha B_2^2 B_3+\beta B_2,\quad C_4= B_2 B_3,\\
       CD&: D_1 = C_1 C_3^2 ,\quad D_2 = C_3^{-1}C_4,\quad D_3 = C_3^{-1}\\
    ED&: E_1 = D_1 D_2^{2} ,\quad E_2 = D_2^{-1}, \quad E_3 = \alpha D_2 + \beta D_3,\quad E_4 =  D_2^{-1}D_3,\\
       EA&: A_1= E_1 E_3^2,\quad A_2=E_3^{-1},\quad A_3= E_4E_3^{-1}.
   \end{aligned}    
   \label{eq:F2transitions}
   \end{equation}
Note that in order to find well-behaved transition functions, the choice of embedding of the algebra of the theta functions into $\mathbb{C}[t,x^{\pm1},y^{\pm1}]$ for each open patch $\mathcal U_i$ is important (see section \ref{sec::resolvedconif}). 
Note that for the third and fourth lines of \eqref{eq:F2transitions} to be realized in $\mathbb{C}[t,x^{\pm1},y^{\pm1}]$, we have to recompute the theta-functions $C_{1,2,3,4}$ and $E_{1,2,3,4}$ at a focal point lying in the face of the scattering diagram of which $D$ is a vertex.

We turn off the non-toric deformation and recover the local $\mathbb{F}_2$ ($[\alpha:\beta]=[1:0]$). We see that the $E$ patch  becomes
\begin{equation}
   \mathcal U_E=\mathbb{C}[E_1,E_2,E_3,E_4]/(E_2 E_3-1)  \cong \mathbb{C}^2 \times \mathbb{C}^*\,.
\end{equation}
Moreover $\mathcal U_E \subset \mathcal U_D \cap \mathcal U_A$, so this patch is redundant and may be eliminated. On the other hand on the $C$ patch,
\begin{equation}
     \mathcal U_C=\mathbb{C}[C_1,C_2,C_3,C_4]/(C_2 C_3- C_4)=\mathbb{C}[C_1,C_2,C_3] \cong \mathbb{C}^3,
\end{equation}
and it becomes the usual toric patch. Setting the fiber coordinates to zero on every patch ($A_1=B_1=C_1=D_1=E_1=0$), we find a compact divisor whose geometry can be read off from the transition functions and is $\mathbb{F}_2$. Notice that in this limit, due to the consistency condition at every junction, the power of $t$ associated to each wall counts the multiplicity of the 5-branes in the web.

Taking the opposite limit of the $\mathbb{P}^1$ fibration ($[\alpha:\beta]=[0:1]$) the $C$ patch becomes $\mathcal U_C\cong \mathbb{C}^2 \times \mathbb{C}^*$ so $\mathcal U_C \subset \mathcal  U_B \cap \mathcal U_D$ and $\mathcal U_E \cong \mathbb{C}^3$. Setting $A_1=B_1=C_1=D_1=E_1=0$ one finds a $\mathbb{P}^1 \times\mathbb{P}^1=\mathbb{F}_0$ compact divisor.  Notice that in this limit, due to the consistency condition at every junction, the power of $t$ associated to each wall counts the multiplicity of the 5-branes in the web after crossing the 7-brane.

There is a way to understand the deformed patches $\mathcal{U}_C,\mathcal{U}_E$ by considering their dual tiles in the toric diagram at the poles of the deformation $\mathbb{P}^1$. At one pole, $\mathcal{U}_C$ is dual to a toric triangle and $\mathcal{U}_E$ does not correspond to any tile, and at the opposite pole the behavior reverses. This is what happens with polytope mutations, where a T-cone (in this case a trivial T-cone) is flipped across the web \cite{Arias-Tamargo:2024fjt}. Thus at a generic point of the $\mathbb{P}^1$ we are bridging between the two tessellations.  

\begin{figure}[h]
    \centering

\scalebox{.9}{
\begin{tikzpicture}[
    x=0.65cm,
    y=0.65cm,
    >=Stealth,
    every node/.style={font=\small},
    redline/.style={red, very thick,dashed},
    greenline/.style={green!55!black, very thick,dashed},
    orangeline/.style={orange, very thick,dashed},
    magentaline/.style={magenta, very thick,dashed},
    blueline/.style={goodblue, very thick,dashed},
    blackline/.style={black, very thick}
]

\coordinate (A) at (-9,-2);
\coordinate (a) at (-6,0);
\coordinate (B) at (-3,4);
\coordinate (b) at (-4,2);
\coordinate (C) at (3,4);
\coordinate (c) at (2,2);
\coordinate (D) at (9,4);
\coordinate (d) at (6,3);
\coordinate (E) at (3,-2);
\coordinate(e) at (2,-0.5);

\draw[kink] (A) -- (E) node[pos=.5,below]{$t$} --(D) node[pos=.5,below]{$t$};
\draw[kink] (A) -- (B)  node[pos=.5, above left]{$t$};
\draw[kink] (B) -- (C)  node[pos=.5,above]{$t$};
\draw[kink] (D) -- (C)  node[pos=.5,above right]{$t$};
\draw[wall] (E) -- (C)  node[pos=.5,right]{$\alpha +\frac{\beta}{y}$};
\draw[wall]  ($(E)+(0,-3)$) --(E)  node[pos=.5,right]{$\alpha t+\frac{\beta}{y}$};
\draw[wall] (C)-- ($(C)+(0,3)$)  node[pos=.5,right]{$\alpha +\frac{\beta}{y} t$};
\draw[kink] (B) -- ++(0,3) node[pos=.5,left]{$t$};
\draw[kink] (A) -- ++(-4,-2) node[pos=.5,above left]{$t$};
\draw[kink] (D) -- ++(4,2) node[pos=.5,below left]{$t$};

\draw[broken, goodgreen] (a)--($(a)+(1,1)$) node[pos=0.5, right]{$A_3$};
\draw[broken, goodgreen] (a)--++(1,0)  node[pos=0.5, below]{$A_2$};
\draw[broken, goodgreen] (a)--++(-2,-1) node[pos=0.5, below]{$A_1$};
\node[focalPoint,goodgreen] at (a) {};

\draw[broken, goodblue] ($(a)+(1,1)$)--(b) node[pos=0.5, right]{$B_3$};
\draw[broken, goodblue] (b)--++(0,1.5) node[pos=0.5, above left]{$B_1$};
\draw[broken, goodblue] (b)-- ($(b)+(1.5,0)$)  node[pos=0.5, above]{$B_2$};
\node[focalPoint,goodblue] at (b) {};

\draw[broken,goodred] ($(c)+(-1.5,0)$)--(c) node[pos=0.5, above]{$C_2$};
\draw[broken,goodred] (c)--($(c)+(1.5,0)$) node[pos=1,above]{$C_3$} ;
\draw[broken,goodred] (c)--($(c)+(1.0,1.0)$) --($(c)+(2.0,1.0)$);
\draw[broken,goodred] ($(c)+(0,-1)$)--(c) node[pos=0.5, left]{$C_4$}--($(c)+(0,1.5)$) node[pos=1, right]{$C_1$};
\node[focalPoint,goodred] at (c) {};

\draw[broken, orange] ($(d)+(-1.5,0)$)--(d)  node[pos=0.5, above]{$D_2$};
\draw[broken, orange] ($(d)+(-1.5,-1.5)$)--(d)  node[pos=0.5, above]{$D_3$};
\draw[broken, orange](d)--++(2,1) node[pos=0.5, right]{$\,D_1$};
\node[focalPoint,orange] at (d) {};

\draw[broken,magenta] ($(e)+(-1.5,0)$)--(e) node[pos=0.5, above]{$E_3$};
\draw[broken,magenta] (e)--($(e)+(1.5,1.5)$) ;
\draw[broken,magenta] (e)--($(e)+(1.0,0)$) --($(e)+(2.0,1.0)$) node[pos=0.5,right]{$E_2$};
\draw[broken,magenta] ($(e)+(0,-1)$)--(e) node[pos=0.5, left]{$E_4$}--($(e)+(0,1.5)$) node[pos=0.75, right]{$E_1$};
\node[focalPoint,magenta] at (e) {};

\fill[green!55!black] (A) circle (2.4pt);
\fill[goodblue] (B) circle (2.4pt);
\fill[goodred] (C) circle (2.4pt);
\fill[orange] (D) circle (2.4pt);
\fill[magenta] (E) circle (2.4pt);

\node[black] at ($(A)+(0,-0.5)$) {$A$};
\node[black] at ($(B)+(0.5,0.5)$) {$B$};
\node[black] at ($(C)+(0.5,0.5)$) {$C$};
\node[black] at ($(D)+(0.5,0.5)$) {$D$};
\node[black] at ($(E)+(0.5,-0.5)$) {$E$};

\end{tikzpicture} 
}
\caption{Scattering diagrams and broken lines for local $\mathbb F_0$ and, after Hanany-Witten transition, local $\mathbb F_2$. }
\label{fig:resolvedF0}
\end{figure}
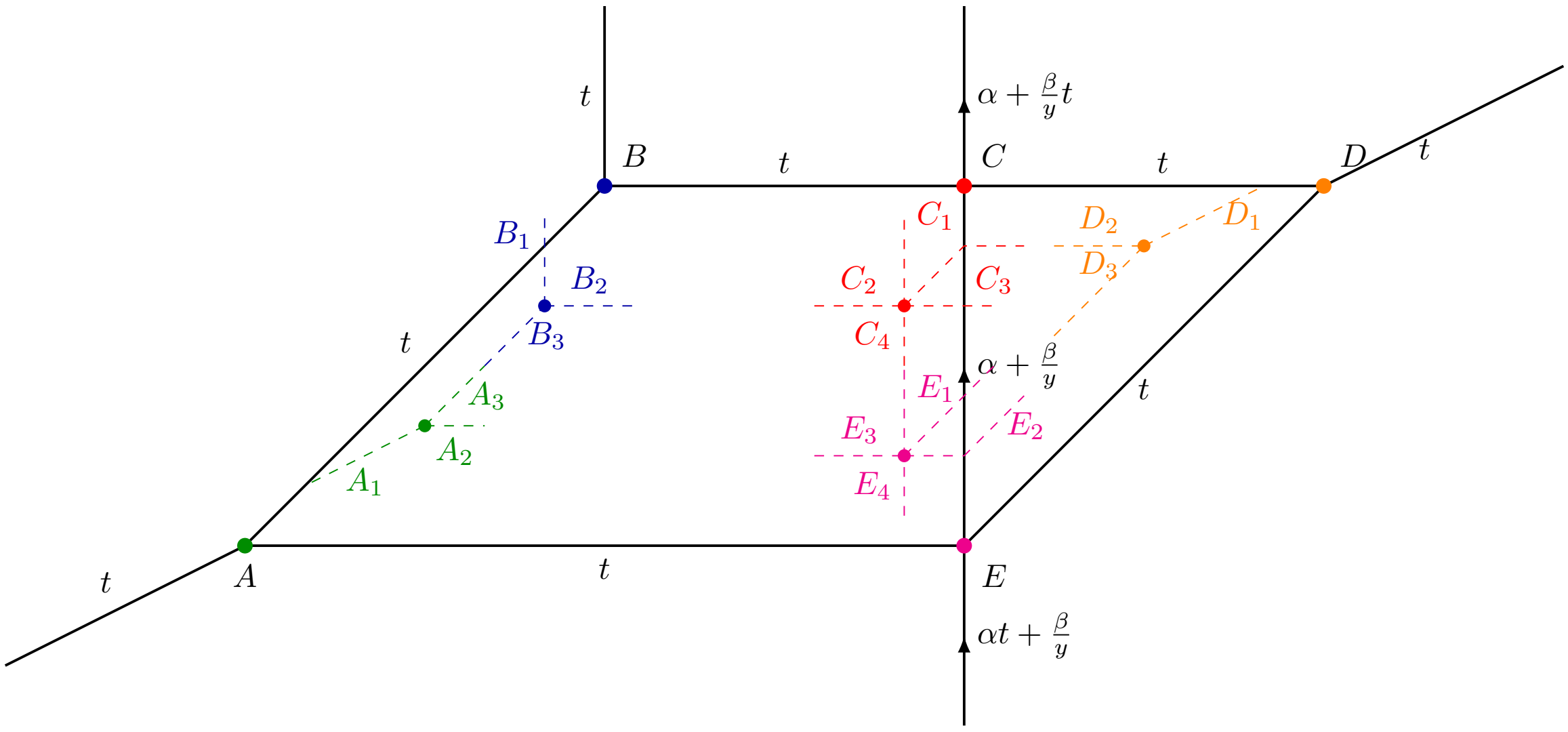

We recover the same result starting from the local $\mathbb{F}_0$, see Figure \ref{fig:resolvedF0}.

\subsection{Change of tessellation} \label{sec::1-3curve}

In the case of toric webs,  maximally resolving the geometry corresponds on the toric diagram to completely tiling it with minimal  area 1/2 triangles, dual to $\mathbb{C}^3$ patches which are glued along the directions of the edges. These toric patches are smooth and cannot be resolved further. GTPs generalize this by including new classes of smooth patches including trapezia and T-cones, which are then glued, dual to new triangulation pieces with area greater than 1/2. 

In the toric case, changing triangulations is achieved through a series of flop transitions. If we have two triangles with vertices $v_1 v_2 v_3$ and $v_2 v_3 v_4$ there is a $\mathbb{P}^1$ generated by the gluing along $v_2 v_3$. If the $\mathbb{P}^1$ has volume 0, we have a singular patch $v_1 v_2 v_3 v_4$ that can be resolved into two triangles  $v_1 v_2 v_4$ and $v_1 v_3 v_4$ (if non-degenerate) with another $\mathbb{P}^1$ defined along  $v_1v_4$. In terms of intersection data, each vertex $v_i$ corresponds to a divisor $\mathcal{D}_i$; performing a flop is just exchanging the $\mathbb{P}^1$ curve given by $\mathcal{D}_2 \cdot \mathcal{D}_3$ with a new $\mathbb{P}^1$ curve $\mathcal{D}_1 \cdot \mathcal{D}_4$ . This transition can be thought of as the volume of one $\mathbb{P}^1$/length of the segment in the scattering diagram ``becoming negative'' after crossing the singular point. Since the volume is parameterized by a combination of extended Coulomb branch parameters, after flopping this quantity flips sign.

We are interested in studying if similar flop transitions happen in GTP tilings with non-minimal triangle pieces. We will consider the concrete example of the local (1,-3) curve, $\mathcal{O}(1) \oplus \mathcal{O}(-3)\to  \mathbb P^1$. 
Its toric polygon is not convex, and we cannot use this geometry to flow to a super-conformal fixed point. Geometrically, the compact $\mathbb P^1$ is not shrinkable. However, a non-toric deformation of it can be obtained by adding two $7$-branes as on the left of Figure \ref{fig:HW_(1,-3)curve}. When we try to remove these $7$-branes by sending them to infinity, one of them has to cross the 5-brane attached to the other. The latter 7-brane can be sent to infinity safely but the first one, now attached to two 5-branes due to the Hanany-Witten effect, needs to be kept at a finite, though arbitrarily large, distance. We are then free to set all the finite lengths in the diagram to $0$ and we obtain the convex GTP shown on the right of Figure~\ref{fig:HW_(1,-3)curve}.

\begin{figure}[h]
    \centering
    \begin{tikzpicture}[x=3.5cm,y=3.5cm]
    \node (A) at (0,0) {\scalebox{.8}{    
        \begin{tikzpicture}[x=.7cm,y=.7cm]
            \draw (0,0) -- (1,1) -- (3,1) -- (5,0);
            \draw (1,1) -- (1,1.5);
            \draw (3,1) -- (2.5,1.5);
            \node[focalPoint] at (1,1.5) {};
            \node[focalPoint] at (2.5,1.5) {};
            \draw[densely dotted] (1,1.5) -- (1,4);
            \draw[densely dotted] (2.5,1.5) to[in=-90,out=135] (1.5,3.95);
        \end{tikzpicture}}};
    \node (B) at (1,0) {\scalebox{.8}{    
        \begin{tikzpicture}[x=.7cm,y=.7cm]
            \draw (0,0) -- (1,1) -- (3,1) -- (5,0);
            \draw (1,1) -- (1,1.5);
            \draw (3,1) -- (1,3) -- (0,3);
            \node[focalPoint] at (1,1.5) {};
            \draw[densely dotted] (1,1.5) -- (1,4);
        \end{tikzpicture}}};
    \node (C) at (2,0) {\scalebox{.8}{
        \begin{tikzpicture}[x=.7cm,y=.7cm]
            \tikzmath{\e = .1;}
            \draw (0,0) -- (1,1) -- (3,1) -- (5,0);
            \draw (1,1) -- (1,3.5);
            \draw (1+\e,3-\e) -- (1+\e,3.5);
            \draw (3,1) -- (1+\e,3-\e) -- (0,3-\e);
            \node[focalPoint] at (1+.5*\e,3.5) {};
            \draw[densely dotted] (1+.5*\e,3.5) -- (1+.5*\e,4);
        \end{tikzpicture}}};
    \node (D) at (3,0) {\scalebox{.8}{
        \begin{tikzpicture}[x=.7cm,y=.7cm]
            \tikzmath{\e = .1;}
            \draw (0,0) -- (1.66666,1.6666) -- (5,0);
            \draw (1.66666,1.6666) -- (1.66666,3);
            \draw (1.66666-\e,1.6666) -- (1.66666-\e,3);
            \draw (0,1.66666) -- (1.66666,1.6666);
            \node[focalPoint] at (1.6666-.5*\e,3) {};
            \draw[densely dotted] (1.6666-.5*\e,3) -- (1.6666,4);
        \end{tikzpicture}
        }};
    \node (a) at (0,-.6) {\scalebox{.9}{
        \begin{tikzpicture}[x=.5cm,y=.5cm] 
            \draw[gridline] (-1,-.5)--(-1,2.5); 
            \draw[gridline] (0,-.5)--(0,2.5); 
            \draw[gridline] (1,-.5)--(1,2.5); 
            \draw[gridline] (-1.5,0)--(1.5,0); 
            \draw[gridline] (-1.5,1)--(1.5,1); 
            \draw[gridline] (-1.5,2)--(1.5,2); 
            \draw[ligne] (-1,1)--(0,0); 
            \draw[ligne] (1,2)--(0,0) -- (0,1); 
            \draw[ligne] (-1,1)--(0,1) -- (1,2); 
            \node[bd] at (1,2) {}; 
            \node[bd] at (-1,1) {}; 
            \node[bd] at (0,0) {}; 
        \end{tikzpicture} }};
    \node (b) at (2,-.6) {\scalebox{.9}{    
        \begin{tikzpicture}[x=.5cm,y=.5cm] 
            \draw[gridline] (-1,-.5)--(-1,2.5); 
            \draw[gridline] (0,-.5)--(0,2.5); 
            \draw[gridline] (1,-.5)--(1,2.5); 
            \draw[gridline] (-1.5,0)--(1.5,0); 
            \draw[gridline] (-1.5,1)--(1.5,1); 
            \draw[gridline] (-1.5,2)--(1.5,2); 
            \draw[ligne] (-1,2)--(1,2); 
            \draw[ligne] (-1,2)--(-1,1); 
            \draw[ligne] (-1,1)--(0,0); 
            \draw[ligne] (1,2)--(0,0) -- (0,1); 
            \draw[ligne] (-1,1)--(0,1) -- (1,2); 
            \node[bd] at (-1,2) {}; 
            \node[bd] at (1,2) {}; 
            \node[bd] at (-1,1) {}; 
            \node[bd] at (0,0) {}; 
            \node[wd] at (0,2) {}; 
        \end{tikzpicture}}};
    \node (c) at (3,-.6) {\scalebox{.9}{
        \begin{tikzpicture}[x=.5cm,y=.5cm] 
            \draw[gridline] (-1,-.5)--(-1,2.5); 
            \draw[gridline] (0,-.5)--(0,2.5); 
            \draw[gridline] (1,-.5)--(1,2.5); 
            \draw[gridline] (-1.5,0)--(1.5,0); 
            \draw[gridline] (-1.5,1)--(1.5,1); 
            \draw[gridline] (-1.5,2)--(1.5,2); 
            \draw[ligne] (-1,2)--(1,2); 
            \draw[ligne] (-1,2)--(-1,1); 
            \draw[ligne] (-1,1)--(0,0); 
            \draw[ligne] (1,2)--(0,0); 
            \node[bd] at (-1,2) {}; 
            \node[bd] at (1,2) {}; 
            \node[bd] at (-1,1) {}; 
            \node[bd] at (0,0) {}; 
            \node[wd] at (0,2) {}; 
        \end{tikzpicture}}};  
    \draw[-latex] ([xshift=-15pt]A.east) -- (B);
    \draw[-latex] ([xshift=-15pt]B.east) -- (C);
    \draw[-latex] ([xshift=-15pt]C.east) -- (D);
    \end{tikzpicture}
    
    \caption{Hanany-Witten transition for a non-toric deformation of the local \hbox{$(1,-3)$-curve}.}
    \label{fig:HW_(1,-3)curve}
\end{figure}
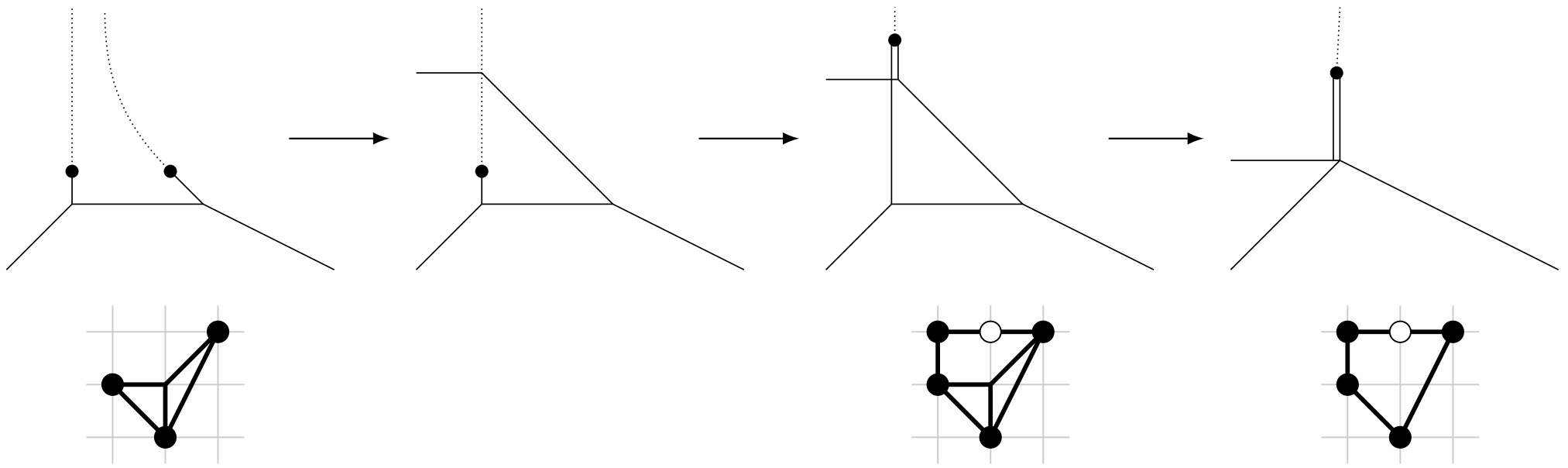

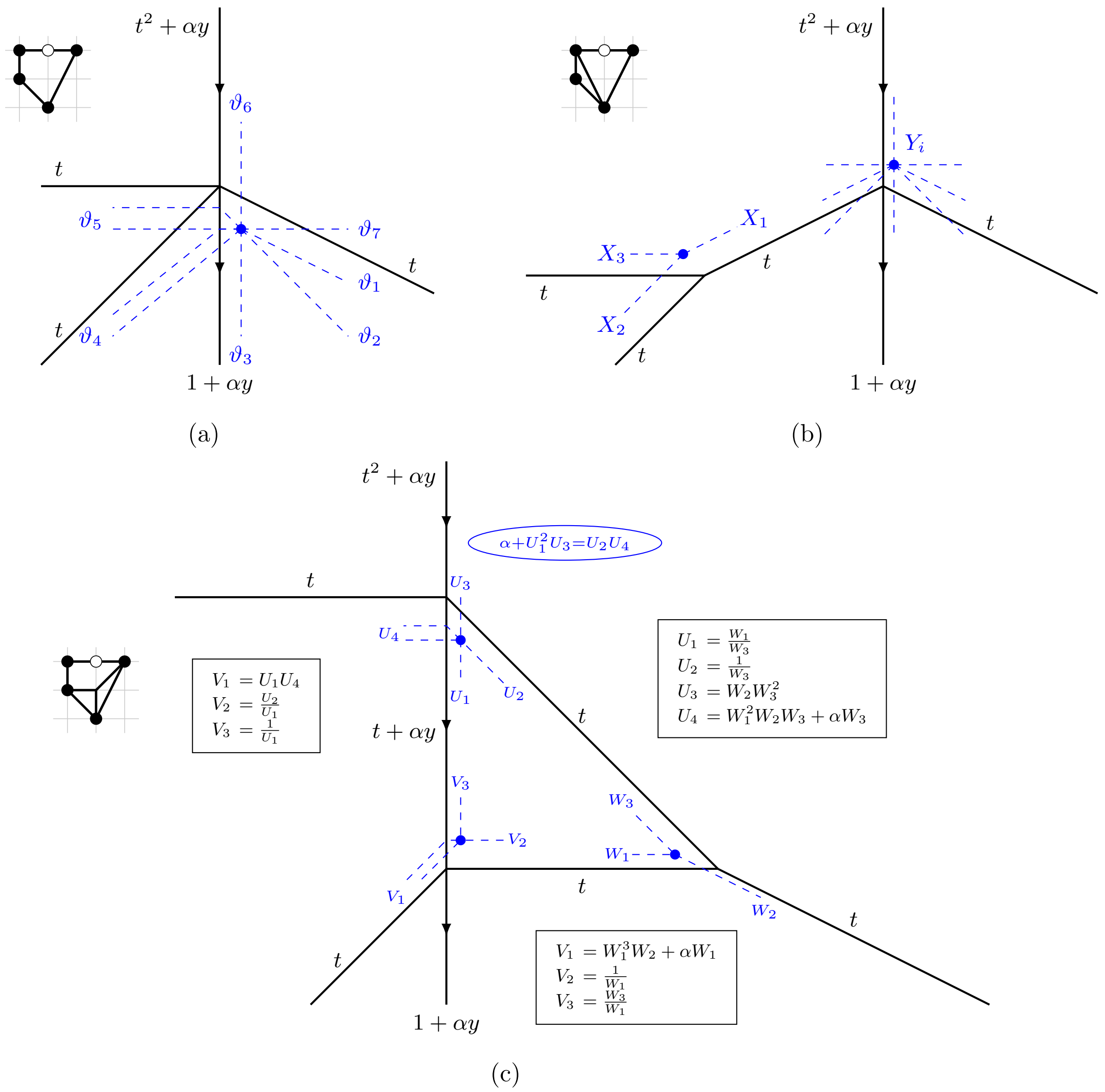
\begin{figure}[t]
    \centering

    \begin{subfigure}[t]{0.495\textwidth}
        \centering
        \scalebox{1}{
        \begin{tikzpicture}
            \tikzmath{\x = 5;\y =.3;\z = 1.5;} 

            \draw[wall] (0,.5*\x) -- node[pos=.1,left] {$t^2 + \alpha y$} (0,0);
            \draw[wall] (0,0) -- node[pos=1,below] {$1 + \alpha  y$} (0,-0.5*\x) ;
            \draw[kink] (-.5*\x,-.5*\x) -- node[pos=.1,above] {$t$}(0,0)  -- node[pos=.9,above] {$t$} (.6*\x, -.3*\x); 
            \draw[kink] (-.5*\x,0) -- node[pos=.1,above] {$t$} (0,0);

            \node[focalPoint,blue] at (\y,-2*\y) {};
            \draw[broken,blue] (\y,-2*\y) -- (\y,\z-2*\y) node[pos=1,above] {$\vartheta_6$};
            \draw[broken,blue] (\y,-2*\y) -- (\y+\z,-2*\y) node[pos=1,right] {$\vartheta_7$};
            \draw[broken,blue] (\y,-2*\y) -- (\y+\z,-2*\y-.5*\z) node[pos=1,right] {$\vartheta_1$};
                \draw[broken,blue] (\y,-2*\y) -- (\y+\z,-2*\y-\z) node[pos=1,right] {$\vartheta_2$};
            \draw[broken,blue] (\y,-2*\y) -- (\y,-\z-2*\y) node[pos=1,below] {$\vartheta_3$};
            \draw[broken,blue] (\y,-2*\y) -- (-\z,-\z-2*\y) node[pos=1,left] {$\vartheta_4$};
            \draw[broken,blue] (\y,-2*\y) -- (-\z,-2*\y);
            \draw[broken,blue] (\y,-2*\y) -- (0,-\y) -- (-\z,-\y) node[pos=1,left,yshift = -5pt] {$\vartheta_5$};
            \draw[broken,blue] (\y,-2*\y) -- (0,-2*\y) -- (-\z,-\z-\y);

            \node at (-2.5,1.5)
                {\raisebox{-.5\height}{\
                 
                  \scalebox{.8}{  
                \begin{tikzpicture}[x=.5cm,y=.5cm] 
                \draw[gridline] (-1,-.5)--(-1,2.5); 
                \draw[gridline] (0,-.5)--(0,2.5); 
                \draw[gridline] (1,-.5)--(1,2.5); 
                \draw[gridline] (-1.5,0)--(1.5,0); 
                \draw[gridline] (-1.5,1)--(1.5,1); 
                \draw[gridline] (-1.5,2)--(1.5,2); 
                \draw[ligne] (-1,2)--(1,2); 
                \draw[ligne] (-1,2)--(-1,1); 
                \draw[ligne] (-1,1)--(0,0); 
                \draw[ligne] (1,2)--(0,0); 
                \node[bd] at (-1,2) {}; 
                \node[bd] at (1,2) {}; 
                \node[bd] at (-1,1) {}; 
                \node[bd] at (0,0) {}; 
                \node[wd] at (0,2) {}; 
                \end{tikzpicture}}}} ;
        
        \end{tikzpicture}}
         \caption{}
        \label{fig:curveScatDiagAsymp}

    \end{subfigure}
    \begin{subfigure}[t]{0.495\textwidth}
        \centering
        \scalebox{1}{
        \begin{tikzpicture}
            \tikzmath{\x = 5;\y =.3;\z = 1;} 

            \draw[wall] (0,.5*\x) -- node[pos=.1,left] {$t^2 + \alpha y$} (0,0);
            \draw[wall] (0,0) -- node[pos=1,below] {$1 + \alpha  y$} (0,-0.5*\x) ;
            \draw[kink] (-.5*\x,-.25*\x) -- node[pos=.35,below] {$t$}(0,0)  -- node[pos=.5,above] {$t$} (.6*\x, -.3*\x); 
            \draw[kink] (-\x,-.25*\x) -- node[pos=.1,below] {$t$} (-.5*\x,-.25*\x)-- node[pos=.7,below] {$t$} (-.75*\x,-.5*\x);
        
            \node[focalPoint,blue] at (-.5*\x-\y,-.25*\x+\y) {};
            \draw[broken,blue]  (-.5*\x-\y,-.25*\x+\y) -- node[pos=1,fill=white,inner sep=1pt] {$X_3$} (-.5*\x-\y-\z,-.25*\x+\y) ;
            \draw[broken,blue]  (-.5*\x-\y,-.25*\x+\y) -- node[pos=1,fill=white,inner sep=1pt] {$X_2$} (-.5*\x-\y-\z,-.25*\x+\y-\z) ;
            \draw[broken,blue]  (-.5*\x-\y,-.25*\x+\y) -- node[pos=1,fill=white,inner sep=1pt] {$X_1$} (-.5*\x-\y+\z,-.25*\x+\y+.5*\z) ;
        
            \node[focalPoint,blue] at (.5*\y,\y) {};
            \node[blue] at (.5*\y+.3*\z,\y+.3*\z) {$Y_i$};
            \draw[broken,blue]  (.5*\y,\y) --  (.5*\y,\y) ;
            \draw[broken,blue]  (.5*\y,\y) --  (.5*\y,\y+\z) ;
            \draw[broken,blue]  (.5*\y,\y) --  (.5*\y,\y-\z) ;
            \draw[broken,blue]  (.5*\y,\y) --  (.5*\y+\z,\y) ;
            \draw[broken,blue]  (.5*\y,\y) --  (.5*\y-\z,\y) ;
            \draw[broken,blue]  (.5*\y,\y) --  (.5*\y+\z,\y-.5*\z) ;
            \draw[broken,blue]  (.5*\y,\y) --  (.5*\y+\z,\y-\z) ;
            \draw[broken,blue]  (.5*\y,\y) --  (.5*\y-\z,\y-\z) ;
            \draw[broken,blue]  (.5*\y,\y) --  (.5*\y-\z,\y-.5*\z) ;

            \node at (-4,1.5)
                {\raisebox{-.5\height}{\
                 
                  \scalebox{.8}{  
                \begin{tikzpicture}[x=.5cm,y=.5cm] 
                \draw[gridline] (-1,-.5)--(-1,2.5); 
                \draw[gridline] (0,-.5)--(0,2.5); 
                \draw[gridline] (1,-.5)--(1,2.5); 
                \draw[gridline] (-1.5,0)--(1.5,0); 
                \draw[gridline] (-1.5,1)--(1.5,1); 
                \draw[gridline] (-1.5,2)--(1.5,2); 
                \draw[ligne] (-1,2)--(1,2); 
                \draw[ligne] (-1,2)--(-1,1); 
                \draw[ligne] (-1,1)--(0,0); 
                \draw[ligne] (1,2)--(0,0); 
                \draw[ligne] (-1,2)--(0,0); 
                \node[bd] at (-1,2) {}; 
                \node[bd] at (1,2) {}; 
                \node[bd] at (-1,1) {}; 
                \node[bd] at (0,0) {}; 
                \node[wd] at (0,2) {}; 
                \end{tikzpicture}}}} ;
                    
        \end{tikzpicture}}
         \caption{}
        \label{fig:curveScatDiagFlopped}
        
    \end{subfigure}

    \begin{subfigure}[t]{0.95\textwidth}
        \centering
        \scalebox{1}{
        \begin{tikzpicture}
            \tikzmath{\x = 3.8;\y =.2;\z = .6;} 
            
            \draw[wall] (0,1.5*\x) -- node[pos=.1,left] {$t^2 + \alpha y$} (0,\x);
            \draw[wall] (0,\x) -- node[pos=.5,left] {$t + \alpha y$} (0,0);
            \draw[wall] (0,0) -- node[pos=1,below] {$1 + \alpha  y$} (0,-.5*\x) ;
            \draw[kink] (-.5*\x,-.5*\x) -- node[pos=.2,above] {$t$}(0,0) -- node[pos=.5,below] {$t$} (\x,0) -- node[pos=.5,above] {$t$} (2*\x, -.5*\x); 
            \draw[kink] (-\x,\x) -- node[pos=.5,above] {$t$} (0,\x) -- node[pos=.5,above] {$t$}(\x,0);

            \node[focalPoint,blue] at (\y,\x - 3*\y) {};
            \draw[broken,blue] (\y,\x - 3*\y + \z) -- (\y,\x - 3*\y) -- (-\z,\x - 3*\y);
            \draw[broken,blue] (\y+\z,\x - 3*\y-\z) -- (\y,\x - 3*\y) -- (\y,\x - 3*\y-\z);
            \draw[broken,blue] (\y,\x - 3*\y) -- (0,\x - 2*\y) -- (-\z,\x - 2*\y);
            \node[blue] at (\y,\x - 3*\y + \z+.2) {$\scriptstyle U_3$};
            \node[blue] at (-\z-.2,\x - 2.6*\y) {$\scriptstyle U_4$};
            \node[blue] at (\y,\x - 3*\y-\z-.2) {$\scriptstyle U_1$};
            \node[blue] at (\y+\z+.15,\x - 3*\y-\z-.15) {$\scriptstyle U_2$};
        
            \node[ellipse,draw,blue,inner sep =1pt] at (.2*\x+.9,1.2*\x) {$\scriptstyle\alpha +  U_1^2U_3 = U_2U_4$};
            
            \node[focalPoint,blue] at (\y,2*\y) {};
            \draw[broken,blue] (\y,2*\y+\z) -- (\y,2*\y)-- (0,2*\y) -- (-\z,2*\y-\z) ;
            \draw[broken,blue] (\z+\y,2*\y)-- (\y,2*\y)--  (\y-\z,2*\y-\z);

            \node[blue]  at  (.5*\y-\z-.2,2*\y-\z -.2) {$\scriptstyle V_1$};
            \node[blue]  at  (\y,2*\y+\z+.2) {$\scriptstyle V_3$};
            \node[blue]  at  (\y+\z +.2,2*\y) {$\scriptstyle V_2$};
        
            \node[focalPoint,blue] at (\x-3*\y, \y) {};
            \draw[broken,blue] (\x-3*\y-\z, \y) -- (\x-3*\y, \y) -- (\x-3*\y-\z, \y+\z);
            \draw[broken,blue] (\x-3*\y, \y) -- (\x-3*\y+2*\z, \y-\z);
            \node[blue] at (\x-3*\y-\z-.2, \y) {$\scriptstyle W_1$};
            \node[blue] at (\x-3*\y-\z-.15, \y+\z+.15) {$\scriptstyle W_3$};
            \node[blue] at (\x-3*\y+2*\z+.05, \y-\z-.2) {$\scriptstyle W_2$};
        
        
            \node[draw] at (-.7*\x,.6*\x) {\scalebox{.8}{
                $\begin{array}{rl}
                    V_1 &=U_1 U_4 \\
                    V_2 &= \frac{U_2}{U_1}\\
                    V_3 &= \frac{1}{U_1}\\
                \end{array}$
                }};
            \node[draw] at (1.2*\x,.7*\x) {\scalebox{.8}{
                $\begin{array}{rl}
                    U_1 &= \frac{W_1}{W_3}\\
                    U_2 &=\frac{1}{W_3}\\
                    U_3 &= W_2 W_3^2 \\ 
                    U_4 &= W_1^2 W_2W_3 + \alpha W_3 
                \end{array}$
                }};
            \node[draw] at (.7*\x,-.4*\x) {\scalebox{.8}{
                $\begin{array}{rl}
                    V_1 &= W_1^3 W_2 +\alpha W_1 \\
                    V_2 &= \frac{1}{W_1} \\
                    V_3 &= \frac{W_3}{W_1} \\
                \end{array}$
                }};
        
            \node at (-5,2.5)
                {\raisebox{-.5\height}{\
                 
                  \scalebox{.8}{  
                \begin{tikzpicture}[x=.5cm,y=.5cm] 
                \draw[gridline] (-1,-.5)--(-1,2.5); 
                \draw[gridline] (0,-.5)--(0,2.5); 
                \draw[gridline] (1,-.5)--(1,2.5); 
                \draw[gridline] (-1.5,0)--(1.5,0); 
                \draw[gridline] (-1.5,1)--(1.5,1); 
                \draw[gridline] (-1.5,2)--(1.5,2); 
                \draw[ligne] (-1,2)--(1,2); 
                \draw[ligne] (-1,2)--(-1,1); 
                \draw[ligne] (-1,1)--(0,0); 
                \draw[ligne] (1,2)--(0,0) -- (0,1); 
                \draw[ligne] (-1,1)--(0,1) -- (1,2); 
                \node[bd] at (-1,2) {}; 
                \node[bd] at (1,2) {}; 
                \node[bd] at (-1,1) {}; 
                \node[bd] at (0,0) {}; 
                \node[wd] at (0,2) {}; 
                \end{tikzpicture}}}} ;
                    
        \end{tikzpicture}}
         \caption{}
        \label{fig:curveScatDiag}

    \end{subfigure}
        
    \caption{Scattering diagrams for three tessellations of the GTP associated with the local $(1,-3)$-curve. They give different presentations of the same threefold.}
    \label{fig:(1,-3)curve_diags}
\end{figure}

This GTP admits three tessellations, the trivial tessellation building the affine threefold~$\mathcal X_\alpha$ (Figure \ref{fig:curveScatDiagAsymp}), one involving a T-cone (Figure \ref{fig:curveScatDiagFlopped}) and the last one  involving a trapezium (a smoothed SPP, Figure \ref{fig:curveScatDiag}). At $\alpha\neq 0$, these are maximal due to the s-rule. At the toric limit $\alpha = 0$, the threefolds associated with Figures \ref{fig:curveScatDiag} and \ref{fig:curveScatDiagFlopped} are singular partial resolutions of $\mathcal X_0$. We now turn to the more interesting $\alpha \neq 0$ regime. This is realized by the presence of a 7-brane at finite distance, whose position transverse to the monodromy cut is given by mass deformation parameters in the extended Coulomb branch. The three tessellations are related in the scattering diagram by the distance between the intersection with the wall and the triple kink junctions, which flips sign between (b) and (c). The question becomes, can this distance be understood as the volume of a $\mathbb{P}^1$ curve?  On the same note, looking at Figure \ref{fig:curveScatDiag} we see that there is an open face, which could suggest the presence of a compact divisor, according to the toric expectation. We will tackle these two questions by using the explicit charts and transition functions to look for this compact holomorphic curve and divisor.

Let us start by studying $\mathcal X_\alpha$ (Figure \ref{fig:curveScatDiagAsymp}). While it is not a complete intersection, it nonetheless admits a nice geometric description: of the $7$ theta-functions shown in Figure \ref{fig:curveScatDiagAsymp}, $\vartheta_6$ can be eliminated using $\vartheta_5 \vartheta_7 = t^2 + \alpha\vartheta_6$ while the others can be arranged in a matrix:
\begin{equation}\label{eq:rank1_presentation}
    M = \begin{pmatrix}
        \vartheta_4 & \vartheta_3  & \vartheta_2 \\ \vartheta_3 + \alpha & \vartheta_2 & \vartheta_1 \\ \vartheta_5 & t & \vartheta_7
    \end{pmatrix}
\end{equation}
in such a way that the relations between them are precisely the vanishing of all $2\times 2$ minors of $M$. Therefore, $\mathcal X_\alpha$ is defined by the condition $\operatorname{rank}(M)\leq 1$. In fact, as $\vartheta_3$ and $\vartheta_3 +\alpha$ are both coefficients of $M$, the matrix is never zero and  $\operatorname{rank}(M)=1$.

Equation \eqref{eq:rank1_presentation} also gives a construction of $\mathcal X_\alpha$ as a GIT quotient of a complete intersection in $\mathbb C^6$ by $\mathbb C^*$, namely $yv-zu = yu - xv + \alpha =0$, where $x,y,z$ have weight $+1$ and $u,v,w$ have weight $-1$.

The open charts corresponding to the $5$ external legs are the loci where the theta-function in the direction of the leg is non-vanishing. 

The threefolds in Figures \ref{fig:curveScatDiagFlopped} and \ref{fig:curveScatDiag}  are obtained by gluing affine charts $(\mathcal U_X,\mathcal U_Y)$ and $(\mathcal U_U,\mathcal U_V,\mathcal U_W)$ respectively.  We will show that the natural maps $\mathcal U_{U,\ldots,Y} \to \mathcal X$ are inclusions of open subsets, grouped to form two open coverings. Thus, all three diagrams yield the same threefold.

Let us first turn to Figure \ref{fig:curveScatDiag}. There, the three affine charts correspond to the subset of $\mathcal X$ where a row of $M$ does not vanish (equivalently, where one of the homogeneous coordinates $u,v,w$ is non-zero). One can then set the non-vanishing coordinate to $1$, thus fixing the $\mathbb C^*$ action. The remaining coordinates, after relabeling, recover the theta-functions shown in  Figure \ref{fig:curveScatDiag}. By computing theta-functions explicitly, we find: 

\begin{align*}
    M &= \begin{pmatrix}
        \vartheta_4 & \vartheta_3  & \vartheta_2 \\ \vartheta_3 + \alpha & \vartheta_2 & \vartheta_1 \\ \vartheta_5 & t & \vartheta_7
    \end{pmatrix} = \begin{pmatrix}
        u\\v\\w
    \end{pmatrix}\begin{pmatrix}
        x\\y\\z
    \end{pmatrix}^T  \\
    &= \begin{pmatrix}
        U_1 U_4 & U_1^2 U_3 & U_1 U_2 U_3 \\
        U_2 U_4 & U_1 U_2 U_3 & U_2^2 U_3 \\
        U_4 & U_1 U_3 & U_2 U_3
    \end{pmatrix} = \begin{pmatrix}
        U_1 \\ U_2 \\1
    \end{pmatrix}\begin{pmatrix}
        U_4 \\ U_1 U_3 \\ U_2 U_3
    \end{pmatrix}^T\tag{$\mathcal U_U$}\\
    &= \begin{pmatrix}
        V_1 & V_1 V_2-\alpha & V_1 V_2^2-\alpha V_2 \\
        V_1 V_2 & V_1 V_2^2-\alpha V_2 & V_1 V_2^3-\alpha V_2^2 \\
        V_1 V_3 & V_1 V_2 V_3-\alpha V_3 & V_1 V_2^2 V_3-\alpha V_2 V_3
    \end{pmatrix}= \begin{pmatrix}
        1 \\V_2\\V_3
    \end{pmatrix}\begin{pmatrix}
        V_1 \\ V_1V_2 - \alpha \\ V_2(V_1 V_2 - \alpha)
    \end{pmatrix}^T \tag{$\mathcal U_V$}\\
    &= \begin{pmatrix}
        W_2 W_1^3+\alpha W_1 & W_1^2 W_2 & W_1 W_2 \\
        W_1^2W_2 +\alpha  & W_1 W_2 & W_2 \\
        W_2 W_3 W_1^2+\alpha W_3 & W_1 W_2 W_3 & W_2 W_3
    \end{pmatrix} = \begin{pmatrix}
        W_1 \\ 1 \\ W_3 
    \end{pmatrix} \begin{pmatrix}
        W_1 ^2 W_2 +\alpha \\ W_1 W_2 \\ W_2
    \end{pmatrix}^T \tag{$\mathcal U_W$}
\end{align*}

As $xv-yu = \alpha$, we see that $u\neq 0$ or $v\neq 0$ everywhere, i.e. $\mathcal X_\alpha = \mathcal U_V\cup \mathcal U_W$ . The fact that $\mathcal U_U$ is included in the union of the two other charts can be interpreted using the Hanany-Witten transition shown in Figure \ref{fig:HW_(1,-3)curve}: the chart $\mathcal U_U$ corresponds to the brane junction created in the process of moving the $7$-branes to infinity. 

This means that there is an isomorphism between the SCFT affine geometry $\mathcal X_\alpha$ and the geometry given by Figure \ref{fig:curveScatDiag}. As a consequence, there are no $\mathbb{P}^1$ curves, so there is nothing to flop. Similarly, the open face in this resolution does not correspond to any compact divisor.

We can also open the scattering diagram as in Figure \ref{fig:curveScatDiagFlopped}. There are two affine charts $\mathcal U_X\simeq \mathbb C^3$ and $\mathcal U_Y$ the T-cone geometry with parameters $(p,q) = (2,1)$ shown in Figure \ref{fig:Tcone_(2,1)} (see also Appendix \ref{sec:Tcones}). The two charts are glued together to give back the affine $\mathcal X_\alpha$. Computing the theta-functions explicitly, we see that the natural maps $\mathcal U_{X,Y}\to \mathcal X_\alpha$ are given by:
\begin{align*}
     M &= \begin{pmatrix}X_2 \\ X_1 X_2^2 +\alpha \\ X_3\end{pmatrix}\begin{pmatrix}1 \\ X_1 X_2 \\ X_1 (X_1 X_2^2 +\alpha)\end{pmatrix}^T \tag{$\mathcal U_X$}\\
     &= \begin{pmatrix}
         Y_4  & Y_3 & Y_2 \\
         Y_3 + \alpha & Y_2 & Y_1 \\
         Y_6 & t & Y_8
     \end{pmatrix}\tag{$\mathcal U_Y$}
\end{align*}

Therefore, $\mathcal U_X$ is the open subspace where the first column of $M$ is non-vanishing, while $\mathcal U_Y$ is the subspace where at least one of the second or third columns of $M$ is non-vanishing.
The first point can be derived as before: where $x\neq 0$, we can fix the $\mathbb C^*$ action by setting $x=1$. To check the latter assertion, notice that, in the GIT quotient coordinates, the locus where at least one of the second or third columns of $M$ is non-vanishing is $\{y\neq 0\}\cup \{z\neq 0\}$. Let us show that on this subset we can fix the $\mathbb C^*$ action down to a $\mathbb Z_2$ subgroup by imposing the conditions $y=u$ and $z=v$.

Where $z\neq 0$, we have $\alpha z = v ( xz - y^2 )$
so $v\neq 0$. As $z$ and $v$ are non-zero, we can use the $\mathbb C^*$ action to set $z=v$. Then $yv=zu$ implies $y=u$. Likewise, where $y\neq 0$, $\alpha y = u (xz-y^2)$ so $u\neq 0$. Use the $\mathbb C^*$ action to set $y=u$, and $z=v$ follows.

We can then set $\tilde u=v=z,\tilde v=x,\tilde z=y=u,\tilde w=w$. They satisfy the relation $\tilde u\tilde v=\tilde z^2 +\alpha$ and are odd under the remaining $\mathbb Z_2$ subgroup: this is indeed the geometric description of the T-cone with $(p,q)= (2,1)$. In these variables, $M$ can be obtained from the matrix in Equation \eqref{eq:Tcone_rank_relation} by removing one row and one column.

Once again we find that the geometry of the opened web in the alternative tessellation was affine all along. There is no $\mathbb{P}^1$ curve to flop. All three tessellations describe the same affine geometry. Its corresponding 5d theory has no Coulomb branch\footnote{In fact, using the method of \cite{Bourget:2026hsg}, it can be shown that the topological string partition function on $\mathcal X$ is $1$, indicating that it engineers the trivial theory.}.

\subsection{Symington polygon} \label{sec::symington}

In order to understand the subtleties involved with non-toric deformations it is useful to take advantage of the fact that a 7-brane can be freely moved along the direction parallel to its monodromy cut and move it to a compact interior/non-compact exterior face in such way that no 5-branes are attached to it. The 7-brane still affects the web by its monodromy cut, that causes the 5-branes to bend accordingly. This can be effectively done in the scattering diagram by erasing the ingoing-outgoing walls corresponding to the 7-brane trajectory. In such way, spurious open faces and junctions are removed: they were only there due to Hanany-Witten brane creation.\\
The dual operation on the GTP polytope is called the Symington polygon (see \cite{Symington:2002,LeungSymington:2010} and Section 5 of \cite{Alexeev:2024bko} for more details). Symington polytopes are constructed from polyhedral decompositions of standard integer polytopes by cutting some faces and identifying their interior edges. This introduces an integral affine structure with singularities, which are exactly the singularities introduced by the 7-branes. Similarly to how the toric diagram can be thought of as being the base of the torus fibration, the Symington polygon extends this to the case where singularities induced by the 7-branes are present. The cuts correspond to the intersection of a (bunch of) 5-branes with the monodromy of a 7-brane, which applies an $\text{SL}(2,\mathbb Z)$ transformation and causes it to break. 

To build the Symington polytope from a GTP, start with a tessellation of the GTP and identify a T-cone attached to the 7-brane we want to move. This T-cone may be primitive or non-primitive \footnote{A non-primitive T-cone is a block of tessellation pieces whose lattice base is equal to its height and is anchored to an apex either in the interior of the GTP or on its boundary. The pieces are either  primitive T-cones (including the trivial ones) or trapezoidal pieces built out of them. For example, the bottom tessellation of Figure \ref{fig:(1,-3)curve_diags} includes a non-primitive T-cone composed of an SPP trapezium and a trivial T-cone.}. If we want to move the 7-brane to an exterior face, the apex of the T-cone will be in the boundary, while if we want to move the 7-brane to an interior face, its apex will be in the interior. We cut the T-cone and identify the non-base edges and vertices. Note that this is very similar to the 5d tangram algorithm proposed in \cite{CarrenoBolla:2024fxy}, which is used to find full supersymmetric tessellations of GTPs.

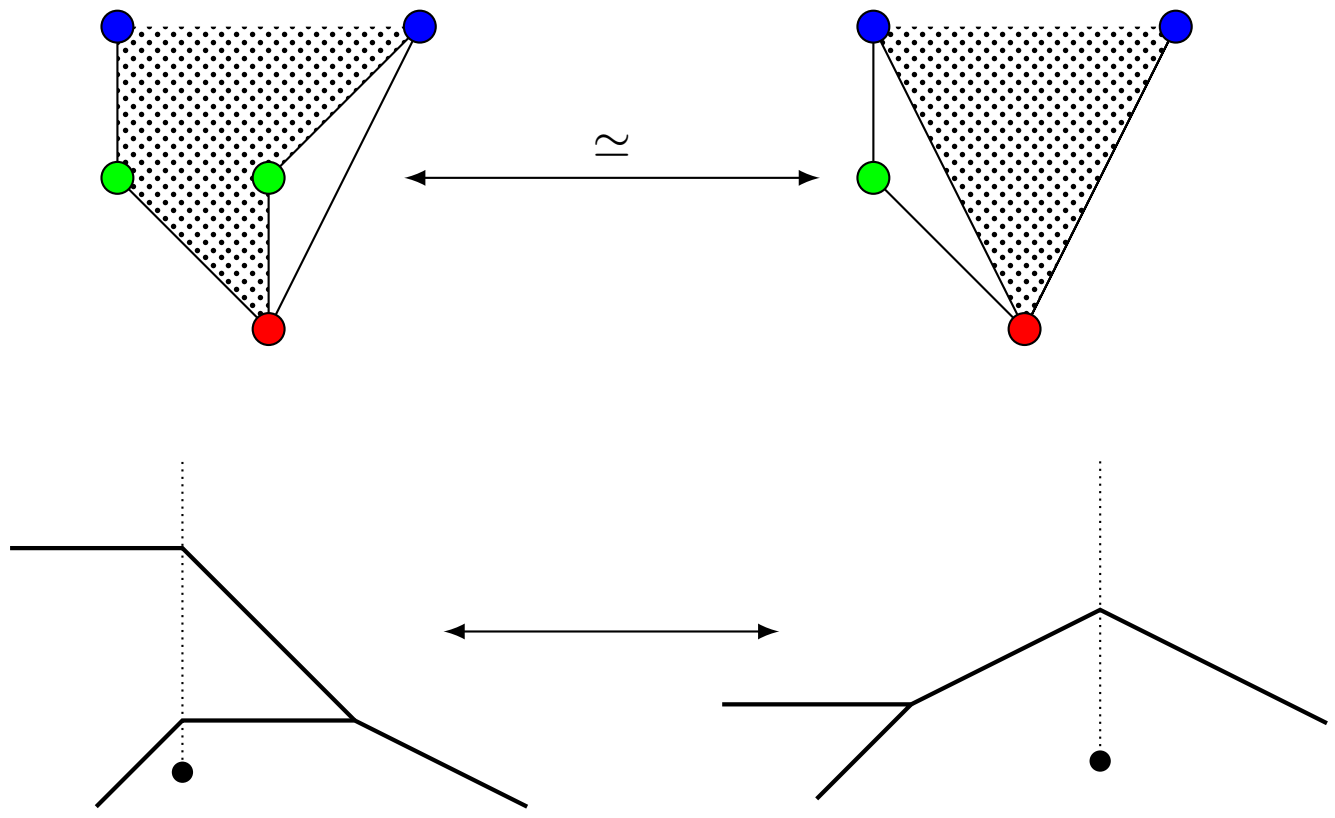
\begin{figure}[h]
    \centering
    
    \begin{tikzpicture}
    \node (A) at (0,0) {
        \begin{tikzpicture}
            \draw (-1,2) -- (-1,1) -- (0,0) -- (1,2) -- (0,1) -- (0,0);
            \fill[pattern=crosshatch dots ] (-1,2) -- (-1,1) -- (0,0) -- (0,1) -- (1,2);
            \filldraw[fill=red] (0,0) circle (3pt);
            \filldraw[fill=blue] (1,2) circle (3pt);
            \filldraw[fill=blue] (-1,2) circle (3pt);
            \filldraw[fill=green] (0,1) circle (3pt);
            \filldraw[fill=green] (-1,1) circle (3pt);
        \end{tikzpicture}};
    \node (B) at (5,0) {
        \begin{tikzpicture}
            \draw (0,0) -- (-1,2) -- (-1,1) -- (0,0) -- (1,2)  -- (0,0);
            \fill[pattern=crosshatch dots ] (-1,2) -- (0,0) -- (1,2);
            \filldraw[fill=red] (0,0) circle (3pt);
            \filldraw[fill=blue] (1,2) circle (3pt);
            \filldraw[fill=blue] (-1,2) circle (3pt);
            \filldraw[fill=green] (-1,1) circle (3pt);
        \end{tikzpicture}} ;

    \node (b) at (5,-3) {
        \begin{tikzpicture}[scale=0.5]
            \tikzmath{\x = 5;\y =.3;\z = 1;} 
            \draw[kink] (-.5*\x,-.25*\x) -- node[pos=.35,below] {}(0,0)  -- node[pos=.5,above] {} (.6*\x, -.3*\x); 
            \draw[kink] (-\x,-.25*\x) -- node[pos=.1,below] {} (-.5*\x,-.25*\x)-- node[pos=.7,below] {} (-.75*\x,-.5*\x);  
            \draw[densely dotted] (0,-2) -- (0,2);
            \node[focalPoint] at (0,-2) {};
        \end{tikzpicture}};

    \node (a) at (0,-3){ 
        \begin{tikzpicture}[scale=0.3]
            \tikzmath{\x = 3.8;\y =.2;\z = .6;} 
            
            \draw[densely dotted] (0,1.5*\x) -- node[pos=.1,left] {} (0,\x);
            \draw[densely dotted] (0,\x) -- node[pos=.5,left] {} (0,0);
            \draw[densely dotted] (0,0) -- node[pos=1,below] {} (0,-.3*\x) ;
    
             \node[focalPoint] at (0,-.3*\x) {};
            \draw[kink] (-.5*\x,-.5*\x) -- node[pos=.2,above] {}(0,0) -- node[pos=.5,below] {} (\x,0) -- node[pos=.5,above] {} (2*\x, -.5*\x); 
            \draw[kink] (-\x,\x) -- node[pos=.5,above] {} (0,\x) -- node[pos=.5,above] {}(\x,0);       
        \end{tikzpicture}};

    \draw[latex - latex] ([xshift=-10pt]A.east) -- node[pos=.5,above] {$\simeq$} ([xshift=-3pt]B.west);
    \draw[latex - latex] ([xshift=-20pt]a.east) -- node[pos=.5,above] {} ([xshift=15pt]b.west);
    \end{tikzpicture}

    \caption{At the top, Symington polygons corresponding to each tessellation. At the bottom, dual brane webs with all 5-branes detached from the 7-branes. The two Symington polygons are equivalent, related by moving the branch cut. The two brane webs are related by moving the 7-branes left to right.}
    \label{fig:Symington}
\end{figure}

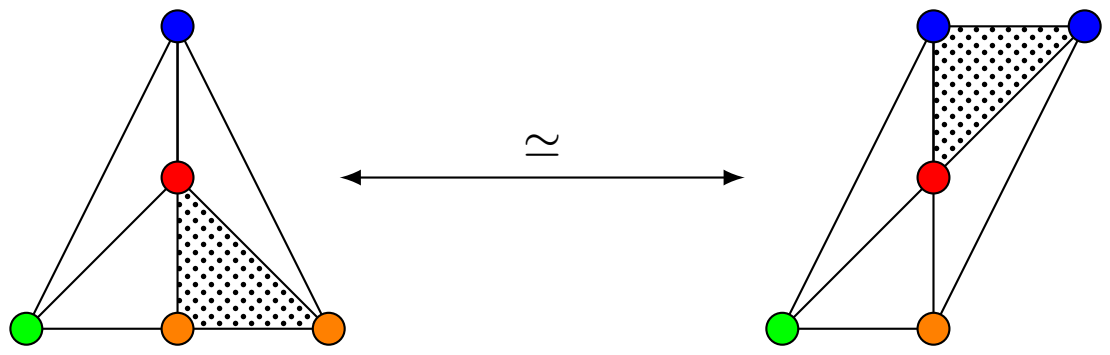
\begin{figure}[h]
    \centering
    \begin{tikzpicture}
        \node (A) at (0,0) {
            \begin{tikzpicture}
                \draw (-1,-1) -- (0,-1) -- (1,-1) -- (0,1) -- (-1,-1) ;
                \draw (-1,-1) -- (0,0) -- (0,1) ;
                \draw (1,-1) -- (0,0) -- (0,1) ;
                \draw (0,-1) -- (0,0);
                \fill[pattern=crosshatch dots ]  (1,-1)--(0,0)--(0,-1);
                
                \filldraw[fill=red] (0,0) circle (3pt);
                \filldraw[fill=blue] (0,1) circle (3pt);
                \filldraw[fill=green] (-1,-1) circle (3pt);
                \filldraw[fill=orange] (0,-1) circle (3pt);
                \filldraw[fill=orange] (1,-1) circle (3pt);
            \end{tikzpicture}};

        \node (B) at (5,0){
            \begin{tikzpicture}
                \draw (-1,-1) -- (0,-1) -- (1,1) -- (0,1) -- (-1,-1) ;
                \draw (-1,-1) -- (0,0) -- (0,1) ;
                \draw (1,1) -- (0,0) -- (0,1) ;
                \draw (0,-1) -- (0,0);
                \fill[pattern=crosshatch dots ]  (1,1)--(0,0)--(0,1);
                
                \filldraw[fill=red] (0,0) circle (3pt);
                \filldraw[fill=orange] (0,-1) circle (3pt);
                \filldraw[fill=green] (-1,-1) circle (3pt);
                \filldraw[fill=blue] (0,1) circle (3pt);
                \filldraw[fill=blue] (1,1) circle (3pt);
            \end{tikzpicture}};

        \draw[latex - latex] ([xshift=-5pt]A.east) -- node[pos=.5,above] {$\simeq$} ([xshift=0pt]B.west);
    \end{tikzpicture}
    \caption{Symington polygons associated to the HW move between $\mathbb{F}_0$ and $\mathbb{F}_2$. The 7-brane is moved to the compact face in each case. The two Symington polygons are equivalent, related by moving the branch cut.}
    \label{fig:SymingF2F0}
\end{figure}

We depict the Symington polygons associated to each tessellation of the local (1,-3) curve in Figure \ref{fig:Symington}. All interior faces are removed, which makes manifest the absence of compact divisors. As an additional example, we show the Symington polygon for the earlier introduced Hanany-Witten move $\mathbb{F}_0 \to \mathbb{F}_2$ GTP in Figure \ref{fig:SymingF2F0}.

\subsection{Divisors, curves and triple intersection}
\label{sec:divisors}

We have shown how the scattering diagram of an web away from the origin of its extended Coulomb branch gives an explicit description of the resolved geometry in terms of open affine patches and transition functions. This is a very explicit description from which geometric data can be extracted. In this section, we will study in a concrete example divisors, which are dual to ECB moduli. Intersecting these divisors yields curves, on which M2 branes can wrap resulting in 5d BPS states. Finally, from the triple intersection of the divisors in the 3-fold the prepotential of the 5d theory can be read off. This extends the standard toric methods to this class of non-toric geometries. We will study this with the GTP realizing the local  $\mathbb F_3$  geometry.

The scattering diagram and transition functions for the GTP geometry $\mathcal X_3$ associated with\footnote{The relation between $\mathcal X_3$ and local $\mathbb F_3$ is similar to the one between the local $(1,-3)$-curve and its non-toric deformation : the toric polygon for local $\mathbb F_3$ is non-toric; when sending 7-branes to infinity to make it convex, Hanany-Witten transitions occur that turn it into a GTP. See Section \ref{sec::1-3curve} and Figure \ref{fig:HW_(1,-3)curve}.} local $\mathbb F_3$ is given in Figure \ref{fig:GTP_F3_diag}. See \cite{Bourget:2026hsg,Kho:2026zwc} for papers studying the Hanany-Witten transition between local $\mathbb F_1$ and $\mathcal X_3$ using other methods.  $\mathcal X_3$ has a single compact divisor $\mathcal D_0$, as well as five non-compact distinguished divisors $\mathcal D_i \,(i=1,2,3)$ and $\mathcal E,\mathcal E'$ (for "exuberant" \cite{ABBGS}), whose local equations we list, for each affine patch, in Table \ref{tab:divisorX3}. We remark that there is no compact divisor associated to the face given by $(\mathcal{U}_U, \mathcal U_V, \mathcal{U}_W)$ as expected from the results of the previous section.

The non-compact divisors intersect $\mathcal D_0$ along compact curves $\mathcal C_i$, $\mathcal C_e,\mathcal C_e'$ all isomorphic to $\mathbb P^1$. We can explicitly compute  their intersection numbers and normal bundles in $\mathcal X$ (and in $\mathcal D_0$).

Let us briefly comment about $\mathcal C_1$. Focusing on the subdiagram corresponding to the local $(-3,1)$-curve, this would have been a candidate for a compact curve. However, it is missing one pole, which is recovered by adding the patch $\mathcal{U}_X$. Thus, the $\mathbb{P}^1$ skips through the wall.

\begin{figure}
\scalebox{.7}{
\begin{tikzpicture}
    \tikzmath{\x = 3.3;\y =.2;\z = .6;} 
    
    \draw[wall] (0,1.5*\x) -- node[pos=.1,left] {$t^2 + \alpha y$} (0,\x);
    \draw[wall] (0,\x) -- node[pos=.5,left] {$t + \alpha y$} (0,0);
    \draw[wall] (0,0) -- node[pos=.5,right] {$1 + \alpha  y$} (0,-\x) ;
    \draw[wall] (0,-\x) -- node[pos=.9,right] {$1 + \alpha t y$} (0,-1.5*\x);
    \draw[kink] (-2*\x,-1.5*\x)-- node[pos=.5,above] {$t$} (-\x,-\x) -- node[pos=.5,above] {$t$}(0,0) -- node[pos=.5,below] {$t$} (\x,0) -- node[pos=.5,above] {$t$} (3*\x, -\x)--node[pos=.5,above] {$t$} (4.5*\x,-1.5*\x);  
    \draw[kink] (-\x,\x) -- node[pos=.5,above] {$t$} (0,\x) -- node[pos=.5,above] {$t$}(\x,0);
    \draw[kink] (-\x,-\x) --node[pos=.5,below] {$t$}  (0,-\x) -- node[pos=.5,above] {$t$} (3*\x,-\x);

    \node[focalPoint,blue] at (\y,\x - 3*\y) {};
    \draw[broken,blue] (\y,\x - 3*\y + \z) -- (\y,\x - 3*\y) -- (-\z,\x - 3*\y);
    \draw[broken,blue] (\y+\z,\x - 3*\y-\z) -- (\y,\x - 3*\y) -- (\y,\x - 3*\y-\z);
    \draw[broken,blue] (\y,\x - 3*\y) -- (0,\x - 2*\y) -- (-\z,\x - 2*\y);
    \node[blue] at (\y,\x - 3*\y + \z+.2) {$\scriptstyle U_3$};
    \node[blue] at (-\z-.2,\x - 2.6*\y) {$\scriptstyle U_4$};
    \node[blue] at (\y,\x - 3*\y-\z-.2) {$\scriptstyle U_1$};
    \node[blue] at (\y+\z+.15,\x - 3*\y-\z-.15) {$\scriptstyle U_2$};

    \node[ellipse,draw,blue,inner sep =1pt] at (.2*\x+.9,1.2*\x) {$\scriptstyle\alpha +  U_1^2U_3 = U_2U_4$};
    
    \node[focalPoint,blue] at (\y,2*\y) {};
    \draw[broken,blue] (\y,2*\y+\z) -- (\y,2*\y)-- (0,2*\y) -- (-\z,2*\y-\z) ;
    \draw[broken,blue] (\z+\y,2*\y)-- (\y,2*\y)--  (\y-\z,2*\y-\z);
    \draw[broken,blue]  (\y,2*\y)--  (\y,2*\y-\z);
    \node[blue]  at (\y,2*\y-\z-.2) {$\scriptstyle V_4$};
    \node[blue]  at  (.5*\y-\z-.2,2*\y-\z -.2) {$\scriptstyle V_1$};
    \node[blue]  at  (\y,2*\y+\z+.2) {$\scriptstyle V_3$};
    \node[blue]  at  (\y+\z +.2,2*\y) {$\scriptstyle V_2$};

    \node[ellipse,draw,blue,inner sep =2pt] at (-.2*\x-.9,.3) {$\scriptstyle V_1 V_2 = V_4 + \alpha$};

    \node[focalPoint,blue] at (\x-3*\y, \y) {};
    \draw[broken,blue] (\x-3*\y-\z, \y) -- (\x-3*\y, \y) -- (\x-3*\y-\z, \y+\z);
    \draw[broken,blue] (\x-3*\y, \y) -- (\x-3*\y+2*\z, \y-\z);
    \node[blue] at (\x-3*\y-\z-.2, \y) {$\scriptstyle W_1$};
    \node[blue] at (\x-3*\y-\z-.15, \y+\z+.15) {$\scriptstyle W_3$};
    \node[blue] at (\x-3*\y+2*\z+.05, \y-\z-.2) {$\scriptstyle W_2$};


    \node[focalPoint,blue] at (-\x+3*\y, -\x+\y) {};
    \draw[broken,blue](-\x+3*\y-2*\z, -\x+\y-\z)  -- (-\x+3*\y, -\x+\y)  -- (-\x+3*\y+\z, -\x+\y+\z);
    \draw[broken,blue] (-\x+3*\y, -\x+\y)  -- (-\x+3*\y+\z, -\x+\y);
    \node[blue] at (-\x+3*\y-2*\z-.05, -\x+\y-\z-.2) {$\scriptstyle X_3$};
    \node[blue] at (-\x+3*\y+\z+.15, -\x+\y+\z+.15) {$\scriptstyle X_2$};
    \node[blue] at (-\x+3*\y+\z+.2, -\x+\y) {$\scriptstyle X_1$};


    \node[focalPoint,blue] at (\y, -\x+\y) {};
    \draw[broken,blue] (\y, -\x+\y+\z) -- (\y, -\x+\y) -- (\y, -\x+\y-\z);
    \draw[broken,blue] (-\z, -\x+\y) -- (\y, -\x+\y) -- (\y+\z, -\x+\y);
    \draw[broken,blue] (\y, -\x+\y) -- (0, -\x+2*\y) -- (-\z,-\x+2*\y);
    \node[blue] at (\y, -\x+\y+\z+.2) {$\scriptstyle Y_3$};
    \node[blue] at (\y, -\x+\y-\z-.2) {$\scriptstyle  Y_4$};
    \node[blue] at (\y+\z+.2, -\x+\y){$\scriptstyle Y_1$};
    \node[blue] at (-\z-.2, -\x+1.5*\y) {$\scriptstyle Y_2 $};

    \node[ellipse,draw,blue,inner sep =2pt] at (.2*\x+1.1,-\x +.6) {$\scriptstyle Y_1 Y_2 = 1+ \alpha Y_3$};

    \node[focalPoint,blue] (Zcenter) at (3*\x -5*\y, -\x+\y) {};
    \node[blue,inner sep =1pt] (Z1) at (3*\x -5*\y-2*\z, -\x+\y+\z) {$\scriptstyle Z_1$};
    \node[blue,inner sep =1pt] (Z2) at (3*\x -5*\y -2* \z, -\x+\y) {$\scriptstyle Z_2$};
    \node[blue,inner sep=1pt] (Z3) at (3*\x -5*\y+3*\z, -\x+\y-\z) {$\scriptstyle Z_3$}; 
    \draw[broken,blue] (Z3) -- (Zcenter) -- (Z1);
    \draw[broken,blue] (Zcenter) -- (Z2);
    

    \node[draw] at (-.7*\x,.6*\x) {\scalebox{.8}{
        $\begin{array}{rl}
            V_1 &=U_1 U_4 \\
            V_2 &= \frac{U_2}{U_1}\\
            V_3 &= \frac{1}{U_1}\\
            V_4 &= U_1^2 U_3
        \end{array}$
        }};
    \node[draw] at (1.2*\x,.7*\x) {\scalebox{.8}{
        $\begin{array}{rl}
            U_1 &= \frac{W_1}{W_3}\\
            U_2 &=\frac{1}{W_3}\\
            U_3 &= W_2 W_3^2 \\ 
            U_4 &= W_1^2 W_2W_3 + \alpha W_3 
        \end{array}$
        }};
    \node[draw] at (1.9*\x,.1*\x) {\scalebox{.8}{
        $\begin{array}{rl}
            V_1 &= W_1^3 W_2 +\alpha W_1 \\
            V_2 &= \frac{1}{W_1} \\
            V_3 &= \frac{W_3}{W_1} \\
            V_4 &= W_1^2 W_2
        \end{array}$
        }};
    \node[draw] at (-1.05*\x,-.4*\x) {\scalebox{.8}{
        $\begin{array}{rl}
            V_1 &= \frac{1}{X_2} \\ 
            V_2 &= X_1+ \alpha X_2 \\
            V_3 &= X_2^2 X_3\\
            V_4 &= X_1 / X_2
        \end{array}$
        }};
    \node[draw] at (-.7*\x,-1.4*\x) {\scalebox{.8}{
        $\begin{array}{rl}
            Y_1 &= X_1 + \alpha X_2\\
            Y_2 &= \frac{1}{X_1} \\
            Y_3 &= \frac{X_2}{X_1}\\
            Y_4 &= X_1^2X_3
        \end{array}$
        }};
    \node[draw] at  (1.9*\x,-1.4*\x)  {\scalebox{.8}{
        $\begin{array}{rl}
            Y_1 &= \frac{1}{Z_2}\\
            Y_2 &= Z_2 + \alpha \frac{Z_1}{Z_2} \\
            Y_3 &= \frac{Z_1}{Z_2^2}\\
            Y_4 &= Z_2^3 Z_3
        \end{array}$
        }};
    \node[draw] at (2.6*\x,-.4*\x) {\scalebox{.8}{
        $\begin{array}{rl}
            W_1  &= Z_2 \\
            W_2 &= \frac{1}{Z_1} \\
            W_3 &= Z_1^2 Z_3
        \end{array}$
        }};
    \node[draw] at (1.2*\x,-1.4*\x)  {\scalebox{.8}{
        $\begin{array}{rl}
            V_1 &= \frac{Y_2}{Y_3} \\
            V_2 &= Y_1 \\ 
            V_3 &= Y_3^2 Y_4 \\
            V_4 &= \frac{1}{Y_3}
        \end{array}$
        }};
\end{tikzpicture}}
\caption{Scattering diagram for the $\mathbb{F}_3$ GTP}
\label{fig:GTP_F3_diag}
\end{figure}
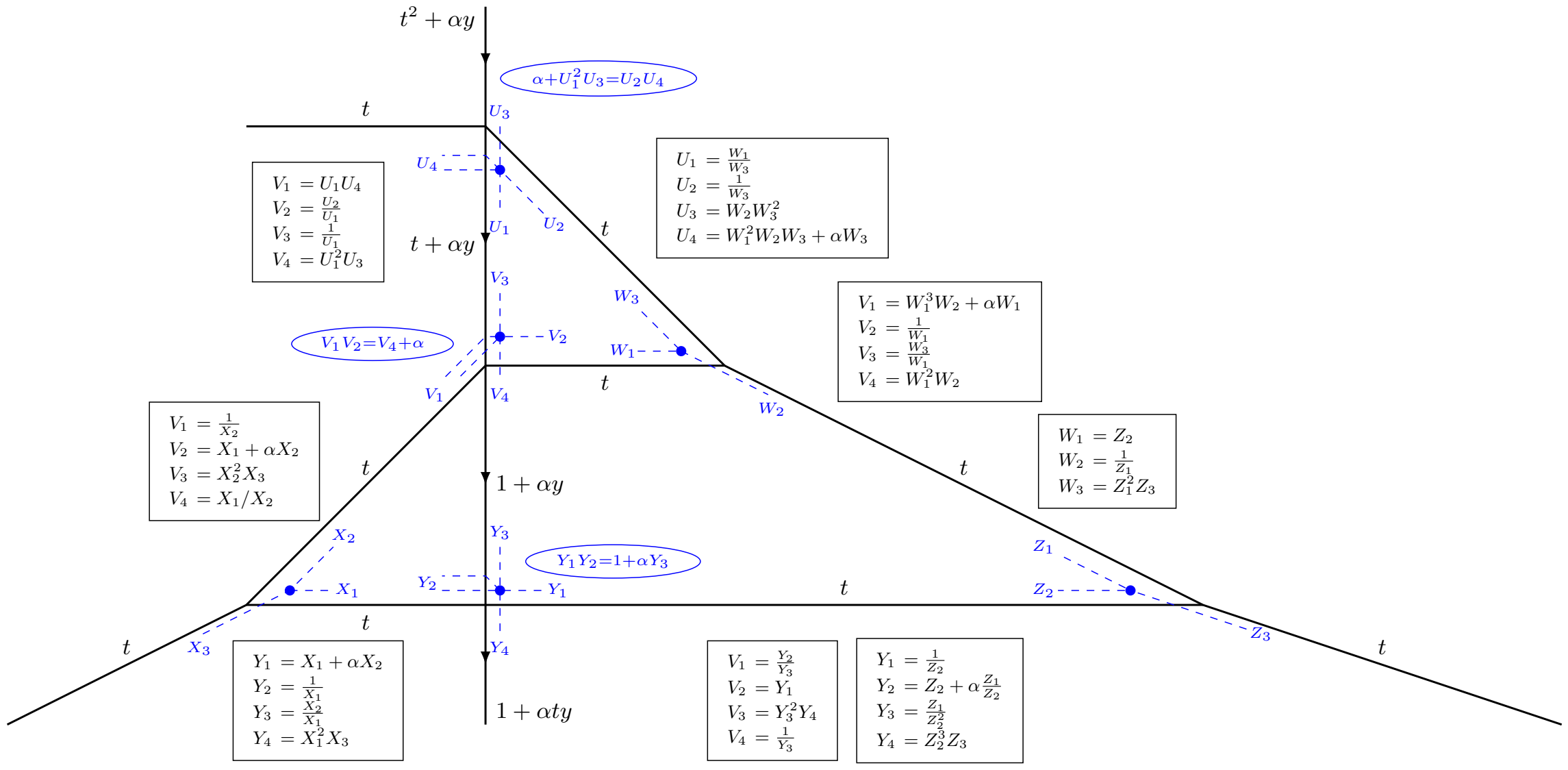

\begin{figure}[h]
\centering
\begin{tikzpicture}
    \fill[color=red, fill opacity=.4] (0,0) -- (1,0) -- (3,-1) -- (-1,-1);
    \fill[color=blue,fill opacity=.4] (-3,1) -- (0,1) -- (1,0) -- (3,-1) -- (7,-2) -- (7,2) -- (-3,2);
    \fill[color= green,fill opacity=.4] (-3,1) -- (0,1) -- (1,0) -- (0,0) -- (-1,-1) -- (-3,-2);
    \fill[color=orange,fill opacity=.4] (-3,-2)-- (-1,-1) -- (3,-1) -- (7,-2);
    \draw (-3,1) -- (0,1) -- (1,0) -- (3,-1) -- (7,-2);
    \draw (3,-1) -- (-1,-1) -- (0,0) -- (1,0);
    \draw (-1,-1) -- (-3,-2);
    \draw (0,2) -- (0,-2);
    \node at (0.8,-0.5) {$\mathcal D_0$};
    \node at (-1.5,0) {$\mathcal D_1$};
    \node at (3.5,0.5) {$\mathcal D_2$};
    \node at (1.5,-1.5) {$\mathcal D_3$};
\end{tikzpicture}
\begin{tikzpicture}
    \draw (-1,2) -- (-1,1) -- (0,-1) -- (1,2) -- (0,1) -- (0,0) -- (0,-1);
    \draw (1,2) -- (0,0) -- (-1,1);
    \fill[pattern=crosshatch dots ] (-1,2) -- (-1,1) -- (0,0) -- (0,1) -- (1,2);
    \draw[very thick] (0,0) -- (1.5,0);
    \draw[very thick] (0,0) -- (-1.5,0);
    \node at (-2,0) {$\mathcal E'$};
    \node at (2,0) {$\mathcal E$};
    \filldraw[fill=red] (0,0) circle (3pt);
    \filldraw[fill=blue] (1,2) circle (3pt);
    \filldraw[fill=blue] (-1,2) circle (3pt);
    \filldraw[fill=green] (0,1) circle (3pt);
    \filldraw[fill=green] (-1,1) circle (3pt);
    \filldraw[fill=orange] (0,-1) circle (3pt);
    \node (D0) at (1.5,-1) {$\mathcal D_0$};
    \draw[-latex,bend left=10] (D0) to (.1,-.1);
    \node at (-1.5,1) {$\mathcal D_1$};
    \node (D1) at (1.5,1) {$\mathcal D_1$};
    \draw[-latex, bend left=15] (D1) to (.15,1);
    \node at (1.5,2) {$\mathcal D_2$};
    \node at (-1.5,2) {$\mathcal D_2$};
    \node at (0,-1.5) {$\mathcal D_3$};
\end{tikzpicture}

\caption{The natural divisors of the deformed $F_3$ GTP geometry and the Symington polygon.}
\label{fig:F3_GTP_divisors}
\end{figure}
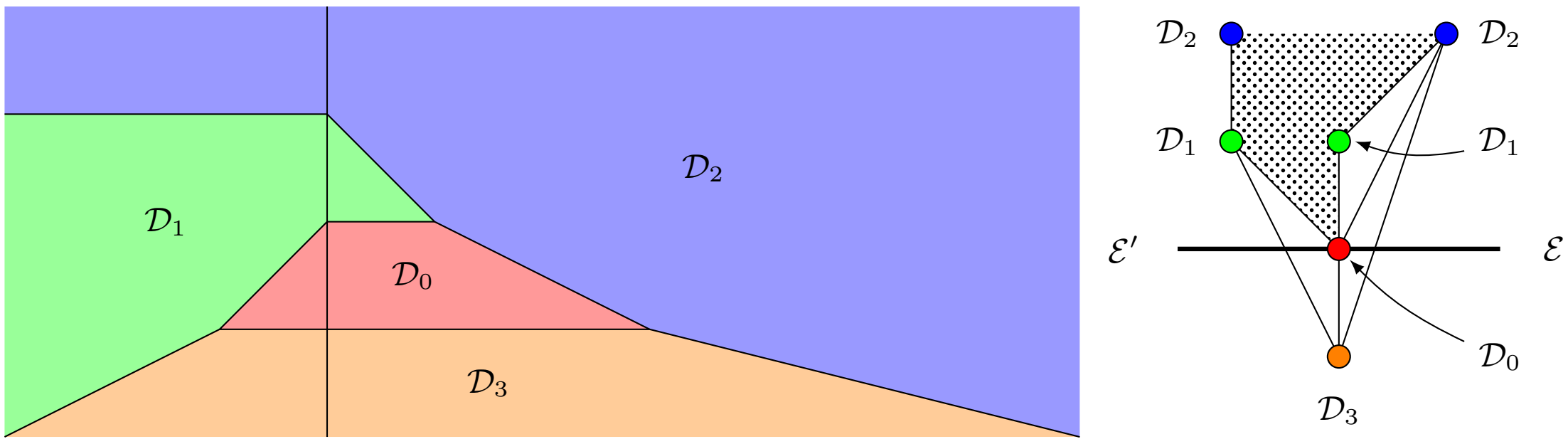

\begin{table}[h]
    \centering
    \begin{tabular}{c||c|c|c|c|c|c}
         &  $\mathcal U_U$&  $\mathcal U_V$&    $\mathcal U_W$&$\mathcal U_X$&  $\mathcal U_Y$&  $\mathcal U_Z$\\ \hline\hline
       $\mathcal D_0$  & & $V_3 = 0$ & $W_3 = 0$& $X_3 = 0$ & $Y_4 = 0$ & $Z_3 = 0$ \\ \hline
       $\mathcal D_1$ & $U_3 = 0$ & $V_4 = 0$& $W_2 = 0$ & $X_1= 0$&& \\\hline
       $\mathcal D_2$ & $U_1 = 0$ & & $W_1 = 0$ & && $Z_2=0$ \\\hline 
       $\mathcal D_3$ & &&&$X_2 = 0$ & $Y_3 = 0$ & $Z_1=0$ \\\hline
       $\mathcal E$ & $U_4 = 0$ & $V_1 = 0$ & $W_1^2 W_2  +\alpha = 0$& &$Y_2 = 0$ & $Z_2^2 + \alpha Z_1 = 0$ \\\hline
       $\mathcal E'$ & $U_2 = 0$ &  $V_2=0$&&$X_1+\alpha X_2=0$& $Y_1=0$&
    \end{tabular}
    \caption{Equations for the divisors of $\mathcal X_3$. An empty box means that the intersection of the divisor with the coordinate chart is empty.}
    \label{tab:divisorX3}
\end{table}
\begin{table}[h]
    \centering
    \begin{tabular}{c||c|c}
         & in $\mathcal D_0$ & in $\mathcal X_3$ \\\hline\hline
       $\mathcal C_1$  & $\mathcal O(-1)$& $\mathcal O(-1) \oplus \mathcal O(-1)$ \\\hline
       $\mathcal C_2$ & $\mathcal O$ & $\mathcal O\oplus \mathcal O(-2)$\\\hline
       $\mathcal C_3$ & $\mathcal O(3)$ & $\mathcal O(3)\oplus \mathcal O(-5)$ \\\hline
       $\mathcal C_e$ & $\mathcal O(1)$ &$\mathcal O(1) \oplus \mathcal O(-3)$ \\\hline
       $\mathcal C_e'$ & $\mathcal O$&  $\mathcal O\oplus \mathcal O(-2)$
    \end{tabular}
    \caption{Normal bundles to the curves}
    \label{tab:normalBundlesX3}
\end{table}

We can identify the intersections of the curves from Table \ref{tab:normalBundlesX3}: $\mathcal C_1$ and $\mathcal C_2$ have a simple intersection at the origin of $\mathcal U_W$, $\mathcal C_2$ and $\mathcal C_3$ at the origin of $\mathcal U_Z$, etc. $\mathcal C_e$ does not intersect $\mathcal C_1$ and intersects $\mathcal C_2$ and $\mathcal C_3$ at the origin of $\mathcal U_Z$ with multiplicity $1$ and $2$ respectively. $\mathcal C_e'$ intersects $\mathcal C_1$ and  $\mathcal C_3$ at the origin of $\mathcal U_X$,  intersects $\mathcal C_e$ at the point $(V_1,V_2,V_3,V_4)=(0,0,0,-\alpha),\; (Y_1,Y_2,Y_3,Y_4)=(0,0,-\frac1{\alpha},0) $ and does not intersect $\mathcal C_2$.

Meromorphic functions induce relations between the divisors. Several of these functions, expressed in the different coordinate charts, are shown in table \ref{tab:relationsDivisorsX3}. Note that the function $t$ (the variable in the algebra $\mathbb C[x^{\pm1},y^{\pm1},t]$) induces the relation $\mathcal D_0 + \mathcal D_1 + \mathcal D_2  +\mathcal D_3= 0$. Using this linear relation, we find that 
\begin{equation}
   \mathcal{D}_0^3=\mathcal D_0^2\cdot(-\mathcal{D}_1-\mathcal{D}_2-\mathcal{D}_3)=8 \,.
\end{equation}

\begin{table}[h]
    \centering
    \begin{tabular}{c||c|c|c|c|c|c}
         &  $\mathcal U_U$&  $\mathcal U_V$&    $\mathcal U_W$&$\mathcal U_X$&  $\mathcal U_Y$&  $\mathcal U_Z$\\ \hline\hline
        $\sum_{i=0}^3 \mathcal D_i = 0$ & $U_1U_3$ & $V_3V_4$ & $W_1W_2W_3$ & $X_1X_2X_3$ & $Y_3Y_4$ & $Z_1Z_2Z_3$ \\\hline
        $\mathcal E + \mathcal D_0 + \mathcal D_3=0$ & $U_4$ & $V_1V_3$ & $W_3(W_1^2 W_2 + \alpha)$ & $X_2X_3$ & $Y_2Y_3Y_4$ & $Z_1Z_3(Z_2^2 + \alpha Z_1)$ \\ \hline 
        $\mathcal D_0 - \mathcal D_2+ 2\mathcal D_3 =  0$ & $\frac{1}{U_1}$ & $V_3$ & $\frac{W_3}{W_1}$ & $X_2^2 X_3$ & $Y_3^2 Y_4$ & $\frac{Z_3Z_1^2}{Z_2}$\\\hline
        $\mathcal D_1+\mathcal D_2-\mathcal E$& $\frac{U_3 U_1}{U_4}$& $\frac{V_4}{V_1}$&$\frac{W_1 W_2}{W_1^2 W_2+\alpha}$&$X_1$&$\frac{1}{Y_2}$&$\frac{Z_2}{Z_2^2+\alpha Z_1}$\\\hline
         $\mathcal E'-\mathcal{D}_2$& $\frac{U_2 }{U_1}$& $V_2$&$\frac{1}{W_1}$&$X_1+\alpha X_2$&$Y_1$&$\frac{1}{Z_2}$\\\hline
          $\mathcal D_3-\mathcal D_2-\mathcal E$& $\frac{1}{U_4 U_1}$& $\frac{1}{V_1}$&$\frac{1}{W_1(W_1^2 W_2+\alpha)}$&$X_2$&$\frac{Y_3}{Y_2}$&$\frac{Z_1}{Z_2(Z_2^2+\alpha Z_1)}$
    \end{tabular}
    \caption{Meromorphic functions realizing relations between divisors}
    \label{tab:relationsDivisorsX3}
\end{table}

With all this information we can compute the triple intersection numbers of divisors (one of which needs to be the compact $\mathcal D_0$). This information can be used to compute the prepotential of the GTP in this particular resolution/Kähler moduli phase. Indeed, if $C_{IJK}$ is the triple intersection matrix, the prepotential is 
\begin{equation}
    \mathcal{F}=\frac{1}{6} C_{IJK} t^I t^J t^K\,,
\end{equation}
where $t^I$ are the extended Kähler moduli associated to the divisor $\mathcal{D}_I$. These are defined as
\begin{equation}
    t^I= \int_{\tilde{\mathcal{C}}_I} \omega\,
\end{equation}
where $\omega$ is the Kähler form and $\tilde{\mathcal{C}}_I$ is the curve Poincaré-dual to the divisor $\mathcal{D}_I$.

Using the relations in Table \ref{tab:relationsDivisorsX3} we choose $\mathcal{D}_0$ and $\mathcal{D}_2$ as independent divisors up to linear equivalence. Denoting $t^0=\phi, t^2=-\frac{1}{2}m_0$ we find
\begin{equation}
    \mathcal{F}=\frac{4 \phi ^3}{3}+\frac{1}{2} m_0 \phi^2\,.
\end{equation}
This is the prepotential of the $SU(2)_\pi$ gauge theory phase of $\tilde{E}_1$, the SCFT engineered on the local $\mathbb{F}_1$ in the normalization of \cite{Intriligator_1997}. We can reach the other non-Lagrangian phase by flopping $\mathcal{C}_1$.
\begin{table}[h]
    \centering
    \begin{tabular}{c||c|c|c|c|c|c}
        $\mathcal D_0$  & $\mathcal D_0$ & $\mathcal D_1$ & $\mathcal D_2$ &$ \mathcal D_3$ & $\mathcal E$ & $\mathcal E'$ \\ \hline\hline
        $\mathcal D_0$ & 8 &-1 &-2 & -5 & -3&-2\\ \hline
        $\mathcal D_1$ &-1 &-1 &1 &1 & 0&1 \\ \hline
        $\mathcal D_2$ &  -2&1 &0 &1 & 1&0\\ \hline
        $\mathcal D_3$ & -5 & 1&1 &3 & 2&1\\ \hline
        $\mathcal E$ &  -3 & 0&1 &2 & 1 & 1\\\hline
        $\mathcal E'$ &  -2 & 1&0&1 & 1 &0
    \end{tabular}
    \caption{Intersection numbers of divisors in $\mathcal X_3$} 
    \label{tab:intesectionsX3}
\end{table}

Finally we can prove that there is a globally defined nowhere vanishing 3-form. Indeed, using the transition functions for each patch $\mathcal U_i$ and $\mathcal U_j$ with $\mathcal U_i \cap \mathcal U_j\neq\emptyset$, for the volume forms $\omega_i, \omega_j$, we find $\omega_i= \omega_j$ orienting the variables in every patch counterclockwise with respect to Figure \ref{fig:GTP_F3_diag}, except for $\mathcal U_U$, which is oriented clockwise. Therefore $K_{\mathcal X}=0$ and the geometry is Calabi-Yau. Notably, the transition functions are independent of $\alpha$: for example, if we focus on the transition between $\mathcal U_X$ and $\mathcal U_Y$
\begin{equation}
    \omega_Y=\frac{d Y_1 \wedge d Y_3 \wedge d Y_4}{Y_1}\to (X_1+\alpha X_2)\frac{d X_1 \wedge d X_2 \wedge d X_3}{X_1+\alpha X_2}= \omega_X\,.
\end{equation}

\subsection{Euler characteristic}

\begin{figure}[h]
    \centering
    \begin{tikzpicture}
        \node at (-2,0) {\raisebox{-.5\height}{ \scalebox{.8}{  
        \begin{tikzpicture}[x=.5cm,y=.5cm] 
            \draw[gridline] (-1,-1.5)--(-1,2.5); 
            \draw[gridline] (0,-1.5)--(0,2.5); 
            \draw[gridline] (1,-1.5)--(1,2.5); 
            \draw[gridline] (-1.5,-1)--(1.5,-1); 
            \draw[gridline] (-1.5,0)--(1.5,0); 
            \draw[gridline] (-1.5,1)--(1.5,1); 
            \draw[gridline] (-1.5,2)--(1.5,2); 
            \draw[ligne] (1,-1)--(0,-1); 
            \draw[ligne] (0,-1)--(-1,2); 
            \draw[ligne] (-1,2)--(1,-1); 
            \draw[ligne] (-1,2)--(0,0); 
            \draw[ligne] (0,-1)--(0,0);
            \draw[ligne] (1,-1)--(0,0);
            \node[bd] at (0,-1) {}; 
            \node[bd] at (-1,2) {}; 
            \node[bd] at (1,-1) {};
            \node[bd] at (0,0) {}; 
        \end{tikzpicture}}}};

        \node at (2,0){\raisebox{-.5\height}{ \scalebox{.8}{  
            \begin{tikzpicture}[x=.5cm,y=.5cm] 
                \draw[gridline] (-1,-1.5)--(-1,2.5); 
                \draw[gridline] (0,-1.5)--(0,2.5); 
                \draw[gridline] (1,-1.5)--(1,2.5); 
                \draw[gridline] (-1.5,-1)--(1.5,-1); 
                \draw[gridline] (-1.5,0)--(1.5,0); 
                \draw[gridline] (-1.5,1)--(1.5,1); 
                \draw[gridline] (-1.5,2)--(1.5,2); 
                \draw[ligne] (-1,2)--(0,-1); 
                \draw[ligne] (0,-1)--(1,2); 
                \draw[ligne] (1,2)--(-1,2); 
                \draw[ligne] (-1,2)--(0,0); 
                \draw[ligne] (0,-1)--(0,0);
                \draw[ligne] (1,2)--(0,0);
                \node[bd] at (0,-1) {}; 
                \node[bd] at (1,2) {}; 
                \node[bd] at (0,0) {}; 
                \node[bd] at (-1,2) {}; 
                \node[wd] at (0,2) {}; 
            \end{tikzpicture}}}};

        \end{tikzpicture}
    \caption{GTPs for local $\mathbb{P}^2$ before and after a Hanany-Witten move.}
    \label{fig:GTPlocalP2}
\end{figure}
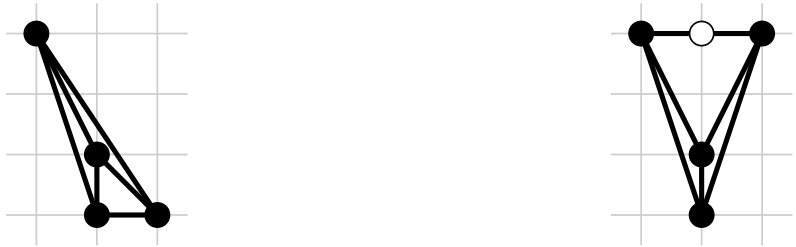

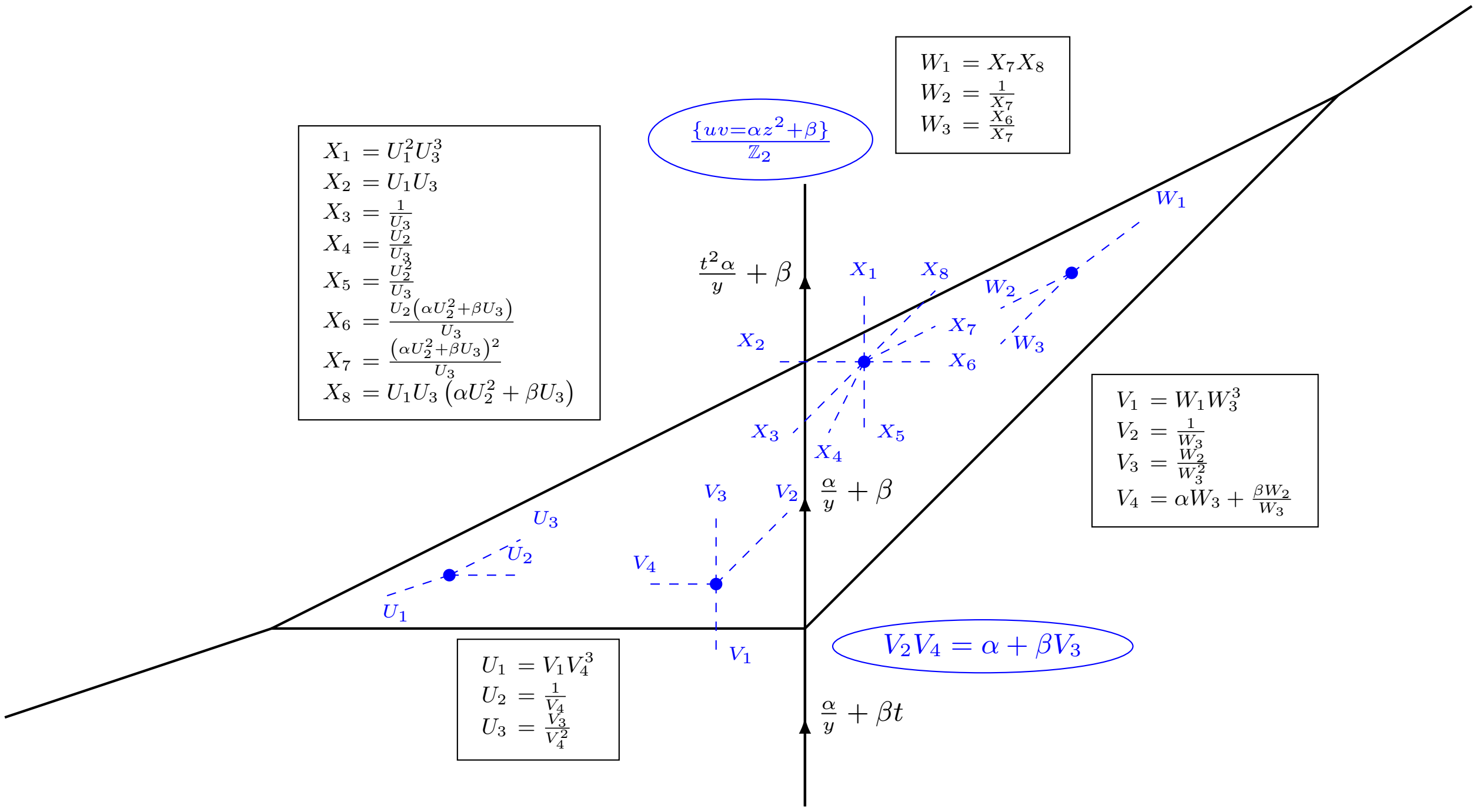
\begin{figure}[h]
\centering
    \begin{tikzpicture}
        \tikzmath{\x=2;\y=.8;}
        \draw[kink] (-1.5*\x,-0.5*\x) --  (0,0) -- (3*\x,0) -- (6*\x,3*\x)--(0,0);
        \draw[kink]  (3*\x,0) -- node[midway, right] {$\frac{\alpha}{y}+ \beta t$}(3*\x,-\x) ;
        \draw[kink] (6*\x,3*\x) --(6.75*\x,3.5*\x);
        \draw[wall] (3*\x,0) -- node[midway, right]{$\frac{\alpha}{y}+ \beta $}(3*\x,1.5*\x) ;
        \draw[wall] (3*\x,-\x) -- (3*\x,0);
        
        \draw[wall] (3*\x,1.5*\x) -- node[midway, left]{$\frac{t^2 \alpha}{y}+ \beta $} (3*\x,2.5*\x);

        \node[focalPoint,blue] at (\x, 0.3*\x) {};
        
        \draw[broken,blue] (\x, 0.3*\x)-- node[pos=1.,above]{$\scriptstyle U_2$}(\x+\y, 0.3*\x);
        
        \draw[broken,blue] (\x, 0.3*\x)-- node[pos=1.,above right]{$\scriptstyle U_3$}(\x+\y, 0.3*\x+0.5*\y);
        
        \draw[broken,blue] (\x, 0.3*\x)-- node[pos=0.75,below]{$\scriptstyle U_1$}(\x-\y, 0.3*\x-0.33*\y);
        
        
        \node[focalPoint,blue] at (4.5*\x,2*\x) {};
        
        \draw[broken,blue] (4.5*\x,2*\x)-- node[pos=1.,above right]{$\scriptstyle W_1$} (4.5*\x+\y,2*\x+0.75*\y);
        \draw[broken,blue] (4.5*\x,2*\x)-- node[pos=1.,above ]{$\scriptstyle W_2$} (4.5*\x-\y,2*\x-0.5*\y);
        \draw[broken,blue] (4.5*\x,2*\x)-- node[pos=1., right ]{$\scriptstyle W_3$} (4.5*\x-\y,2*\x-\y);
        
        
        \node[focalPoint,blue] at (2.5*\x,0.25*\x) {};
        \draw[broken,blue] (2.5*\x,0.25*\x)-- node[pos=1.,above]{$\scriptstyle V_3$} (2.5*\x,0.25*\x+\y);
        
        \draw[broken,blue] (2.5*\x,0.25*\x)-- node[pos=1.,right]{$\scriptstyle V_1$} (2.5*\x,0.25*\x-\y);
        \draw[broken,blue] (2.5*\x,0.25*\x)-- node[pos=1.,above]{$\scriptstyle V_2$} (2.5*\x+\y,0.25*\x+\y);
        
        \draw[broken,blue] (2.5*\x,0.25*\x)-- node[pos=1.,above]{$\scriptstyle V_4$} (2.5*\x-\y,0.25*\x);
        
        \node[ellipse,draw,blue,inner sep =2pt] at (4*\x,-0.1*\x) {$V_2 V_4=\alpha +\beta  V_3$};
        
        \node[ellipse,draw,blue,inner sep =2pt] at (2.75*\x,2.75*\x) {$\frac{\{u v=\alpha z^2+\beta\}}{\mathbb{Z}_2}$};

        \node[focalPoint,blue] at (3.333*\x,1.5*\x) {};
        \draw[broken,blue] (3.333*\x,1.5*\x)-- node[pos=1.,above]{$\scriptstyle X_1$}  (3.333*\x,1.5*\x+ \y);
        \draw[broken,blue] (3.333*\x,1.5*\x)-- node[pos=1.,above]{$\scriptstyle X_8$}  (3.333*\x+\y,1.5*\x+ \y);
        \draw[broken,blue] (3.333*\x,1.5*\x)-- node[pos=1.,right]{$\scriptstyle X_7$}  (3.333*\x+\y,1.5*\x+ 0.5*\y);
        \draw[broken,blue] (3.333*\x,1.5*\x)-- node[pos=1.,right]{$\scriptstyle X_6$}  (3.333*\x+\y,1.5*\x);
        \draw[broken,blue] (3.333*\x,1.5*\x)-- node[pos=1.,right]{$\scriptstyle X_5$}  (3.333*\x,1.5*\x-\y);
        \draw[broken,blue] (3.333*\x,1.5*\x)-- node[pos=1.,left]{$\scriptstyle X_3$}  (3.333*\x-\y,1.5*\x-\y);
        \draw[broken,blue] (3.333*\x,1.5*\x)-- node[pos=1.,below]{$\scriptstyle X_4$}  (3.333*\x-0.5*\y,1.5*\x-\y);
        \draw[broken,blue] (3.333*\x,1.5*\x)-- node[pos=1.,above left]{$\scriptstyle X_2$}  (3.333*\x-1.2*\y,1.5*\x);
        
        \node[draw] at (1.5*\x,-0.4*\x)  {\scalebox{.8}{
            $\begin{array}{rl}
            U_1 &= V_1 V_4^3 \\
            U_2 &= \frac{1}{V_4} \\ 
            U_3 &= \frac{V_3 }{ V_4^2} \\
            \end{array}$
            }};
        
        \node[draw] at (5.25*\x,\x)  {\scalebox{.8}{
            $\begin{array}{rl}
            V_1 &= W_1 W_3^3 \\
            V_2 &= \frac{1}{W_3} \\ 
            V_3 &= \frac{W_2}{W_3^2} \\
            V_4 &=\alpha  W_3+\frac{\beta  W_2}{W_3}\\
            \end{array}$
            }};
        
        \node[draw] at (4.*\x,3*\x)  {\scalebox{.8}{
            $\begin{array}{rl}
            W_1 &=X_7 X_8 \\
            W_2 &= \frac{1}{X_7} \\ 
            W_3 &= \frac{X_6}{X_7} \\
            \end{array}$
            }};
        
        \node[draw] at (\x,2*\x)  {\scalebox{.8}{
            $\begin{array}{rl}
            X_1 &=U_1^2 U_3^3 \\
            X_2 &= U_1 U_3 \\ 
            X_3 &=  \frac{1}{U_3} \\
            X_4&=\frac{U_2}{U_3}\\
            X_5&= \frac{U_2^2}{U_3}\\
            X_6&= \frac{U_2 \left(\alpha  U_2^2+\beta 
            U_3\right)}{U_3}\\
            X_7&=\frac{\left(\alpha  U_2^2+\beta 
            U_3\right){}^2}{U_3}\\
            X_8&=U_1 U_3 \left(\alpha  U_2^2+\beta 
            U_3\right)
            \end{array}$
            }};
    \end{tikzpicture}
    \caption{Scattering diagram for $\mathcal{O}(-3) \to \mathbb{P}^2$ with a 7-brane. For simplicity we only draw the broken lines that do not bend.}
    \label{fig::localP2}
\end{figure}

In this section, we will study the topology of generalized toric varieties after the deformation. We will focus on the example of the local $\mathbb{P}^2$. We can end a 7-brane on one of the external 5-branes and perform a Hanany-Witten move, which results in the toric diagram of the complex cone over $Y^{3,3}$ with a white dot. We depict the GTPs in Figure \ref{fig:GTPlocalP2} and the scattering diagram in Figure \ref{fig::localP2}. 

We are interested in computing the Euler characteristic $\chi$ of the 3-fold. For that, we will take advantage of our explicit patches and transition functions and apply the inclusion-exclusion formula for finite open coverings 
\begin{equation}\label{eq:Euler_MayerVietoris}
    \chi(\mathcal{X})=\sum_i \chi(\mathcal{U}_i)- \sum_{i<j} \chi(\mathcal{U}_i \cap \mathcal{U}_j)+ \sum_{i<j<k} \chi(\mathcal{U}_i \cap \mathcal{U}_j \cap \mathcal{U}_k)- \cdots
\end{equation}

We apply this formula to the covering by open charts associated with vertices. An intersection of three or more of these, or two associated with vertices not linked by an edge, is a toric threefold isomorphic to $\mathbb C\times (\mathbb C^*)^2$ or $(\mathbb C^*)^3$. In particular its Euler characteristic is zero. Therefore, the only terms appearing in \eqref{eq:Euler_MayerVietoris} correspond to edges and vertices of the scattering diagram.

We will start computing the Euler characteristic of the four patches $\mathcal{U}_U, \mathcal{U}_V,\mathcal{U}_W,\mathcal{U}_X$. The patches $\mathcal{U}_U,\mathcal{U}_W$ are isomorphic to $\mathbb{C}^3$, so $\chi(\mathcal{U}_U)=\chi(\mathcal{U}_W)=1$. The topology of the other two patches changes depending on the position in the $\mathbb{P}_{[\alpha:\beta]}^1$ parameter space.

We will first study the topology of these patches at a generic point of the $\mathbb{P}^1$. In the case of $\mathcal{U}_V$ we can eliminate $V_3$ using the relation, so $\mathcal{U}_V\cong \mathbb{C}^3$, $\chi(\mathcal{U}_V)=1$. 
 
On the other hand, $\mathcal U_X$ is a $(2,1)$ T-cone. Recall that it is defined as the total space of the canonical bundle of a free $\mathbb{Z}_2$ orbifold over the following hypersurface
\begin{equation}
   V:\quad  u v=\alpha z^2+\beta
\end{equation}
At generic values of $z$ and $[\alpha:\beta ]$, we find $u v=\lambda\neq 0$, so this fiber is $\mathbb{C}^*$. At the special fibers $z=\pm \sqrt{-\beta/\alpha}$, $uv=0$, so the fiber is two lines intersecting at the origin $\mathbb{C}\vee \mathbb{C}$. All in all,
\begin{equation}
\begin{aligned}
     \chi(V)&= \chi\left(\mathbb{C}\backslash\left\{\pm \sqrt{-\beta/\alpha} \right\}\right) \chi(\mathbb{C^*})+\chi(\left\{ \sqrt{-\beta/\alpha} \right\}) \chi(\mathbb{C}\vee \mathbb{C})+\chi(\left\{ -\sqrt{-\beta/\alpha} \right\}) \chi(\mathbb{C}\vee \mathbb{C})\\
     &=(-1)\cdot 0+1\cdot 1+ 1\cdot 1=2
\end{aligned}
\end{equation}
After applying the free action and taking the total space of the bundle
\begin{equation}
    \chi(\mathcal{U}_X)=\frac{\chi(V)}{|\mathbb{Z}_2|}=1.
\end{equation}
It is straightforward to generalize this argument for the case of a general $(p,q)$ T-cone, where the orbifold is over the hypersurface $u v=\alpha z^p+\beta$, which has $p$ fibers. After applying the $\mathbb{Z}_p$ action, we find $\chi(T_{(p,q)})=1$.

Let us now consider the poles of the $\mathbb{P}^1$. Starting with $[\alpha:\beta]=[1:0]$, in the $\mathcal U_V$ patch the relation becomes $V_2 V_4=1$ and $\chi(\mathcal{U}_V)=0$. In the patch $\mathcal{U}_X$, the hypersurface $V$ reduces to the $A_1$ singularity $u v=z^2$, which is contractible, so

\begin{equation}
    \chi(\mathcal{U}_X)=\chi(V)=1
\end{equation}

Going to the opposite pole at $[\alpha:\beta]=[0:1]$, the relation still gives $\chi(\mathcal U_V)=1$. In $\mathcal{U}_X$, $uv=1$ and $V\cong \mathbb{C} \times \mathbb{C}^*$. In conclusion

\begin{equation}
    \chi(\mathcal{U}_V)= \left\{ \begin{array}{cc}
        1& [\alpha:\beta]\neq [1:0], \\
      0 & [\alpha:\beta]= [1:0],
    \end{array}\right. \quad \chi(\mathcal{U}_X)= \left\{ \begin{array}{cc}
        0& [\alpha:\beta]=[0:1], \\
      1 & [\alpha:\beta]\neq [0:1].
    \end{array}\right. 
\end{equation}

Let us continue computing the intersections $\mathcal{U}_i \cap \mathcal{U}_j$. Using the transition functions, most of the intersections reduce to $\mathbb{C}^2 \times \mathbb{C}^*$, so their Euler characteristic vanishes. However, the case $\mathcal{U}_V \cap \mathcal{U}_X$ is different due to the relations. First, consider a generic point in the $\mathbb{P}^1$. $\mathcal{U}_V \cap \mathcal{U}_X$  becomes
\begin{equation}
    V_3\neq 0, \quad V_2 V_4=\alpha +\beta  V_3
\end{equation}
For a generic value of $V_3$, $V_2 V_4=c \neq 0$, therefore its fiber is $\mathbb{C}\times \mathbb{C}^*$. However, if $V_3=-\alpha/\beta$, then $V_2 V_4=0$, and the fiber is $\mathbb{C}\vee \mathbb{C}$, therefore $\chi(\mathcal{U}_V \cap \mathcal{U}_X)=1$. If $[\alpha:\beta]=[1:0]$, the relation reduces to $ V_2 V_4=1$, so $\chi(\mathcal{U}_V \cap \mathcal{U}_X)=0$. On the other hand, if  $[\alpha:\beta]=[0:1]$, then $V_2 V_4=V_3 \neq 0$, so again $\chi(\mathcal{U}_V \cap \mathcal{U}_X)=0$.

Combining all of this information, we see that

\begin{equation}
     \chi(\mathcal{X})=\chi(\mathcal{U}_U)+ \chi(\mathcal{U}_V)+ \chi(\mathcal{U}_W)+ \chi(\mathcal{U}_X)- \chi(\mathcal{U}_V \cap \mathcal{U}_X)=\left\{ \begin{array}{c}
     1+0+1+1-0=3,\quad [\alpha:\beta]=[1:0]\\
     1+1+1+1-1=3,\quad [\alpha:\beta] \text{ generic}\\
    1+1+1+0-0=3,\quad [\alpha:\beta]=[0:1]
     \end{array}\right.
\end{equation}
This matches with the expectation that this variety is just a non-toric presentation of local $\mathbb{P}^2$ (see example 7.2 of \cite{Alexeev:2024bko}).

For the  $\mathbb{F}_1 / \mathbb{F}_3$ GTP whose scattering diagram is shown in Figure \ref{fig:GTP_F3_diag}, the charts corresponding to the vertices all have characteristic $1$, except for $\chi(\mathcal U_U) = 0$. This is a manifestation of the fact that $\mathcal{U}_U \subset \mathcal{U}_V \cup \mathcal{U}_W $ so this patch is redundant. The pairwise intersections corresponding the edges of the diagram have Euler characteristic $0$, except for $\chi(\mathcal U_V\cap \mathcal U_Y) = 1$. All in all, the Mayer-Vietoris formula \ref{eq:Euler_MayerVietoris} gives: 
\begin{equation}
    \chi(\mathcal X_\alpha) = 5 - 1 = 4
\end{equation}

All in all, we have shown that primitive T-cones have $\chi=1$, while the trapezoid 4-junctions that appear in tessellating non-primitive T-cones contribute $\chi=0$. Note that this is not an exhaustive list of all possible vertices. 

\begin{figure}
    \centering
    \begin{tikzpicture}
        \node at (0,0){
            \begin{tikzpicture}
                \draw (0,-1) -- (1,2) -- (-1,2) --(0,-1)--(0,0)--(-1,2) ;
                \draw  (0,0) -- (0,-1) ;
                \draw (0,0) -- (1,2) ;
                \fill[pattern=crosshatch dots ]  (1,2)--(0,0)--(-1,2);
                
                \filldraw[fill=red] (0,0) circle (3pt);
                \filldraw[fill=blue] (1,2) circle (3pt);
                \filldraw[fill=blue] (-1,2) circle (3pt);
                \filldraw[fill=green] (0,-1) circle (3pt);
            \end{tikzpicture}};
        
        \node at (4,0){
            \begin{tikzpicture}
                \draw (0,-1) -- (1,-1) -- (-1,2) --(0,-1)--(0,0)--(-1,2) ;
                \draw  (0,0) -- (0,-1) ;
                \draw (0,0) -- (1,-1) ;
                \fill[pattern=crosshatch dots ]  (0,-1)--(0,0)--(1,-1);
                
                \filldraw[fill=red] (0,0) circle (3pt);
                \filldraw[fill=green] (1,-1) circle (3pt);
                \filldraw[fill=blue] (-1,2) circle (3pt);
                \filldraw[fill=green] (0,-1) circle (3pt);
            \end{tikzpicture}};
    \end{tikzpicture}
    \caption{Symington polygons associated to the HW move of local $\mathbb{P}^2$. By counting the number of minimal triangles plus cut out T-cones we recover $\chi=3$ on either case.}
    \label{fig:SymingP2}
\end{figure}
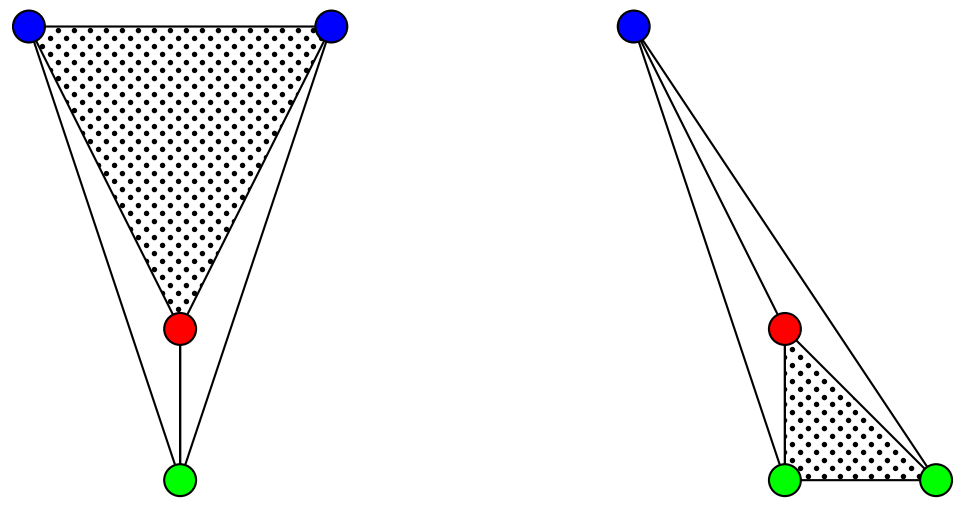

These observation allow us to propose the following conjecture regarding the Euler characteristic of GTPs. The starting observation is that in the toric case, the Euler characteristic simply counts the number of minimal triangles in the fully resolved toric diagram. The analogous combinatorial object to study in the case of GTPs is the Symington polygon (see Section \ref{sec::symington}), which has all T-cones (be they primitive or non-primitive) cut out. We propose that the Euler characteristic of a GTP is simply given by the number of minimal triangles in the Symington polygon plus the number of 7-branes/ cut out T-cones. We depict in Figure \ref{fig:SymingP2} this reasoning in the local $\mathbb P^2$ case.

As remarked above, only the charts and their double intersections contribute to the Euler characteristic, corresponding to vertices and edges in the scattering diagram respectively. At a generic point of the extended Coulomb branch, edges will always be in the form (up to $\text{SL}(2,\mathbb{Z})$ transformation) $f = t^p + \alpha \,x \,t^q$ with $p,q$ non-negative integers and at least one of them zero. Their dual double intersections have $\chi=0$ unless both $p=q=0$, in such a case $\chi=1$.

We leave the extension of these results to infinite scattering diagrams for future work.

\section{More than one $7$-brane: simultaneous Hanany-Witten transitions}\label{sec:multiple-seven-branes}

When several $7$-branes are brought to finite distance, the consistent scattering diagram knows about all the ways in which the $7$-branes can be sent to infinity, possibly after going through the web of 5-branes and undergoing Hanany-Witten transitions. Configurations of 7-branes may be classified into three classes, which determine the behavior of their scattering diagrams. Form a quiver whose nodes correspond to the $7$-branes and, given a pair of nodes, the number of edges is the symplectic pairing of the charges of the two branes. If this quiver is mutation equivalent to  the Dynkin diagram of a finite-dimensional (resp. affine) Lie algebra, then the configuration is called finite (resp. tame). Otherwise, it is called wild \cite{Fomin:2002nsg}. The associated scattering diagram is finite if the configuration of branes is finite, infinite otherwise. A tame diagram will contain accumulation points of walls (see Figure \ref{fig:7brane_crossing_tame:scattering_diag}) and a wild diagram will contain dense regions of walls (see Figure \ref{fig:7brane_crossing_wild}).

Below, we discuss the example of a pair of $7$-branes without any $5$-branes present, as their inclusion doesn't change the discussion significantly. This configuration is finite (resp. tame, resp. wild) if the symplectic pairing of the charges of the branes is $1$ (resp. $2$, resp. $3$ or more). We discuss the consistent scattering diagram and construct a toric variety which naturally parametrizes deformation of the associated Calabi-Yau threefold. The toric points of this parameter space correspond to ways of bringing the $7$-branes to infinity. We then turn to some examples with more than two $7$-branes.

\subsection{Two $7$-branes}

\paragraph{Trivial case}

If the two $7$-branes have the same charge, say $(1,0)$, then the situation is trivial, as the monodromies of the two $7$-branes do not interact. Each $7$-branes produces a wall $(\left(\begin{smallmatrix}1\\0\end{smallmatrix}\right)\mathbb R_+,\alpha_i + \beta_ix^{\pm})$, where the sign of the exponent on $x$ is $+$ if the monodromy cut is placed on the right and $-$ if it is placed on the left. The parameter space, with homogeneous coordinates $(\alpha_i,\beta_i)_{i=1,2}$ is simply $\mathbb P^1 \times \mathbb P^1$. The two walls can be separated or merged together freely.

The quiver contains two disconnected nodes:
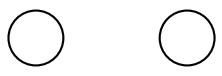
\begin{equation}
    \begin{tikzpicture}
    \node[circle,draw] (A) at (0,0) {};
    \node[circle,draw] (B) at (1,0) {};    
\end{tikzpicture}
\end{equation}

If one suspends a $5$-brane between the two $7$-branes, then the two walls are forced to be superposed. As discussed in Section \ref{sec:s-rule}, this configuration is supersymmetric but neither  consistent nor scatterable. The scattering diagram contains a pole in $t$.

\paragraph{Finite case}

\begin{figure}[h]
    \centering
    \begin{subfigure}[t]{.495\textwidth}
    \begin{tikzpicture}
        \tikzmath{\x=3;}
        \node (A) at (-\x,0) {$a+b c x$};
        \node (E) at (-\x,-\x) {$e + b d x y$};
        \node (C) at (0,-\x) {$c + a d y$};
        \node (B) at (\x,0) {$ae+bx$};
        \node (D) at (0,\x) {$ce+dy$};
        \draw[wall] (B) -- (0,0);
        \draw[wall] (D) -- (0,0);
        \draw[wall] (0,0) -- (A);
        \draw[wall] (0,0) -- (C);
        \draw[wall] (0,0) -- (E);
    \end{tikzpicture}
    \caption{}
    \label{fig:7brane_crossing_finite:scattering_diag}
    \end{subfigure}
    \begin{subfigure}[t]{.495\textwidth}
    \begin{tikzpicture}
        \tikzmath{\x=3;}
        \node (A) at (-\x,0) {$a$};
        \node (E) at (-\x,-\x) {$e$};
        \node (C) at (0,-\x) {$c$};
        \node (B) at (\x,0) {$b$};
        \node (D) at (0,\x) {$d$};
        \draw (B) -- (0,0);
        \draw (D) -- (0,0);
        \draw (0,0) -- (A);
        \draw (0,0) -- (C);
        \draw (0,0) -- (E);
        \node at (2,2) {\scalebox{.5}{\begin{tikzpicture}
            \fill (1,0) circle (3pt);
            \fill (0,1) circle (3pt);
            \draw[dashed] (1,0) -- (2,0);
            \draw[dashed] (0,1) -- (0,2);
            \draw[very thick,->] (1.2,.3) -- (1.9,.3);
            \draw[very thick,->] (.3,1.2) -- (.3,1.9);
            \end{tikzpicture}}};
        \node at (2,-2) {\scalebox{.5}{\begin{tikzpicture}
            \fill (1,0) circle (3pt);
            \fill (0,-1) circle (3pt);
            \draw[dashed] (1,0) -- (2,0);
            \draw[dashed] (0,-1) -- (0,2);
            \draw[very thick,->] (1.2,.3) -- (1.9,.3);
            \draw[very thick,->] (0,-1.2) -- (0,-1.9);
            \end{tikzpicture}}};
        \node at (-2,2) {\scalebox{.5}{\begin{tikzpicture}
            \fill (-1,0) circle (3pt);
            \fill (0,1) circle (3pt);
            \draw[dashed] (-1,0) -- (2,0);
            \draw[dashed] (0,1) -- (0,2);
            \draw[very thick,->] (-1.2,0) -- (-1.9,0);
            \draw[very thick,->] (.3,1.2) -- (.3,1.9);
            \end{tikzpicture}}};
        \node at (-2.5,-1.2) {\scalebox{.5}{\begin{tikzpicture}
            \fill (-1,0) circle (3pt);
            \fill (-1,-1) circle (3pt);
            \draw[dashed] (-1,0) -- (2,0);
            \draw[dashed] (-1,-1) -- (0,0) -- (0,2);
            \draw[very thick,->] (-1.2,0) -- (-1.9,0);
            \draw[very thick,->] (-1.2,-1.2) -- (-1.7,-1.7);
            \end{tikzpicture}}};
        \node at (-1.1,-2.5) {\scalebox{.5}{\begin{tikzpicture}
            \fill (0,-1) circle (3pt);
            \fill (-1,-1) circle (3pt);
            \draw[dashed] (0,-1) -- (0,2);
            \draw[dashed] (-1,-1) -- (0,0) -- (2,0);
            \draw[very thick,->] (-1.2,-1.2) -- (-1.7,-1.7);
            \draw[very thick,->] (0,-1.2) -- (0,-1.9);
            \end{tikzpicture}}};
    \end{tikzpicture}
    \caption{}   
    \label{fig:7brane_crossing_finite:fan}
    \end{subfigure}
    \caption{The consistent diagram (a) for two $7$-branes crossing with Dirac pairing 1. There are $5$ variables parametrizing the deformations. They are homogeneous coordinates associated with the rays in the toric fan (b). The toric points correspond to ways of sending the $7$-branes to infinity.}
    \label{fig:7brane_crossing_finite}
\end{figure}
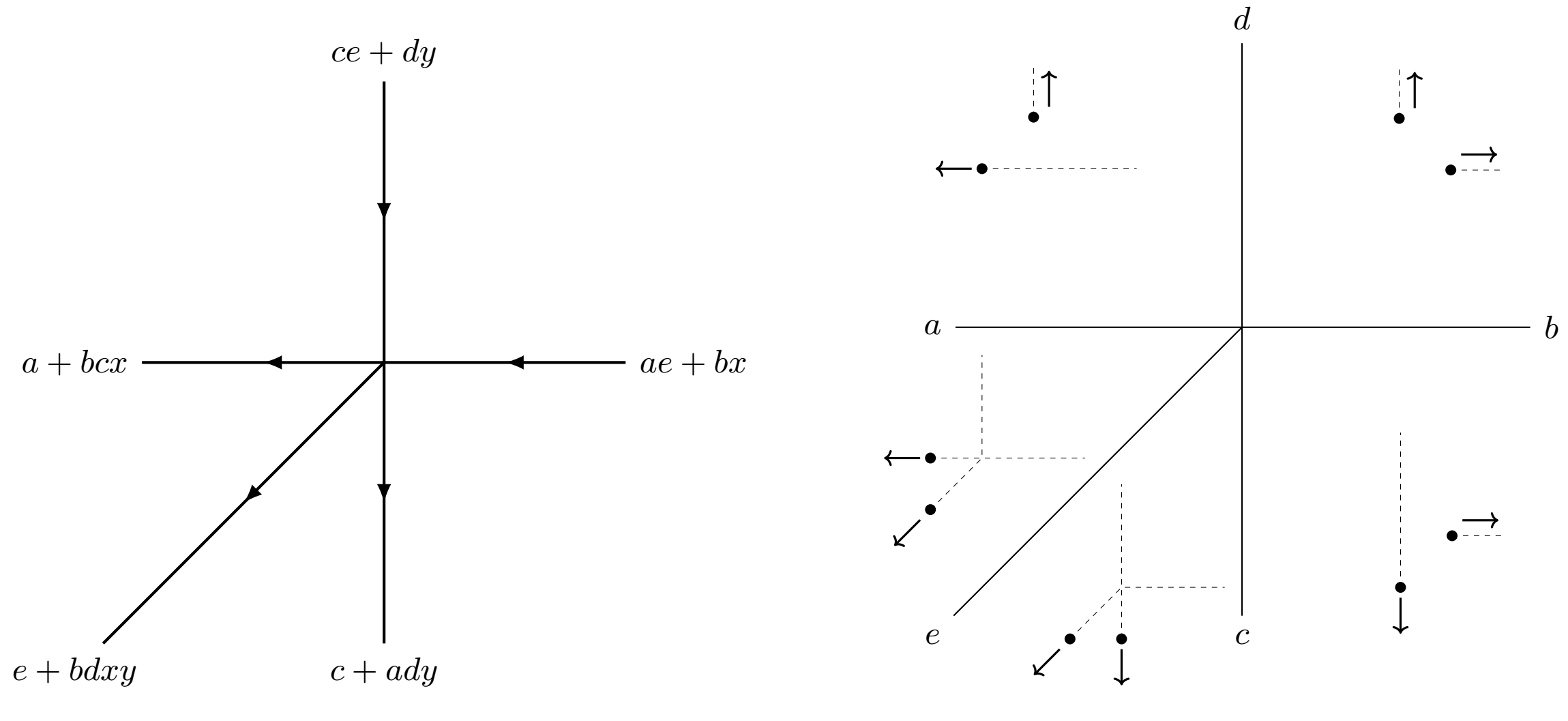

Consider now a configuration of two $7$-branes of charge $(1,0)$ and $(0,1)$. The quiver is the finite Dynkin diagram $A_2$:
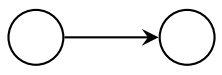
\begin{equation}
    \begin{tikzpicture}
    \node[circle,draw] (A) at (0,0) {};
    \node[circle,draw] (B) at (1,0) {};    
    \draw[-stealth] (A) -- (B);
\end{tikzpicture}
\end{equation}

The consistent scattering diagram is shown in figure \ref{fig:7brane_crossing_finite:scattering_diag}. We use $5$ homogeneous coordinates on the parameter space for this transition. Three of these ($a,c$ and $e$) are introduced on kinks : when non-zero, they can be rescaled independently without changing the threefold. The last two $b,d$ are introduced in the coefficients of the incoming walls associated with the $7$-branes. The factors of $a,c,e$ in the wall-crossing functions are chosen so that after merging with the kinks, the deforming term has just a power of $b$ or $d$. In the end, we see that $\frac{b}{a e}$ and $\frac{d}{ce}$ are actual coordinates on the parameter space. Therefore, the three $\mathbb C^*$ act on the homogeneous coordinates with weights:
\begin{equation}
\begin{array}{c|ccccc}
& a& b&c&d&e \\ \hline
\mathbb C^*_1     & 1 & 1 & 0 & 0 & 0 \\
\mathbb C^*_2     & 0 & 0 & 1 & 1 & 0 \\
\mathbb C^*_3     & 0 & 1 & 0 & 1 & 1 \\
\end{array}
\end{equation}
These describe a smooth toric blow-up of $\mathbb P^1 \times \mathbb P^1$, with the fan shown in figure \ref{fig:7brane_crossing_finite:fan}. The five two-dimensional cones in the fan (i.e. the toric points in the parameter space) correspond to the different ways to bring the $7$-branes to infinity. They are given by the vanishing of the homogeneous coordinates associated with the two neighboring rays. The remaining homogeneous coordinate can be rescaled to $1$.

The fact that this parameter space is not $\mathbb P^1 \times \mathbb P^1$ (the product of the parameter spaces associated with the $7$-branes taken in isolation), but a blow-up of it reflects the fact that there is a ordering choice when doing the two Hanany-Witten transitions simultaneously. The blow-up replaces the toric point corresponding to the double transition by two toric points, corresponding to the two orderings.

\paragraph{Tame case}

\begin{figure}
    \centering
    \begin{subfigure}[t]{.8\textwidth}
        \centering   
        \begin{tikzpicture}[decoration={markings,mark=at position 0.5 with {\arrow{latex}}}] 
            \tikzmath{\x = 2;}
            \node (E) at (\x,\x) {$a b^2 d + e x y$};
            \node (F) at (\x,-\x) {$a^2 bc + f xy^{-1}$};
            \node (C) at (-\x,\x) {$c + b^3 d^2 f xy^{-1}$};
            \node (D) at (-\x,-\x) {$d  +a^3 c^2 e xy$};
            \node (A) at (-3*\x,\x) {$a + b^4 d^3 ef^2 x^3 y^{-1} $};
            \node (B) at (-3*\x,-\x) {$b + a^4 c^3 e^2 f x^3y$};
            \node (G) at (-3.5*\x,0) {$\frac{1}{(1-abcdef \, x^2)^4}$};
            \draw[postaction={decorate}] (E) -- (0,0);
            \draw[postaction={decorate}] (F) -- (0,0);
            \draw[postaction={decorate}] (0,0) -- (A);
            \draw[postaction={decorate}] (0,0) -- (B);
            \draw[postaction={decorate}] (0,0) -- (C);
            \draw[postaction={decorate}] (0,0) -- (D);
            \draw[postaction={decorate}] (0,0) -- (-3*\x,0);
            \foreach \n in {2,...,3}
                \tikzmath{\y = 3*\x /(2*\n+1);}
                \draw[postaction={decorate}] (0,0)--(-3*\x,\y);
            \foreach \n in {4,...,50}
                \tikzmath{\y = 3*\x /(2*\n+1);}
                \draw (0,0)--(-3*\x,\y);
            \foreach \n in {2,...,3}
                \tikzmath{\y = 3*\x /(2*\n+1);}
                \draw[postaction={decorate}] (0,0)--(-3*\x,-\y);
            \foreach \n in {4,...,50}
                \tikzmath{\y = 3*\x /(2*\n+1);}
                \draw (0,0)--(-3*\x,-\y);
        \end{tikzpicture}   
        \caption{}
    \label{fig:7brane_crossing_tame:scattering_diag}
    \end{subfigure}

    \vspace{.5cm}
    \begin{subfigure}[t]{.8\textwidth}
        \centering      
            \begin{tikzpicture}
            \tikzmath{\x =3;}
            \node (E) at (\x,\x) {$e$};
            \node (F) at (\x,-\x) {$f$};
            \node (C) at (-\x,\x) {$c$};
            \node (D) at (-\x,-\x) {$d$};
            \node (A) at (-3*\x,\x) {$a $};
            \node (B) at (-3*\x,-\x) {$b$};
            \draw (E) -- (0,0);
            \draw (F) -- (0,0);
            \draw(0,0) -- (A);
            \draw (0,0) -- (B);
            \draw (0,0) -- (C);
            \draw (0,0) -- (D);
            \fill[pattern color=gray!70,pattern=north east lines] (A) -- (0,0) -- (B);
            \node at (2.5,0) {\scalebox{.5}{\begin{tikzpicture}
                \fill (1,1) circle (3pt);
                \fill (1,-1) circle (3pt);
                \draw[dashed] (1,1) -- (2,2);
                \draw[dashed] (1,-1) -- (2,-2);
                \draw[very thick,->] (1,1.6) -- (1.5,2.1);
                \draw[very thick,->] (1,-1.6) -- (1.5,-2.1);
                \end{tikzpicture}}};
            \node at (0,-2.5) {\scalebox{.5}{\begin{tikzpicture}
                \fill (-1,-1) circle (3pt);
                \fill (1,-1) circle (3pt);
                \draw[dashed] (-1,-1) -- (2,2);
                \draw[dashed] (1,-1) -- (2,-2);
                \draw[very thick,->] (-1.2,-1.2) -- (-1.7,-1.7);
                \draw[very thick,->] (1,-1.6) -- (1.5,-2.1);
                \end{tikzpicture}}};
            \node at (-1.7*\x,-3.1) {\scalebox{.5}{\begin{tikzpicture}
                \fill (-1,-1) circle (3pt);
                \fill (-3,1) circle (3pt);
                \draw[dashed] (-1,-1) -- (2,2);
                \draw[dashed] (-3,1) -- (0,0) -- (2,-2);
                \draw[very thick,->] (-1.2,-1.2) -- (-1.7,-1.7);
                \draw[very thick,->] (-2.6,.6) -- (-3.5,.9);
                \end{tikzpicture}}};
            \node at (0,2.5) {\scalebox{.5}{\begin{tikzpicture}
                \fill (1,1) circle (3pt);
                \fill (-1,1) circle (3pt);
                \draw[dashed] (1,1) -- (2,2);
                \draw[dashed] (-1,1) -- (2,-2);
                \draw[very thick,->] (1,1.6) -- (1.5,2.1);
                \draw[very thick,->] (-1.2,1.2) -- (-1.7,1.7);
                \end{tikzpicture}}};
            \node at (-1.7*\x,3.1) {\scalebox{.5}{\begin{tikzpicture}
                \fill (-1,1) circle (3pt);
                \fill (-3,-1) circle (3pt);
                \draw[dashed] (-1,1) -- (2,-2);
                \draw[dashed] (-3,-1) -- (0,0) -- (2,2);
                \draw[very thick,->] (-1.2,1.2) -- (-1.7,1.7);
                \draw[very thick,->] (-2.6,-.6) -- (-3.5,-.9);
                \end{tikzpicture}}};
            \end{tikzpicture}
        \caption{}
        \label{fig:7brane_crossing_tame:fan}
    \end{subfigure}
    \caption{The consistent diagram (a) for two $7$-branes crossing with Dirac pairing 2. There are $6$ variables parametrizing the deformations. They are homogeneous coordinates associated with the rays in the toric fan (b). The crossed out area is not part of the fan. The toric points correspond to ways of sending the $7$-branes to infinity.}
    \label{fig:7brane_crossing_tame}
\end{figure}
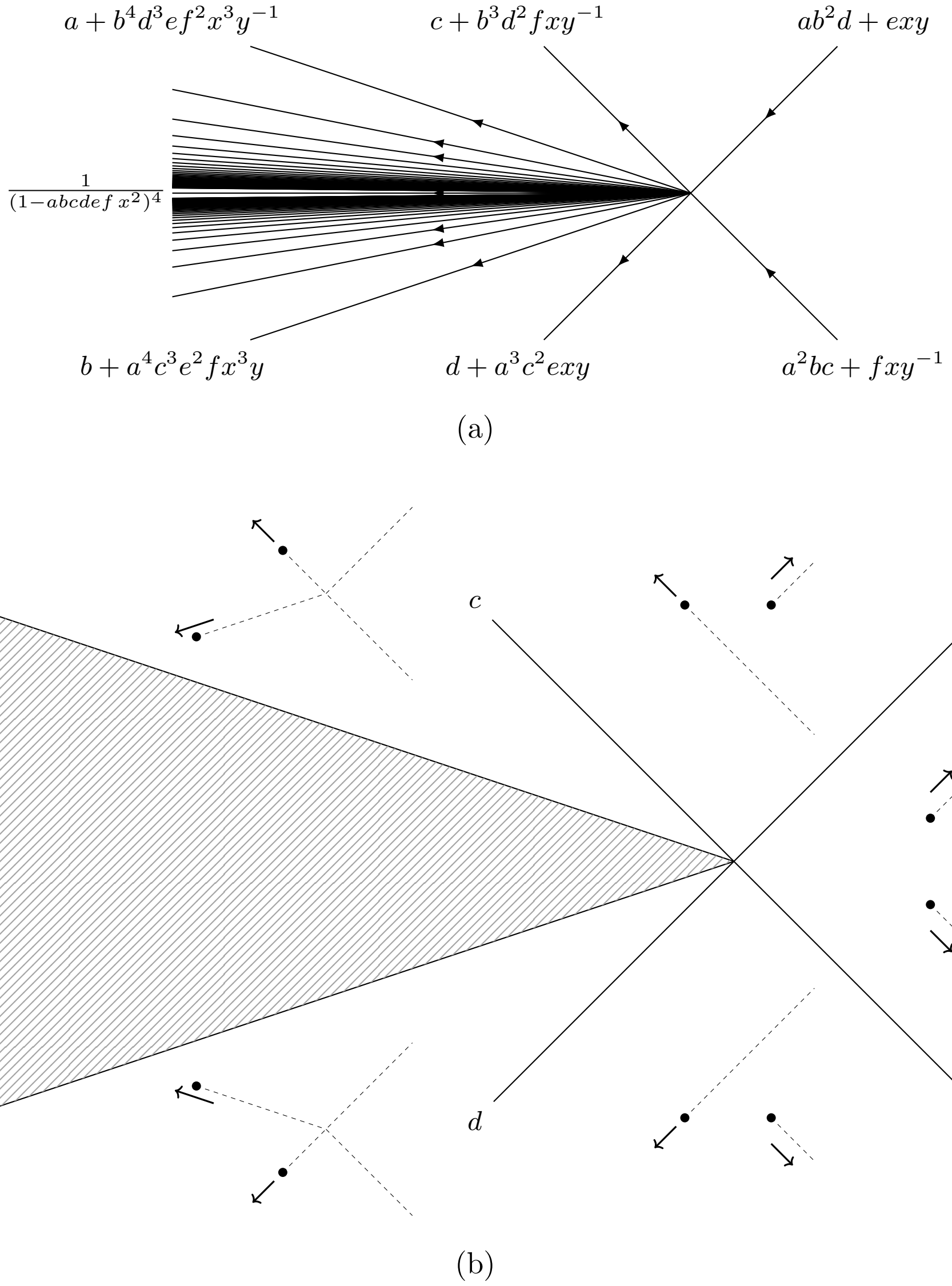

We now consider a tame case, with two $7$-branes of charges $(1,1)$ and $(1,-1)$. The consistent scattering diagram is shown in figure \ref{fig:7brane_crossing_tame:scattering_diag}. Six variables are now needed, four of them introduced on kinks and the last two as part of the incoming walls. These are homogeneous coordinates for the non-compact toric surface constructed from the fan shown in figure \ref{fig:7brane_crossing_tame:fan}.

The quiver is the Kronecker quiver:
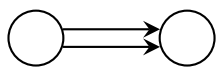
\begin{equation}
\begin{tikzpicture}
    \node[circle,draw] (A) at (0,0) {};
    \node[circle,draw] (B) at (1,0) {};    
    \draw[-stealth] (A.18) -- (B.162);
    \draw[-stealth] (A.-18) -- (B.198);
\end{tikzpicture}
\end{equation}

\paragraph{Wild case}

\begin{figure}
    \centering
    \begin{tikzpicture}[decoration={markings,mark=at position 0.5 with {\arrow{latex}}}] 
            \tikzmath{\x = 2;}
            \node (E) at (\x,3*\x) {$ab^3d + exy^3$};
            \node (F) at (\x,0) {$a^3bc + fx$};
            \node (C) at (-3*\x,0) {$c + b^8d^3fx$};
            \node (D) at (-\x,-3*\x) {};
            \node[below right=-10pt and -10pt of D] {$ d + a^8c^3exy^3$};
            \node (A) at (-3*\x,-9*\x/4) {};
            \node[below left=-10pt and -7pt of A] {$a + b^{21}d^8ef^3x^4y^3 $};
            \node (B) at (-4*\x/3,-3*\x) {};
            \node[below left=-5pt and -10pt of B] {$b + a^{21}c^8e^3fx^4y^9$};
            \draw[postaction={decorate}] (E) -- (0,0);
            \draw[postaction={decorate}] (F) -- (0,0);
            \draw[postaction={decorate}] (0,0) -- (A);
            \draw[postaction={decorate}] (0,0) -- (B);
            \draw[postaction={decorate}] (0,0) -- (C);
            \draw[postaction={decorate}] (0,0) -- (D);
            
                      
            \fill[pattern color=black,pattern=north west lines] (0,0) -- (-3*\x,-2.48754*\x) -- (-3*\x,-3*\x)-- (-1.38197*\x,-3*\x); 
    \end{tikzpicture}
    \caption{The consistent diagram for two $7$-branes crossing with Dirac pairing 3. The diagram is dense in the crossed out area \cite{grafnitz:2023sdp}. There are $6$ variables parametrizing the deformations. They are homogeneous coordinates associated with the rays in a toric fan similar with the one in figure \ref{fig:7brane_crossing_tame}.}
    \label{fig:7brane_crossing_wild}
\end{figure}
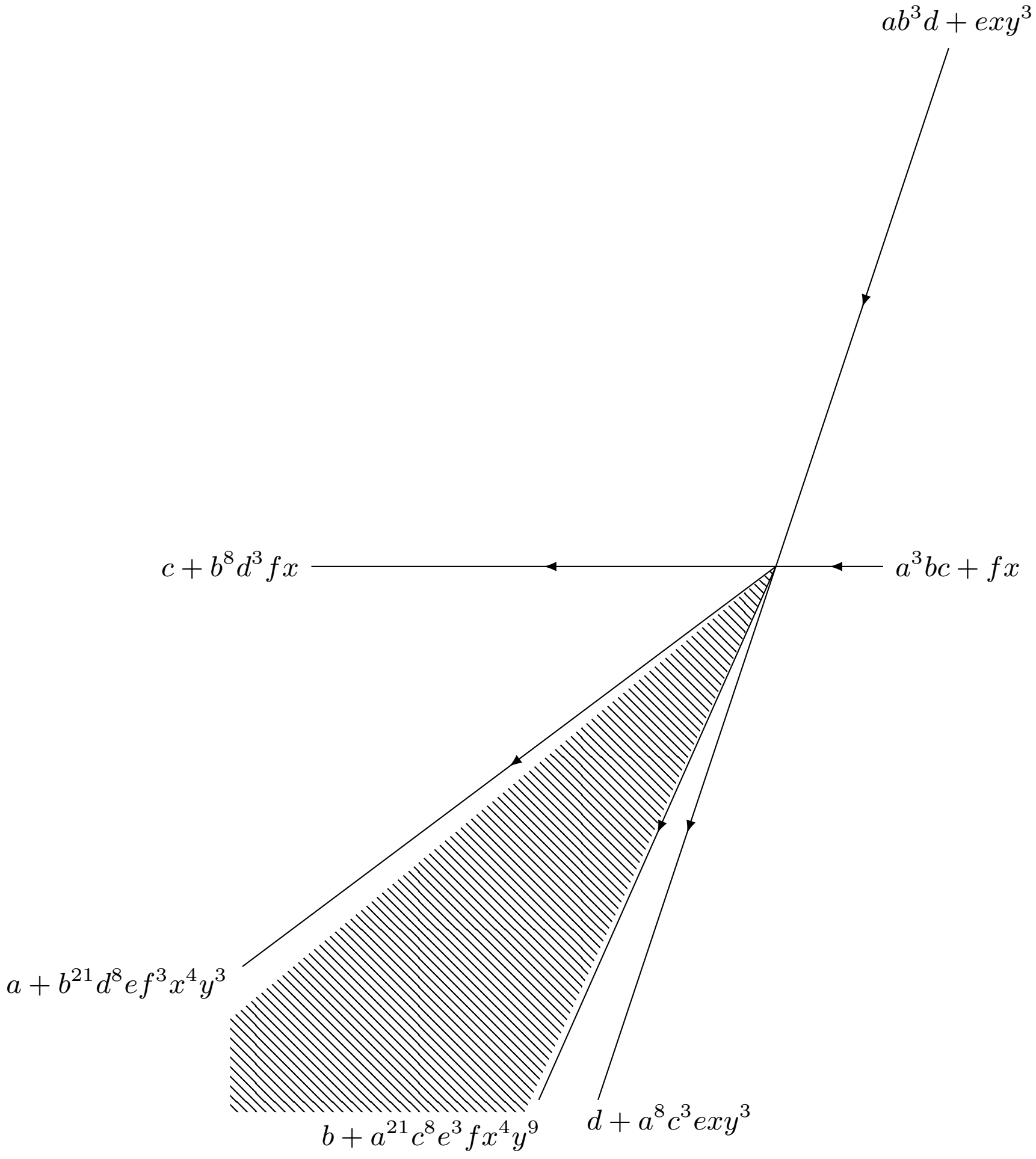

We now consider a wild case, with two $7$-branes of charges $(1,0)$ and $(1,3)$. The consistent scattering diagram is shown in figure \ref{fig:7brane_crossing_wild}. Six homogeneous variables are now needed, four of them introduced on kinks and the last two as part of the incoming walls. The quiver is the three arrow Kronecker quiver:

\begin{equation}
    \begin{tikzpicture}
        \node[circle,draw] (A) at (0,0) {};
        \node[circle,draw] (B) at (1,0) {};
        \draw[-stealth] (A) -- (B);
        \draw[-stealth] (A.20) -- (B.160);
        \draw[-stealth] (A.-20) -- (B.200);
    \end{tikzpicture}
\end{equation}

\subsection{More than two $7$-branes.}

In general, when $n$ $7$-branes are present, the parameter space is a toric $n$-fold. It can be compact in exceptional cases, but is non-compact generically. 

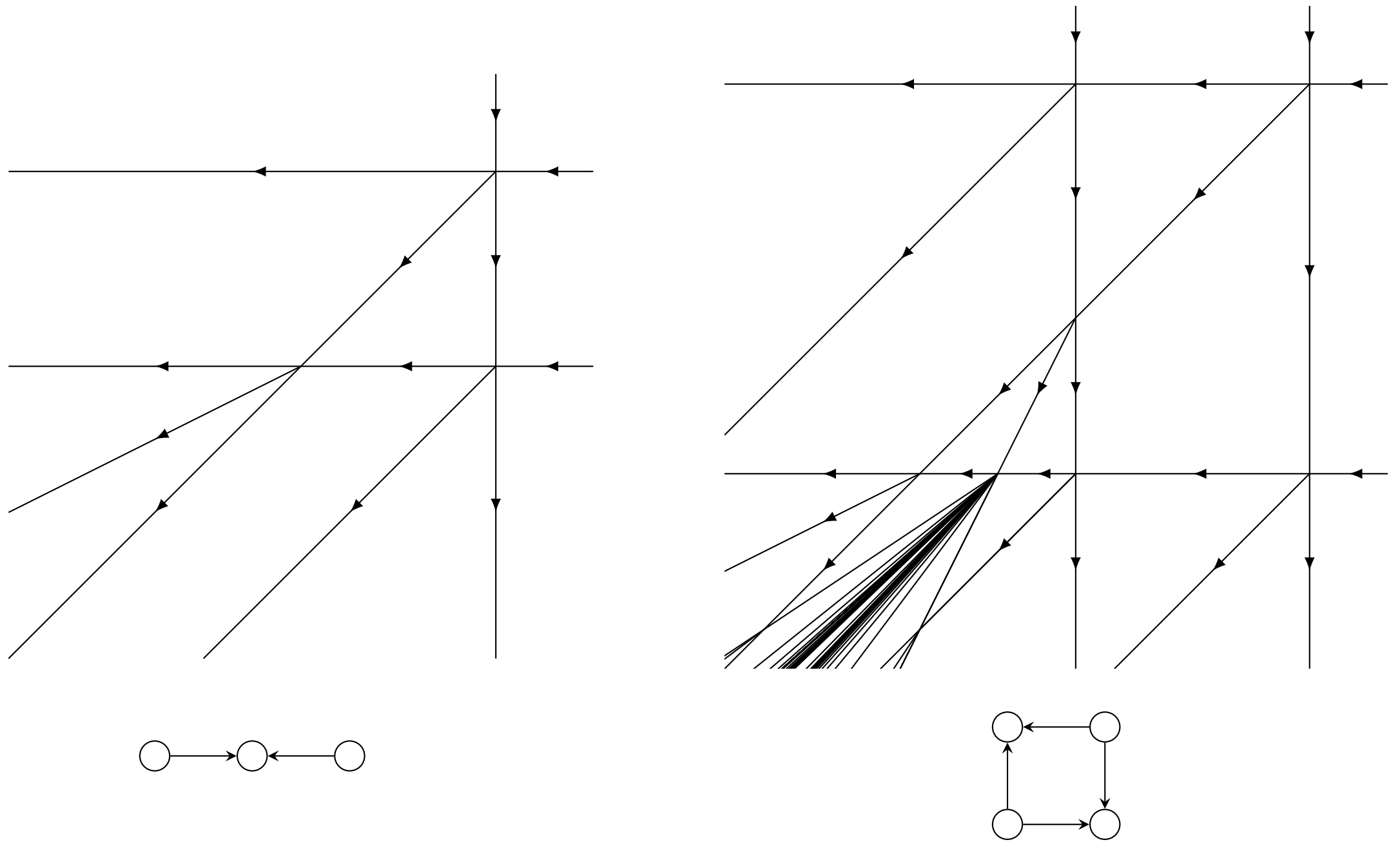
\begin{figure}
\begin{subfigure}{.495\textwidth}
    \centering
    \raisebox{20pt}{
    \begin{tikzpicture}[decoration={markings,mark=at position 0.5 with {\arrow{latex}}}] 
        \draw[postaction={decorate}] (1,0) -- (0,0);
        \draw[postaction={decorate}] (0,1) -- (0,0); 
        \draw[postaction={decorate}] (0,0) -- (0,-2);
        \draw[postaction={decorate}] (1,-2) -- (0,-2);
        \draw[postaction={decorate}] (0,-2) -- (0,-5);
        \draw[postaction={decorate}] (0,0) -- (-2,-2);
        \draw[postaction={decorate}] (0,-2) -- (-3,-5);
        \draw[postaction={decorate}] (-2,-2) -- (-5,-7/2);
        \draw[postaction={decorate}] (0,-2) -- (-2,-2);
        \draw[postaction={decorate}] (-2,-2) -- (-5,-5);
        \draw[postaction={decorate}] (-2,-2) -- (-5,-2);
        \draw[postaction={decorate}] (0,0) -- (-5,0);
    
        \node at (-2.5,-6) {
            \begin{tikzpicture}
                \node[circle,draw] (A) at (0,0) {};
                \node[circle,draw] (B) at (1,0) {};    \node[circle,draw] (C) at (2,0) {};
                \draw[-stealth] (A) -- (B);
                \draw[-stealth] (C) -- (B);    
            \end{tikzpicture}};
    \end{tikzpicture}}
\caption{}\label{fig:3_7branes_finite}
\end{subfigure}    
\begin{subfigure}{.495\textwidth}
\begin{tikzpicture}
    \node at (0,0) {
        \begin{tikzpicture}[scale=.8,
            line cap=round,
            line join=round,decoration={markings,mark=at position 0.5 with {\arrow{latex}}}]  
            \clip (-7.5,-7.5) rectangle (1,1);
            
            \draw[postaction={decorate}] (0,1) -- (0,0); 
            \draw[postaction={decorate}] (0,0) -- (0,-5);
            \draw[postaction={decorate}] (0,-5) -- (0,-7.5);
            \draw[postaction={decorate}] (1,0) -- (0,0);
            \draw[postaction={decorate}] (0,0) -- (-3,0);
            \draw[postaction={decorate}] (-3,0) -- (-7.5,0);
            \draw[postaction={decorate}] (-3,1) -- (-3,0);
            \draw[postaction={decorate}] (-3,0) -- (-3,-3);
            \draw[postaction={decorate}] (-3,-3) -- (-3,-5);
            \draw[postaction={decorate}] (-3,-5) -- (-3,-7.5);
            \draw[postaction={decorate}] (1,-5) -- (0,-5);
            \draw[postaction={decorate}] (0,-5) -- (-3,-5);
            \draw[postaction={decorate}] (-3,-5) -- (-4,-5);
            \draw[postaction={decorate}] (-4,-5) -- (-5,-5);
            \draw[postaction={decorate}] (-5,-5) -- (-7.5,-5);
            \draw[postaction={decorate}] (0,0) -- (-3,-3);
            \draw[postaction={decorate}] (-3,-3) -- (-5,-5);
            \draw[postaction={decorate}] (-5,-5) -- (-7.5,-7.5);
            \draw[postaction={decorate}] (-3,0) -- (-7.5,-4.5);
            \draw[postaction={decorate}] (0,-5) -- (-2.5, -7.5);
            \draw[postaction={decorate}] (-3,-3) -- (-4,-5);
            \draw[postaction={decorate}] (-3,-5) -- (-5,-7);
            \draw[postaction={decorate}] (-5,-5) -- (-7.5,-6.25);
            \newcommand{\Ray}[4]{%
            \draw (#1,#2) --
                  ($(#1,#2)+10*(#3,#4)$);
            }
            \Ray{-3}{-5}{-1}{-1}
            \Ray{-8}{-5}{-2}{-1}
            \Ray{-5}{-7}{-2}{-3}
            \Ray{-3}{-8}{-1}{-2}
            \Ray{-7}{-7}{-4}{-3}
            \Ray{-4}{-5}{-1}{-2}
            
            \foreach \k in {1,...,25}
            {
                \pgfmathtruncatemacro{\dx}{-2*\k+1}
                \pgfmathtruncatemacro{\dy}{-2*\k}
                \Ray{-4}{-5}{\dx}{\dy}
            }
            \foreach \k in {1,...,25}
            {
                \pgfmathtruncatemacro{\dx}{-2*\k-1}
                \pgfmathtruncatemacro{\dy}{-2*\k}
                \Ray{-4}{-5}{\dx}{\dy}
            }
        \end{tikzpicture}};
    \node at (-0,-4.5) {
        \begin{tikzpicture}
            \node[circle,draw] (A) at (0,0) {};
            \node[circle,draw] (D) at (1,0) {};    \node[circle,draw] (C) at (0,1) {};    \node[circle,draw] (B) at (1,1) {};
            \draw[-stealth] (A) -- (C);
            \draw[-stealth] (A) -- (D);
            \draw[-stealth] (B) -- (C);
            \draw[-stealth] (B) -- (D);    
        \end{tikzpicture}};
\end{tikzpicture}

\caption{}\label{fig:3_7branes_tame}
\end{subfigure}
    \caption{Scattering diagrams describing the crossing of two $(1,0)$ $7$-branes and one $(0,1)$ $7$-brane (resp. two $(0,1)$ $7$-branes).}
    \label{fig:3and4_7branes}
\end{figure}

With two 7-branes of charge $(1,0)$ and one of charge $(0,1)$ (see Figure \ref{fig:3_7branes_finite}), the quiver is of finite Dynkin type $A_3$ and we see that the consistent scattering diagram is finite.

With two 7-branes of charge $(1,0)$ and two of charge $(0,1)$ (see Figure \ref{fig:3_7branes_tame}), the quiver is of affine Dynkin type $\tilde A_3$ and we see that the consistent scattering diagram has accumulation points of walls.

Other finite configurations of $7$-branes charges include:
\begin{itemize}
    \item $\{(1,0), (0,1),(-1,-1)\}$. The quiver is cyclic, the affine Dynkin diagram $\tilde A_2$ and mutation equivalent to a linear quiver, of finite Dynkin type $A_3$.
    \item $\{(1,0), 3\times (0,1)\}$. The quiver is of finite Dynkin type $D_4$. 
\end{itemize}

\newpage

\appendix

\section{The Consistent Diagram Algorithm}
\label{app:algoConsistent}

The construction is a modification of the degree-by-degree algorithm in Theorem~1.4 of \cite{Gross_2010}. We assign degree zero to $x,y,t_0$ and degree one to each of $t_1,\ldots,t_r$. Choose a reference ray which is not the support of a wall. We write the total wall-crossing automorphism as $\Phi=\Phi_1\circ\cdots\circ\Phi_s$ where the factors $\Phi_1 , \dots, \Phi_s$ are listed counterclockwise from the reference ray. At any finite degree, only finitely many pure walls contribute to this product.

\begin{enumerate}
    \item \textbf{Initialization.} Set $\mathcal D_0=\mathcal D_{\mathrm{in}}$. Modulo $I$, every incoming pure wall has function $1$, so only the kinks contribute. Their combined automorphism is the identity, as a consequence of the charge conservation in the brane web. Thus the total automorphism of $\mathcal D_0$ satisfies $\Phi_0\equiv\operatorname{Id}\pmod I$.

    \item \textbf{Compute the next obstruction.} Suppose that $\mathcal D_{k-1}$ has been constructed and
    \begin{equation}
        \Phi_{k-1}\equiv\operatorname{Id}\pmod{I^k}.
    \end{equation}
    For a primitive vector $u=(p,q)$, define the derivation
    \begin{equation}
        \partial_u(x^a y^b)=(pb-qa)x^a y^b.
    \end{equation}
    The degree-$k$ part of the remaining automorphism can be written uniquely as
    \begin{equation}
        \Phi_{k-1}
        =\operatorname{Id}
        +\sum_{u\in\mathbb Z^2_{\mathrm{pr}}}h_u\partial_u
        \pmod{I^{k+1}},
        \qquad
        h_u=\sum_{\ell\geq1}c_{u,\ell}x^{-\ell p}y^{-\ell q} \, , 
    \end{equation}
    where both sums are finite.   Each coefficient $c_{u,\ell}$ is a polynomial in $t_1,\ldots,t_r$ with coefficients in $\mathbb Z[t_0^{\pm1}]$, homogeneous of degree $k$ in $t_1,\ldots,t_r$.  

    \item \textbf{Transport each correction to its support.}   Add the outgoing wall with support $\mathbb R_{\geq0}u$ and function
    \begin{equation}
        f_{u,k} =1-\sum_{\ell\geq1}c_{u,\ell}t_0^{\ell\kappa_u} x^{-\ell p}y^{-\ell q}. 
    \end{equation}
    where for each $u$ with $h_u\neq0$,  
    \begin{equation}
        \kappa_u=
        \sum_{\substack{\text{kinks }(\mathbb R_{\geq0}v_a,t_0^{n_a})\\
                        \text{encountered before }u}}
            n_a\det(v_a,u).
    \end{equation}  
    \item \textbf{Update and repeat.} Add all these corrections to obtain $\mathcal D_k$, merging walls with identical support by multiplying their functions. Retain the original wall functions; computations at this step need only be performed modulo $I^{k+1}$. The new total automorphism satisfies
    \begin{equation}
        \Phi_k\equiv\operatorname{Id}\pmod{I^{k+1}}.
    \end{equation}
    Continue with degree $k+1$.
\end{enumerate}

Taking the limit gives a consistent diagram $\mathcal D=\lim_{k\to\infty}\mathcal D_k$. The construction may require infinitely many walls, but determines the diagram to any prescribed finite degree in finitely many steps. At each degree the outgoing pure corrections are forced, giving uniqueness up to equivalence.

\section{The Hirzebruch-Jung algorithm} \label{sec:: HJalgo}
The Hirzebruch-Jung (HJ) algorithm can be applied to extend the theta functions associated to walls $\vartheta_{n,m}$ in a scattering diagram to a generating set that saturates the dual lattice. We will follow Section 10.2 of \cite{cox2011toric}. We first order the walls in clockwise order and pick two consecutive walls whose theta functions are $\vartheta_{n_1,m_1}$ and $\vartheta_{n_2,m_2}$. These two directions define a cone which up to an $\text{SL}(2,\mathbb{Z})$ transformation is $\sigma=\langle (0,1),(d,-k)\rangle $ with $d>k>0$ and $\gcd(d,k)=1$.

We define the Hirzebruch-Jung continued fraction of the rational number $d/k$ as
\begin{equation}
    \frac{d}{k}=b_1-\frac{1}{b_2-\frac{1}{\cdots-\frac{1}{b_r}}}
\end{equation}
Notice the minus signs in contrast to ordinary continued fractions. There is a finite number of $b_i$. To lighten the notation, we introduce the representation $[[b_1,\cdots, b_r]]$. For example, the HJ continued fraction of 9/7 is $[[2,2,2,3]]$.

Let $[[b_1,\cdots, b_r]]$ be the HJ continued fraction representation of $d/k$. We split the cone $\sigma$ into subcones $\sigma_i=\langle u_{i-1},u_{i}\rangle$ with $1 \leq i \leq r+1$. We define the vectors $u_i$ such that 
\begin{equation}
    u_{i-1}+u_{i+1}= b_i\, u_i, \quad 1 \leq i \leq r ,
\end{equation}
where $u_0=(0,1)$ and $u_{r+1}=(d,-k)$. From the solution of this system of equations we find the Hilbert basis of the dual lattice. Applying this procedure to every pair of theta functions results in a saturated dual lattice.

\section{T-cones}\label{sec:Tcones}

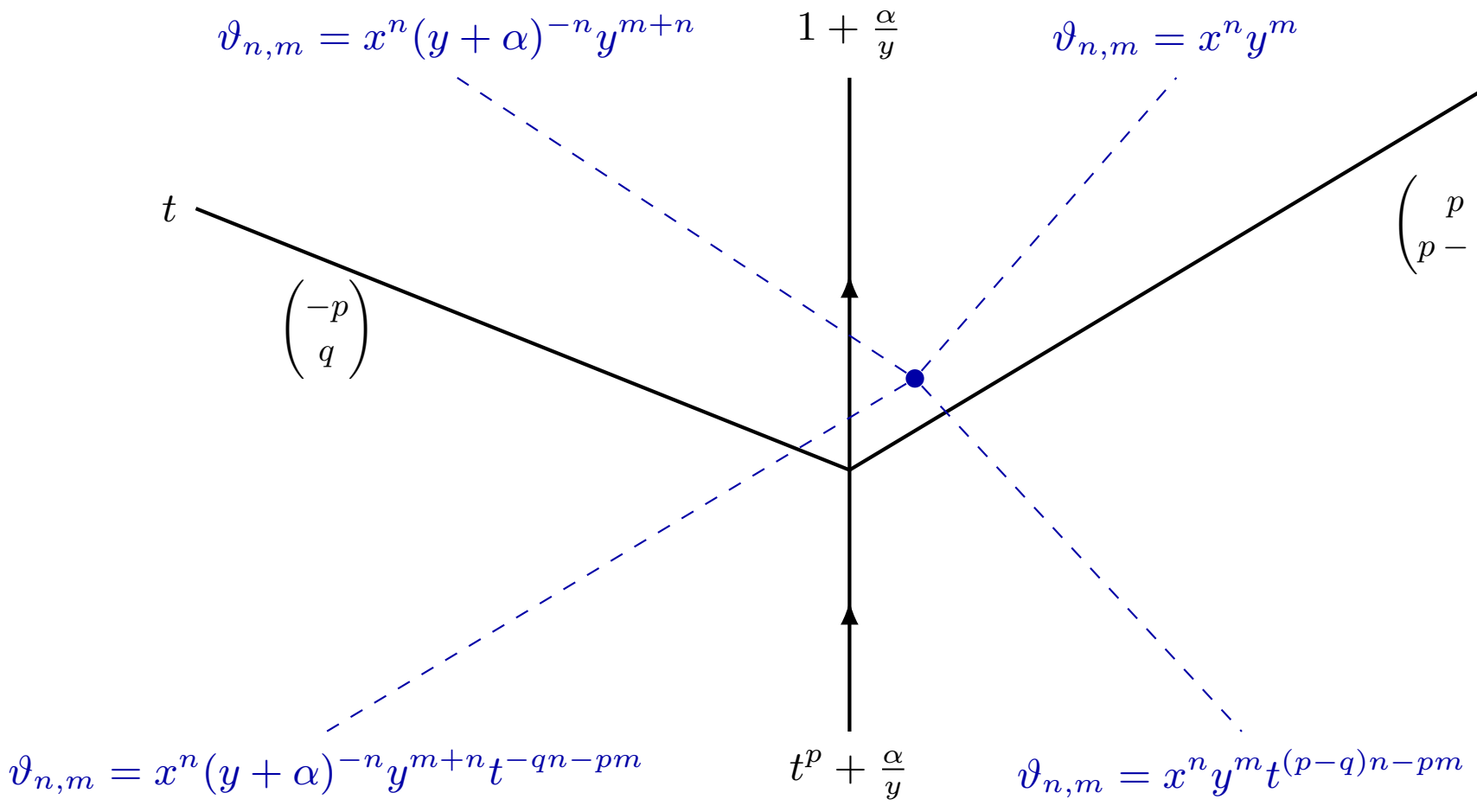
\begin{figure}[h!]
\centering
\begin{tikzpicture}
\draw[wall] (0,-2) to node[pos=0,below] {$t^p + \frac \alpha y$} (0,0);
\draw[wall] (0,0) to node[pos=1,above] {$1 + \frac\alpha y$} (0,3);
\draw[kink] (-5,2) -- node[pos=0,left] {$t$} node[pos=.2,below] {\scalebox{.8}{$\begin{pmatrix} -p  \\q\end{pmatrix}$}} (0,0) -- node[pos=1,right] {$t$}  node[pos=.8,below right] {\scalebox{.8}{$\begin{pmatrix} p  \\p-q\end{pmatrix}$}} (5,3);
\node[focalPoint,goodblue] (p) at (.5,.7) {};
\draw[broken,goodblue] (p) -- node[pos=1,above] {$\vartheta_{n,m}=x^n y^m$} (2.5,3);
\draw[broken,goodblue] (p) -- node[pos=1,above] {$\vartheta_{n,m}=x^n (y+\alpha)^{-n} y^{m+n}$} (-3,3);
\draw[broken,goodblue] (p) -- node[pos=1,below] {$\vartheta_{n,m}=x^n (y+\alpha)^{-n} y^{m+n} t^{-qn-pm}$} (-4,-2);
\draw[broken,goodblue] (p) -- node[pos=1,below] {$\vartheta_{n,m}=x^n y^m t^{(p-q)n-pm}$} (3,-2);
\end{tikzpicture}
\caption{Scattering diagram for an irreducible T-cone. The formulas given for the theta-functions hold in the corresponding quadrant defined by the walls of the diagram.}
\label{fig:Tcone_scatteringDiagram}
\end{figure}

Consider the irreducible T-cone shown in Figure~\ref{fig:Tcone_GTP}, with $p,q$ two integers satisfying $p\geq 2$, $1\leq q\leq p-1$ and $\operatorname{GCD}(p,q)=1$. The corresponding scattering diagram is shown in Figure \ref{fig:Tcone_scatteringDiagram}.
We show that the CY threefold constructed from this scattering diagram is a $\mathbb Z_p$-quotient of the product of $\mathbb C$ and a deformed $A_{p-1}$-singularity $uv = z^p+\alpha$, with the action \eqref{eq:Zp_action}.

In $\mathbb C[t^{1/p}, x^{\pm 1} ,y^{\pm 1/p}]$, we define:
\begin{subequations}
\begin{align}
    u &= x y^{\frac{p-q}{p}}\\
    v&= (y+\alpha)x^{-1} y^{-\frac{p-q}{p}} \\
    z &= y^{\frac{1}{p}}\\
    w &= t y^{-\frac{1}{p}}
\end{align}
\end{subequations}

These polynomials satisfy $uv=z^p + \alpha$, so the subalgebra they generate is the coordinate algebra of the product of $\mathbb C$ and a deformed $A_{p-1}$-singularity. All the generators of $\mathcal A_X$ (the theta-functions and $t$) are monomials in $u,v,w,z$. We will see that these monomials are invariant under a $\mathbb Z_p$ action and that they generate the ring of invariants.

First, we have $zw=t$ and $uv=y+\alpha$. Then, in the four quadrants delimited by the walls, we can express theta-functions as monomials in $(u,v,w,z)$, see Table \ref{tab:Tcone_thetafunctions}.
\begin{table}[h!]
    \centering
\begin{tabular}{c|c|c}
Quadrant & $\vartheta_{n,m}$ in terms of $(x,y,t)$ & $\vartheta_{n,m}$ in terms of $(u,v,w,z)$ \\\hline
 $\begin{array}{c}n\geq 0 \\pm - (p-q)n \geq 0\end{array}$
& $ x^ny^m $&$ u^n z^{pm - (p-q)n} $ \\\hline
 $\begin{array}{c}n\leq 0\\pm+  q n \geq 0\end{array}$
 & $x^n (y+\alpha)^{-n} y^{m+n} $&$v^{-n} z^{pm+qn}$\\\hline
 $\begin{array}{c}n\leq 0\\pm+  q n \leq 0\end{array}$
& $x^n (y+\alpha)^{-n} y^{m+n} t^{-qn-pm} $&$ v^{-n}w^{-qn-pm}$ \\\hline
 $\begin{array}{c}n\geq 0\\pm - (p-q)n \leq 0\end{array}$
& $x^n y^m t^{(p-q)n-pm} $&$ w^{(p-q)n-pm} u^n$ 
\end{tabular}
\caption{}
\label{tab:Tcone_thetafunctions}
\end{table}

Let $1\leq r\leq p-1$ be an integer such that $qr =1\mod p$. We define a $\mathbb Z_p$ action by 
\begin{equation}\label{eq:Zp_action}
    \zeta \cdot (u,v,w,z) = (\zeta u,\zeta^{-1} v,\zeta^{r} w,\zeta^{-r} z) \,.
\end{equation}
Then $uv,zw$ and the monomials in Table \ref{tab:Tcone_thetafunctions} are invariants. Let us show that these monomials generate the ring of invariants. A general monomial $u^av^bw^c z^d$ is invariant if $a - b+r(c-d) = 0\mod p$. Factoring out the highest powers of $uv$ and $zw$ possible, we can assume that $ab=0$ and $cd =0$. This leaves us with four cases: monomials in two variables that can be $(v,z)$, $(u,z)$,  $(u,w)$ or $(v,w)$.

If we take $a=0,c=0$, i.e. we consider a monomial in $(v,z)$, we have $rd=-b\mod p$ hence $d = -qb-pk$ for some integer $k$. The monomial is $v^b z^{-pk-qb}$, which is the one in the second line of Table \ref{tab:Tcone_thetafunctions}, with $-n=b$ and $m=-k$. The two inequalities defining the second quadrant in Figure~\ref{fig:Tcone_scatteringDiagram} are indeed satisfied. The other three cases work out similarly, and correspond to the other lines in Table \ref{tab:Tcone_thetafunctions}.

The theta-functions in the directions of the walls of the diagram are:
\begin{align}
\begin{split}
  \vartheta_{p,p-q} &=u^p \\
  \vartheta_{-p,q} &= v^p  \\
  \vartheta_{0,1} &= z^p \\
  \vartheta_{0,-1} &= w^p 
\end{split}
\end{align}

The open subsets associated with the walls of the diagram are the subsets where the corresponding coordinate is non-vanishing. Theta-functions near the corresponding walls provide natural coordinates. Explicitly, we get: 
\begin{itemize}
     \item The subset $u\neq 0$ is toric $ (\mathbb C)^2\times\mathbb C^*$, with coordinates  $wu^{-r},z u^{r}$ and $u^p\neq 0$.
     \item The subset $v\neq 0$ is toric $(\mathbb C)^2\times\mathbb C^*$, with coordinates  $wv^r, zv^{-r}$ and $v^p\neq 0$
     \item The subset  where $w\neq 0$ is non-toric, with coordinates $\tilde u = uw^{p-q},\tilde v = vw^q, t=wz$ and $\tilde w_+ = w^p\neq 0$, satisfying $\tilde u \tilde v = t^p + \alpha \tilde w_+$.
     \item The subset where $z\neq 0$ is non-toric, with coordinates $\hat u = uz^{q-p},\hat v= v z^{-q},t = zw$ and $\hat w_+ = z^{-p}\neq 0$, satisfying $\hat u \hat v = 1+\alpha \hat w_+$. 
\end{itemize}

From there, one can compute explicitly the transition functions between the different patches. 

\bibliographystyle{JHEP}
\bibliography{bibli.bib}

\end{document}